\ifdefined\pdfoutput
  \pdfoutput=1
\fi          
\documentclass{mem}
\usepackage{iftex}

\ifPDFTeX
  \usepackage[T1]{fontenc}
\fi

\usepackage{amsmath}
\usepackage{natbib}       
\usepackage{hyperref}

\providecommand{\eprint}[2][]{%
  \ifx\\#1\\%
    arXiv:\href{https://arxiv.org/abs/#2}{#2}%
  \else
    #1:\href{https://arxiv.org/abs/#2}{#2}%
  \fi
}

\usepackage{aas_macros}
\usepackage{balance}
\usepackage{flushend}
\usepackage[dvipsnames]{xcolor}
\usepackage[version=4]{mhchem}
\usepackage[draft]{changes}
\usepackage{pdfpages}
\usepackage[utf8]{inputenc}
\usepackage[figuresright]{rotating}
\usepackage[font=small,labelfont=bf]{caption}

\newcommand{\oversim}[2]{\protect{\mbox{\lower0.5ex\vbox{%
  \baselineskip=0pt\lineskip=0.2ex
  \ialign{$\mathsurround=0pt #1\hfil##\hfil$\crcr#2\crcr\sim\crcr}}}}}

\definechangesauthor[name={Ter}, color=orange]{Ter} 
\definechangesauthor[name={Don}, color=ForestGreen]{Don} 
\definechangesauthor[name={Stef}, color=Blue]{Stef} 
\definechangesauthor[name={Pavel}, color=red]{Pavel} 
\definechangesauthor[name={YN}, color=purple]{YN} 
\definechangesauthor[name={AMH}, color=Orchid]{AMH} 
\definechangesauthor[name={MCh}, color=teal]{MCh} 
\definechangesauthor[name={Glenn}, color=magenta]{Glenn} 
\definechangesauthor[name={Alex}, color=cyan]{Alex} 

\idline{1}{1}
\begin{document}

\def\teff{$T\rm_{eff }$}
\def\kms{$\mathrm {km s}^{-1}$}

\pagenumbering{roman}

\onecolumn

\begin{center}
FOREWORD
\end{center}

The \textit{Cosmic Threads: Interlinking the Stellar Initial Mass Function from Star-Birth to Galaxies} workshop was held in the week of March 11–15, 2024, in Sexten, Italy, and brought together experts from various fields to explore the complexity of the Stellar Initial Mass Function (IMF) across different astrophysical scales. The IMF plays a fundamental role in shaping the evolution of stellar populations, star clusters, and galaxies, yet the physical processes underlying its shape and variation remain open questions.

One of the major challenges in studying the IMF is the diversity of approaches and methodologies used across different research communities. Observers, theorists, and computational modelers work with distinct datasets and tools, often leading to fragmented interpretations of the same underlying processes. In particular, research spanning spatial scales from individual star-forming regions to entire galaxies requires cross-disciplinary collaboration, but effective communication is often hindered by differences in terminology and methodology. 

This White Paper is a direct outcome of the discussions at the workshop, aiming to consolidate key findings and highlight open questions in IMF research. By compiling contributions from experts working across different scales and techniques, we seek to provide a unified perspective on the complexity of star formation and the IMF. Furthermore, we emphasize the importance of fostering smaller, focused meetings like this one, which enable researchers to bridge disciplinary gaps and work towards a coherent understanding of the IMF. 

The following sections summarize the insights gathered during the workshop, addressing both the latest advances and the persistent challenges in IMF studies, and outlining future directions for collaborative research. Although we have made every effort to include a diverse range of perspectives and opinions from our colleagues, we acknowledge that the breadth of views in the field is vast, and it is possible that some important voices and insights may be absent. We sincerely apologize for any omissions.

\begin{sidewaysfigure}
\vspace{13.6cm}
\makebox[\textwidth][c]{\includegraphics[width=0.9\textwidth]{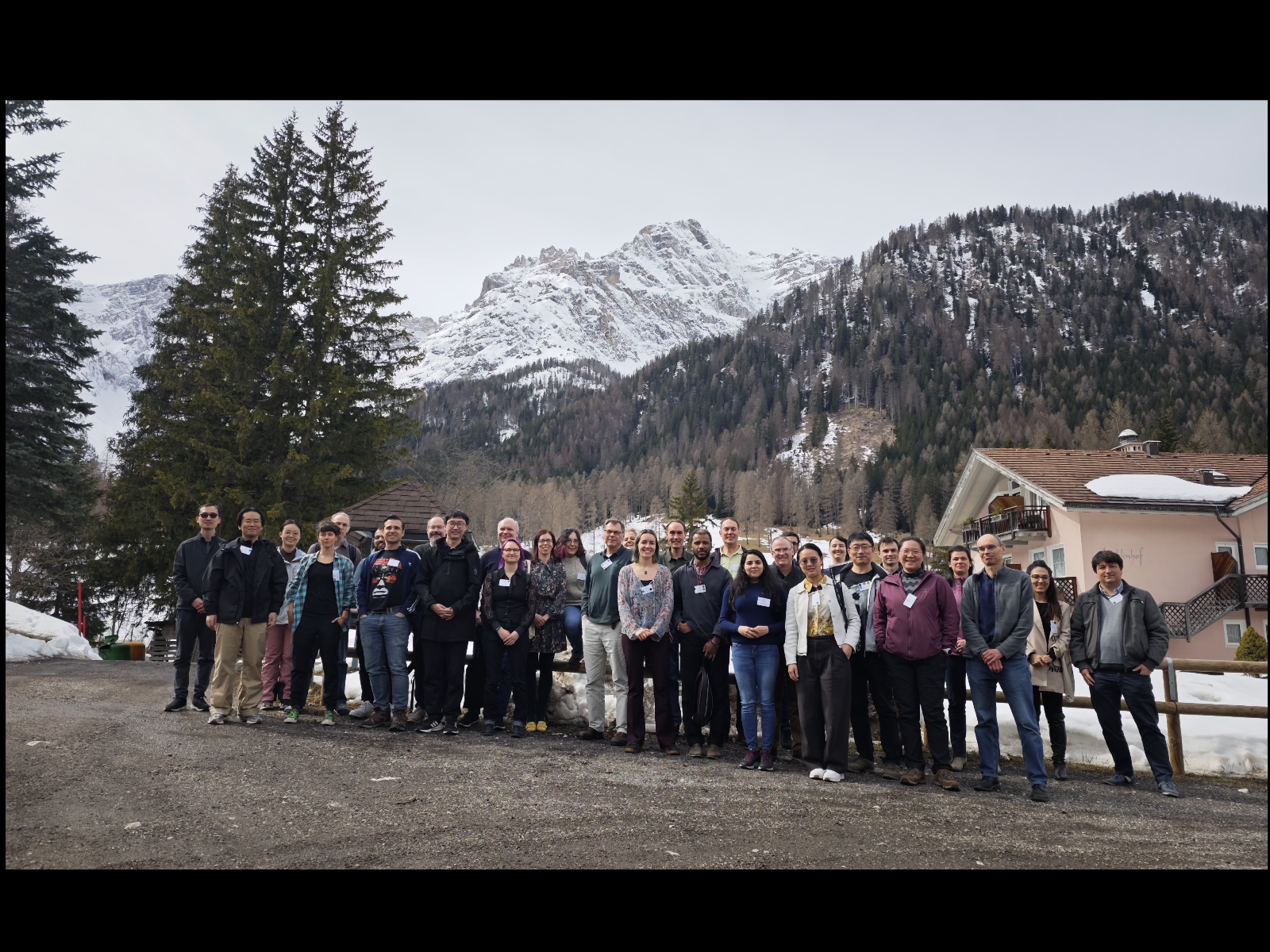}}%
\end{sidewaysfigure}

\begin{sidewaysfigure}
\vspace{13.6cm}
\makebox[\textwidth][c]{\includegraphics[width=0.95\textwidth]{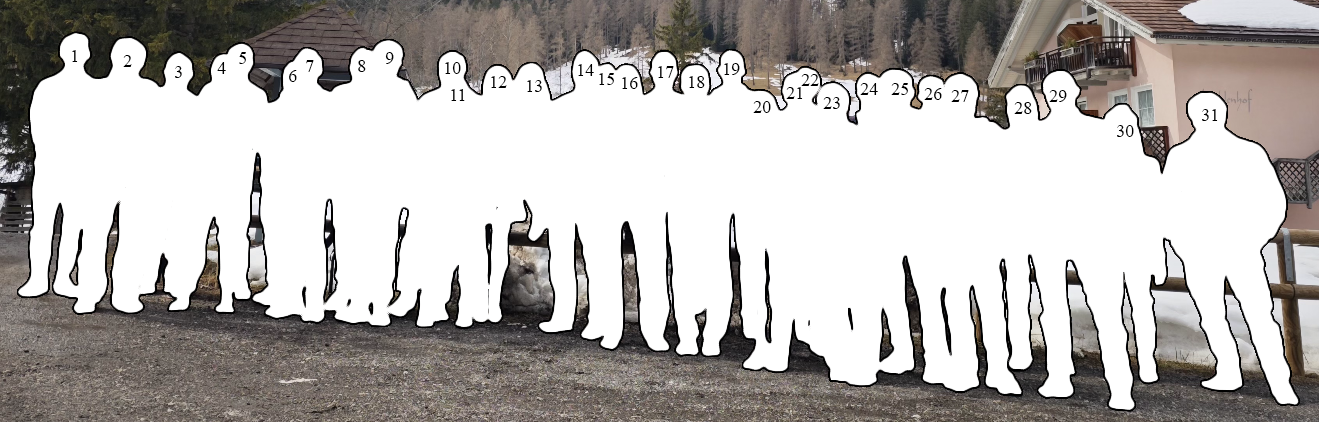}}%
\captionsetup{labelformat=empty}
\caption{ Participants in the \textit{Cosmic Threads: Interlinking the Stellar Initial Mass Function from Star-Birth to Galaxies} workshop. 1. Zhiqiang Yan (Nanjing University, CN); 2. Jiadong Li (MPIA Heidelberg, DE); 3. Ziyi Guo (Nanjing University, CN); 4. Martyna Chru{\'s}li{\'n}ska (MPA Garching, DE); 5. Joop Schaye (Leiden University, NL); 6. Piyush Sharda (Leiden University, NL); 7. Mike Grudi{\'c} (Carnegie Observatories, USA); 8. Morten Andersen (ESO, DE); 9. Long Wang (Sun Yat-sen University, CN); 10. Glenn van de Ven (University of Vienna, AT); 11. Natalia Lah{\'e}n (MPA, DE); 12. Donatella Romano (INAF Bologna, IT); 13. Elena Sabbi (STScI, USA); 14. Pavel Kroupa (Bonn University, DE); 15. Stefania Salvadori (Florence University, IT); 16. Anna Durrant (LJMU, UK); 17. Matthew Bate (Exeter University, UK); 18. Fabien Louvet (CNRS, FR); 19. Guido De Marchi (ESA, NL); 20. Beena Meena (STScI, USA); 21. Andrew Hopkins (Macquarie University, AU); 22. Ignacio Mart{\'i}n-Navarro (IAC, ES); 23. Yueh-Ning Lee (Taiwan University, TW); 24. Tereza Je{\v r}{\'a}bkov{\'a} (ESO, DE); 25. Zhi-Yu Zhang (Nanjing University, CN); 26. Alex Cameron (Oxford University, UK); 27. Xiaoting Fu (PMO Nanjing, CN); 28. Mariya Lyubenova (ESO, DE); 29. Philippe Andr{\'e} (CEA, FR); 30. Alice Concas (ESO, DE); 31. Fabio Fontanot (INAF Trieste, IT). Missing from the picture: Vikrant Jadhav (Bonn University, DE); Francesca Matteucci (Trieste University, IT); Alessio Mucciarelli (Bologna University, IT); Russell Smith (Durham University, UK); Alexandre Vazdekis (IAC, ES).
}
\label{fig:participants}
\end{sidewaysfigure}
\addtocounter{figure}{-1}

\twocolumn

\cleardoublepage \pagenumbering{arabic}

\thispagestyle{empty}

\title{
Cosmic Threads:
}

   \subtitle{Interlinking the Stellar Initial Mass Function from Star-Birth to Galaxies}

\author{Tereza Jerabkova\inst{1}
\and Donatella Romano\inst{2}
\and Pavel Kroupa\inst{3, 4}
\and Philippe Andr\'e\inst{5}
\and Martyna Chru{\'s}li{\'n}ska\inst{6}
\and Fabio Fontanot\inst{7, 8}
\and Andrew Hopkins\inst{9}
\and Vikrant Jadhav\inst{3,10}
\and Natalia Lah\'en\inst{11}
\and Yueh-Ning Lee\inst{12, 13, 14}
\and Alessio Mucciarelli\inst{15, 2}
\and Stefania Salvadori\inst{16, 17}
\and Long Wang\inst{18, 19}
\and Zhiqiang Yan\inst{20, 21}
\and Morten Andersen\inst{6}
\and Anna Durrant\inst{22}
\and Fabien Louvet\inst{23, 24}
\and Mariya Lyubenova\inst{6}
\and Francesca Matteucci\inst{25, 7}
\and Piyush Sharda\inst{26}
\and Glenn van de Ven\inst{27}
\and Alexandre Vazdekis\inst{28}
          }

\institute{
Department of Theoretical Physics and Astrophysics, Faculty of Science, Masaryk University, Kotlá\v{r}ská 2, Brno 611 37, Czech Republic\\
\email{tereza.jerabkova@sci.muni.cz}\\\and
National Institute for Astrophysics, Astrophysics and Space Science Observatory, Via Gobetti 93/3, 40129 Bologna, Italy\\
\email{donatella.romano@inaf.it}\\
\and
Helmholtz-Institut für Strahlen- und Kernphysik, Universität Bonn, Nussallee 14-16, 53115 Bonn, Germany\\
\and
Astronomical Institute, Faculty of Mathematics and Physics, Charles University, V Holešovickách 2, 180\,00 Praha 8, Czech Republic\\
\and
Laboratoire d’Astrophysique (AIM), Université Paris-Saclay, Université Paris Cité, CEA, CNRS, AIM, 91191 Gif-sur-Yvette, France\\
The remaining affiliations can be found at the end of the paper.
}

\authorrunning{Cosmic threads}

\titlerunning{Interlinking the stellar IMF from star-birth to galaxies }

\date{Received: Day Month Year; Accepted: Day Month Year}

\abstract{
The stellar initial mass function (sIMF) describes the distribution of stellar masses formed in a single star formation event in a molecular cloud clump. It is fundamental to astrophysics and cosmology, shaping our understanding of unresolved stellar populations, galactic chemical enrichment and habitable zones, and black hole growth. This White Paper reviews studies on the core mass function, stellar multiplicity, and dynamical processes affecting sIMF determinations, as well as the link between star-forming clumps and the galaxy-wide IMF (gIMF). The evidence gleaned from observed systems for the dependency of the sIMF on the metallicity and density of the clump is portrayed.
We examine evidence from gravitational lensing, stellar and gas kinematics, and spectral diagnostics to assess environmental dependencies of the gIMF. Theoretical perspectives provide further insights into the sIMF’s variability. Beyond summarizing current knowledge, this work aims to establish a shared framework and define strategies for studying a variable IMF in the era of near-infrared integral-field spectroscopy, 30m-class telescopes and major space-based observatories.
}
\maketitle{}

\section{Introduction}
\label{sec:intro}

The stellar initial mass function (sIMF), $\xi(m)$, connects physical processes from subatomic to cluster of galaxies scales, bridging different branches of physics. In particular, its exploration has a profound impact on our understanding of cosmic matter/energy cycles, with implications for cosmology, astrophysics, and nuclear physics.

The stellar IMF represents the number of stars that form within the initial mass range $m$ to $m+{\rm d}m$ during a single star formation event, ${\rm d}N=\xi(m)\,{\rm d}m$. Hereby we need to be aware of the meaning of the IMF: Is the IMF deduced from star counts nearby to the Sun always the same as the stellar IMF of all stars formed together in one star-formation event? How does the latter relate to the IMF of all stars forming at a given time in a region of a galaxy or galaxy-wide? How correct is the assumption that the stellar IMF obtained from star counts nearby to the Sun is representative of the IMF of stars forming in a molecular cloud, or in a galaxy at high redshift? Does $\xi(m)$ vary with the physical conditions of the star-forming gas? Is it a probability density distribution function or an optimal distribution function?  The first would imply that two molecular cloud clumps with the exact same properties give rise to two different ensembles of stars in the embedded star cluster, while the latter implies they would produce the exact same sequence of stellar masses.

\subsection{Historical Overview}
\label{sec:hist_over}

This section is meant to underscore the crucial role of the IMF in astrophysics, from its foundational theoretical formulations to its modern interpretations that challenge traditional views to adapt to new data. Understanding the evolution of these concepts is vital for grasping the broader implications of the IMF for galaxy formation and evolution theories, marking a pivotal chapter in the study of cosmology.

Some of the concepts that are touched upon in the following paragraphs will be covered in greater detail in Sect.~\ref{sec:res_IMF}.

\subsubsection{Edwin Salpeter's foundational work}

Edwin Salpeter laid significant groundwork in this field with his seminal publication \citep{Salpeter1955} seventy years ago \citep{KroupaJerabkova2019}. His research connected quantum mechanics with cosmological phenomena, elucidating how microscopic quantum processes governing nucleosynthesis and energy production in stars manifest as macroscopic stellar properties, such as the rates of stellar births and deaths based on visible stars in our Galaxy. Salpeter's {\it ``Original Mass Function"} expressed the {\it ``relative probability for the creation of stars of mass $m$"}. He ingeniously corrected the then available observed distribution of stellar luminosities of field stars in the solar vicinity for various contaminants, establishing a refined function that significantly contributed to our understanding of stellar evolution and mass distribution. Importantly, Salpeter was ahead of his time in correctly normalizing the IMF by volume and time interval, reflecting the star-formation timescale of a given stellar population --- a detail that, though now often overlooked, was crucial for accurately describing composite stellar populations.

Moving from star counts to an IMF is not trivial \citep[see, e.g., reviews by][]{Scalo1986,Kroupa2002,Chabrier2003,Kroupa+2013, Hopkins2018, Kroupa+2024}. It requires applying corrections for stellar multiplicity and evolutionary effects, making assumptions about the age and structure of the Galactic disc, some knowledge of the star formation rate (SFR) and its evolution with time, considering the diffusion of stellar orbits, and more. Salpeter’s (1955) original power-law approximation, commonly known as the ``Salpeter IMF",
\begin{equation}
   \xi_{\rm L}(m) \approx 0.03 \, (m/M_\odot)^{-x} \, ,
\label{eq:IMF_Salpeter}
\end{equation}
with $x$ = 1.35, is strictly only valid for solar-vicinity field stars with 0.4~$\le m/M_\odot \le$~10 (the range of masses accessible at the time). Note that ${\rm d}N=\xi_{\rm L}(m)\,{\rm d}\,{\rm log}\,m = \xi(m)\,{\rm d}m$ such that $\xi_{\rm L}(m) = \left(m\, {\rm ln}10\right)\,\xi(m)$ and $\alpha=1+x$ with $\xi(m) \propto m^{-\alpha}$. Also often used is the power-law notation $\Gamma = -x$ $=1+\gamma=1-\alpha$ (table~3 in \citealt{Kroupa+2013})\footnote{Truth be told, Salpeter did not use any letter to denote the exponent in his original formula, which simply has the power law value in it \citep[][equation~5]{Salpeter1955}. However, we do not intend to cite here the original work, but rather stick to notation that is widespread in the literature. In this respect, it is also worth pointing out that sometimes the negative sign is given in the power-law definition, but sometimes it appears in the parameter (see Sect.~\ref{sec:terminology}).}.

\subsubsection{Evolution of the IMF concept}

Some of Salpeter's assumptions were revised/refined over the following years, leading to the suggestion that the IMF slope is flatter in its extension to low masses \citep{Kroupa1990} and that a steeper slope, $x \approx$~1.4--1.7, could better describe field stars above 1~$M_\odot$ in the solar neighborhood \citep[][see Sect.~\ref{sec:res_IMF}]{Scalo1986,Kroupa+1993,Mor+2018,Sollima2019}.

In particular, the concept of a static IMF was challenged and refined by subsequent studies. Notable contributions by Miller and Scalo \citep{Miller1979,Scalo1986} included adjustments for main-sequence brightening and the evolving scale height of stellar discs. They introduced a more nuanced interpretation of the IMF, considering time-variable SFRs and enhanced stellar mass-luminosity data. 
The change in stellar structure below a stellar mass of $0.4\,M_\odot$ owing to the hydrogen molecule and the onset of full convection was shown by \cite{Kroupa1990} to significantly affect stellar luminosities and thus star counts.
Corrections for unresolved binaries further led to a major revision of the IMF below about 1~$M_\odot$. Studies of the impact of binary stars on IMF determinations were significantly advanced by the work of Kroupa and collaborators. \cite{Kroupa1991,Kroupa+1993} showed that binary systems play a critical role in the mass distribution of stars. Most recently it has become clear that the unresolved-multiplicity corrections that need to be applied to star counts are not universal and depend on the population of stars under scrutiny \citep{Kroupa2025}. These studies adjusted the previous assumptions about stellar mass and luminosity, providing a more accurate depiction of stellar evolution and interactions within binary systems. Further research by  Tout, Kroupa and colleagues \citep{Tout1996, KT1997, MansfieldKroupa2021, MansfieldKroupa2023}. expanded our understanding of the stellar mass-luminosity relationship, enhancing predictions about stellar lifetimes and evolutionary paths based on their initial masses (and metallicities). These contributions have been essential for refining theoretical models and improving our predictions of stellar behavior. Further advances were achieved by incorporating the effects of unresolved multiple stars and by correcting systematics due to biases such as the Lutz–Kelker and Malmquist biases, which refined the observed luminosity functions \citep[see][]{Kroupa+1993}. These enhancements allowed for a more accurate representation of the IMF across different stellar populations.

\subsubsection{Modern perspectives and ongoing debates}

Today, the discussion around the IMF has shifted towards its potential variability in dependence on factors like the metallicity and density of the environment \citep[e.g.,][]{chon2021,sharda2022,2024MNRAS.527.7306T}, as recent observations seem to suggest\footnote{Please, note that the term ``environment'' here does not refer to the one where galaxies reside. As a matter of fact, the gIMF of quiescent galaxies does not show any significant dependence on the large-scale environment or galactic hierarchy, leaving the gIMF as an intrinsic property of galaxies \citep{rosani2018,eftekhari2019}}.  Hereby it is important to be aware of possible apparent variation that emerges if the incorrect multiplicity correction or incorrect/poorly-known stellar mass--luminosity functions are applied when analysing star-count data. The ongoing debate also considers whether the IMF should be viewed as an invariant probability density distribution function of initial stellar masses or as a more dynamic, self-regulating distribution model \citep[namely, an optimally-sampled density distribution function; ][]{Kroupa+2013,Schulz2015,Yan2023}. On this issue, it is worth emphasizing that for shape-invariant sIMF, 
 if the sIMF is an invariant probability density distribution function then the galaxy-wide IMF is also one with the same form. But
if the stellar IMF is an optimally-sampled distribution function or regulated (e.g. massive star-formation is not stochastic), then the galaxy-wide IMF needs to be explicitly calculated and depends on the physical conditions in a galaxy. Last but not least, evidence that stars form in the cold dense cores of molecular clouds (MCs) suggests that determining the core mass spectrum is key to understanding the stellar IMF origin. However, assessing the correspondence between the core mass function and the stellar IMF remains a challenge.

We shall touch upon all these subjects in the following sections. But first, we need to review the naming conventions and spot nomenclature ambiguities in the way the IMF is discussed in the literature. By fixing the terminology that we shall use throughout this White Paper, we aim to eliminate potential ambiguities and, possibly, to pave the way to the adoption of a common language in the field.

\subsection{Terminology}
\label{sec:terminology}

There are many aspects of the concept referred to as ``the IMF'' which have evolved differing terminologies in the literature. Some of these are merely different conventions (such as the sign convention adopted for a power law slope, or the parameter names). Others are more insidious, as they can lead to confusion or to conflation between different quantities that should be kept distinct. Many of these details are discussed in the reviews by \cite{Kroupa+2013} and \citet{Hopkins2018}. We briefly summarize these issues here for completeness, and to set the context for the remainder of this work.

\subsubsection{Slope and sign conventions}

Regarding conventions around parameter names and sign choices, the power-law slope is sometimes given as $x$, or as $\alpha$ or $\Gamma$, depending on whether a linear or logarithmic functional form is adopted. Different choices regarding the sign are seen, leading, for example, to the Salpeter slope being expressed either as $\alpha_{\rm H}=-2.35$ or $\alpha=2.35$. Depending on this choice, the relationship between $\alpha$ and $\Gamma$ differs, either being $\Gamma=\alpha_{\rm H}+1$ when the sign is included in the parameter, or $\Gamma=\alpha-1$ otherwise.  \citet{Hopkins2018} recommends adopting the forms
\begin{equation}
\frac{{\rm d}N}{{\rm d}m} \propto\left(\frac{m}{M_{\odot}}\right)^{\alpha_{\rm H}}
\end{equation}
and
\begin{equation}
\frac{{\rm d}N}{{\rm d}\log m} \propto\left(\frac{m}{M_{\odot}}\right)^\Gamma
\end{equation}
which lead to the Salpeter slope being expressed as $\alpha_{\rm H}=-2.35$, and to $\Gamma=\alpha_{\rm H}+1$, while the work from 1990 onwards by Kroupa and collaborators strictly applied 
\begin{equation}
\frac{{\rm d}N}{{\rm d}m} \propto\left(\frac{m}{M_{\odot}}\right)^{-\alpha} \,.
\end{equation}
To mitigate 
this problem, the notation $\xi(m)={\rm d}N/{\rm d}m \propto m^\gamma$ and
$\xi_{\rm L}(m)={\rm d}N/{\rm dlog}_{10}m \propto m^{-x} \propto m^\Gamma$
was suggested (table 3 in \citealt{Kroupa+2013}) such that $\gamma=\alpha_{\rm H}=-\alpha=-2.35$ for the Salpeter value and as used in many research papers.

This might seem a pedantic point, but agreeing on a preferred convention minimizes ambiguity in discussion, and avoids confusion when exploring how such parameters may vary, either across the mass range spanned by the IMF, or as the IMF itself differs within and between galaxies, and with time.
 
Apart from the power-law parametrization discussed above, various other shapes have been widely used in the literature. These include log-normal distributions, initially introduced by \cite{Miller1979,Scalo1986} and later refined by \cite{Chabrier2003}, as well as unimodal and bimodal IMFs formulated by \cite{Vazdekis+1996}. Comparing different IMF shapes can often be challenging. To address this, some authors have adopted alternative approaches, such as expressing certain mass fractions within well-defined mass ranges \citep[an approach less dependent on the functional form adopted for the IMF, see][]{LaBarbera13} or employing star-by-star methods \citep{Nacho2024,MartinNavarro+2024}.

Due to the inconsistent conventions adopted by many authors, this White Paper itself may present results in an inconsistent way, despite making a recommendation to the contrary. In particular, in some sections the slope may be given as $\Gamma$, in others as $\alpha$, or $x$. However, we pay attention to clearly specify if the slope is relevant to a logarithmic or linear form of the IMF, and what the Salpeter value is in the given notation.

\subsubsection{IMF types and their temporal variation}
\label{sec:IMFtypes}

This leads to the next point, about confusing or conflating ``types'' of IMF, that may not actually be comparable. IMFs are measured in different ways, and precisely what is measured is not necessarily comparable between different measurements. At a fundamental level, the spatial scale over which an IMF is estimated may lead to different results, since the IMF over an entire galaxy may be rather different to that of a single star cluster or to that for a group of clusters as emphasised by \cite{KroupaWeidner2003} and elaborated on by \cite{Kroupa+2013}. Approaches using cosmic census techniques are in turn sensitive to some effective average of IMFs over entire galaxy populations, which may be different again. Consideration of these issues led \citet{Hopkins2018} to differentiate between stellar, galaxy, and cosmic IMFs, introducing the terms sIMF, gIMF, cIMF. However, since ``cIMF'' is also used for the ``composite IMF'' (see below), in this paper the term ``cosmic IMF'' shall always be spelled out in full.

\begin{figure*}
\centering
\makebox[\textwidth][c]{\includegraphics[width=1.0\textwidth]{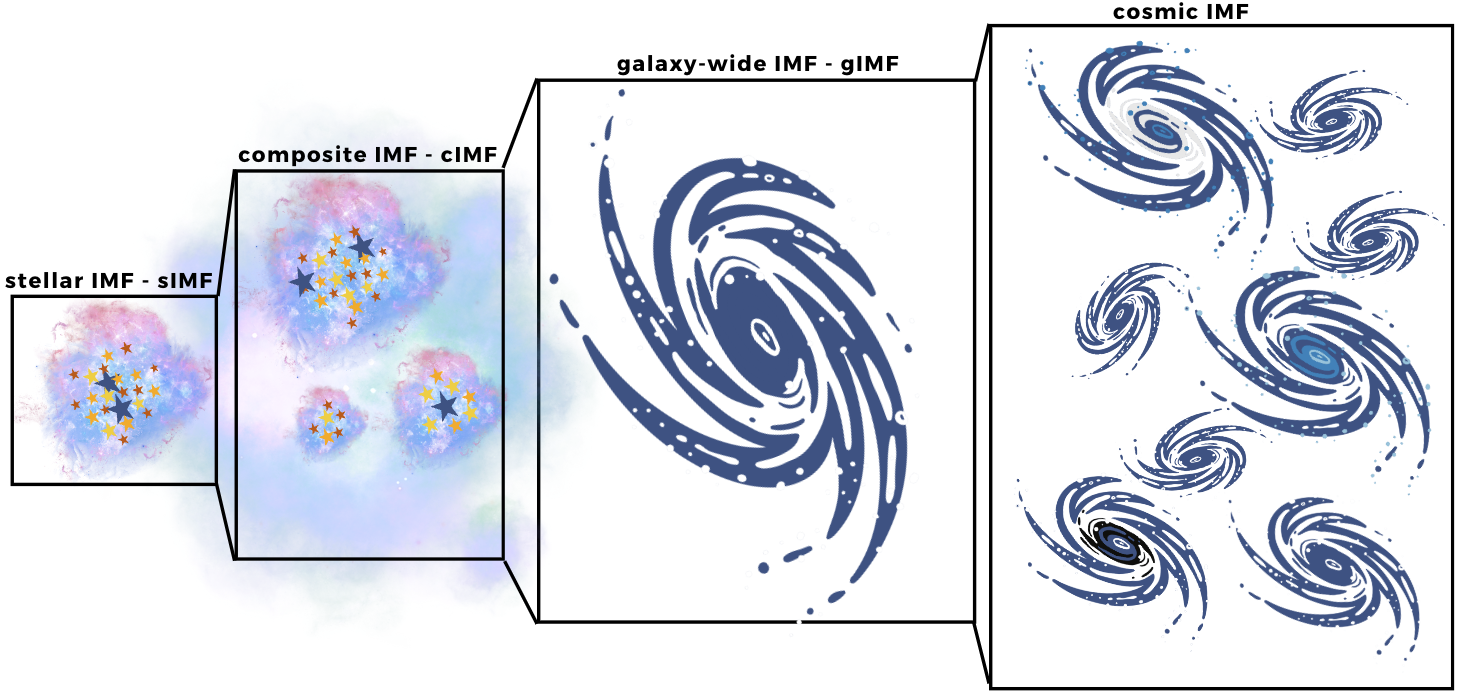}}%
\caption{Schematic illustration of the hierarchical buildup of the initial mass function (IMF). From left to right: the stellar IMF (sIMF), $\xi_{\rm s}(m)$, represents a single star cluster formed in a gravitationally bound part of a molecular cloud; the composite IMF (cIMF), $\xi_{\rm c}(m)$, emerges from the combination of multiple sIMFs within a larger star-forming region; this is then extended to a galaxy-wide IMF (gIMF) composed of all cIMFs across a galaxy, and finally to a cosmological IMF built from the ensemble of galaxies depending on defined constraints (e.g. redshift bin). The figure highlights the scale-dependent nature of the IMF.}
\label{fig:imf}
\end{figure*}

Following the suggestions made by \cite{Jerabkova18} and \cite{Hopkins2018}, to differentiate the different forms of the IMF, the following notation is here proposed and adopted: The {\it stellar IMF} (sIMF), $\xi_{\rm s}(m)$, results from the formation of a single embedded star cluster in a gravitationally bound and collapsing molecular cloud clump that typically has a diameter comparable to a~pc. The {\it composite IMF} (cIMF), $\xi_{\rm c}(m)$, results from adding all sIMFs in a region, see Fig.~\ref{fig:imf}. The cIMF can be the momentary one, in which case the sIMFs in all currently forming embedded clusters in the region are added. Or, it can be the {\it cumulative cIMF} (ccIMF), $\xi_{\rm cc}(m)$, which adds all sIMFs in a region that have ever contributed to the stellar population there. The {\it galaxy-wide IMF} (gIMF), $\xi_{\rm g}(m)$, is the particular case of the cIMF evaluated over a whole galaxy, in which case $\xi_{\rm cg}(m)$ is the total IMF of all initial masses of all stars ever formed in a galaxy up to the time it is being considered. Finally, the term {\it cosmic IMF} (see Sect.~\ref{sec:cosmic}) is always written in full, because the acronym cIMF suggested by \cite{Hopkins2018} is already in use to indicate the composite IMF (see above). 

If the sIMF is an invariant probability density distribution function, as is often assumed, then
$\xi_{\rm g} \propto \xi_{\rm c} \propto \xi_{\rm cc} \propto \xi_{\rm s}$. If the sIMF is an optimally sampled distribution function, then these proportionalities are not given because the shapes of the distribution functions differ, but all IMFs can be calculated from the sIMF. The notation ``IMF'' is used when any of the above is meant. More details can be found in \cite{Kroupa+2013, Yan+2017, Hopkins2018, Kroupa+2024}.

As was just mentioned above, an important factor to consider is the variation in time of the different ``IMFs''. This is one of the least explored aspects of this field, but its significance is becoming more widely recognised, and it is now starting to appear in some models more explicitly \citep[e.g.,][see also the chemo-evolutionary models with varying IMF by \citealt{Vazdekis+1996,vazdekis1997}]{Yan+2017,demasi2018,Jerabkova18,chruslinska2020,Fontanot24,Haslbauer2024}. For an individual star cluster, of course, there is just one ``initial'' mass function, and explicitly no ``time'' over which to vary \citep[although see the point made by][about the ``IMF'' not actually being instantiated at any point in time]{Kroupa+2013}. This is not the case, though, for star-forming regions, galaxies, or galaxy populations, any and all of which may have differing IMFs in different regions at different times that contribute to any measured ``IMF''.

\subsubsection{IMF designation}
\label{sec:introd_IMFdesignation}

There are numerous descriptions used as a shorthand for a ``typical'' or ``reference'' IMF shape, taken as representative of that of the Milky Way (MW). Sometimes, however, these terms are ambiguous. They include (sometimes interchangeably) ``Salpeter'', ``universal'', ``canonical'', ``normal'', or ``standard'' to mean IMFs with the traditional Salpeter slope over some (frequently unspecified) mass range. Sometimes, the naming ``canonical'' IMF is used to designate the IMF that guarantees the best match between observations and theoretical predictions of models for the solar neighborhood, taken as a benchmark for the models themselves \citep[e.g.,][]{portinari2004,romano2017,spitoni2021}. 
\cite{Kroupa+2013} designated the ``canonical IMF" to be the two-part power-law form deduced from observations of star-forming regions and young stellar populations in the Local Group of galaxies (their eq.~55) in the hope of standardizing notation.\footnote{With the progress in this field, \cite{Kroupa+2024} suggest a more precise nomenclature, referring to the ``canonical IMF" as the composite IMF derived from star counts in the local Solar neighborhood and therewith being a bench mark distribution function by being an average constituting the mixture of embedded clusters that gave rise to the Solar neighbourhood stellar ensemble. This distribution aligns with star formation processes in most environments within spiral galaxies.
The ``canonical stellar IMF" is the stellar IMF (as originating from one embedded cluster or molecular cloud clump) with a shape equal to that of the canonical IMF.} Of more concern is the fact that it is often left unspecified whether there is a slope change (also called a ``turnover'', even though the function may not actually turn over) below some mass threshold. There is also disagreement in the literature about what each of these terms means. Conventions adopted by some authors are not always the same as those adopted by others. It is of critical importance in introducing any such reference IMF shape to define mass ranges, and slopes or shapes over each mass range, even if only by citing the original publication.

Allowing for variations in IMF shape leads in turn to a need for descriptors of such variations. This is another source of potential for confusion, though it can be alleviated by using a ratio of integrals, such as the M dwarf mass-to-total mass ratio, which is rather robust against the functional form of the adopted IMF \citep{LaBarbera13}. Most descriptors are relative in their descriptions, using a Salpeter, or more widely accepted MW-style IMF \citep[such as that of][with a change in slope at the low mass end]{Kroupa2001}, as a reference point, and emphasizing a relative over- or under-abundance of stars at the high- or low-mass ends, leading to the commonly seen terms ``bottom heavy'', ``bottom light'', ``top heavy'', or ``top light''. The first point to make is that the reference IMF is not always clear, and neither is the mass range referred to. These descriptors typically refer to the relative abundances of low-mass (mostly but not always) meaning $m < 1$~$M_{\odot}$ and high-mass (mostly but not always) meaning $m > 1$~$M_{\odot}$ stars compared to a Salpeter slope. Sometimes, though, the ``bottom heavy/light'' terminology is relative to a Chabrier or Kroupa IMF, with a different low-mass slope. In any case, loosely using such terms without careful definition is an obvious source of confusion and ambiguity.
More imaginative descriptors extend to include terms such as ``dwarf-rich'', ``diet Salpeter'', ``heavyweight'', ``ski slope'', and more. There is a need for clarity in defining any such nomenclature to avoid a growing list of potentially ambiguous terminology. The simplest approach is to be explicit about defining the mass range or ranges, and associated slope or parametrization that is meant.

In a related vein, approaches that constrain IMF properties through the use of mass-to-light ratios \citep[e.g.,][]{Cappellari12,Smith2020,Mehrgan2024} are potentially very powerful tools, as they do not heavily rely on population synthesis models as many other methods do. These techniques also use a reference IMF, and refer to a ``mismatch'' parameter, which is a ratio of mass-to-light values arising from integrals over an IMF, often confusingly also represented by $\alpha$. One has to be careful here, as such a parameter would render similar values for flatter (due to an excess of remnants) and bottom-heavier (excess of M-dwarfs) IMFs, except when the mass-to-light ratio is calculated taking into account only the alive stars in the population \citep{Ferreras+2013}. There is scope for work to quantify the degree of degeneracy between various IMF shapes and the resulting integrated mass-to-light values and the corresponding mismatch parameter, to explore its relative sensitivity to variations at the high- or low-mass ends of the IMF.

\subsection{Contents}

The IMF concept has evolved significantly over time, reflecting the complexity of the star formation process and our poor, though continuously growing, ability to characterize it in various cosmic environments, from star clusters to galaxies and galaxy clusters. These distinctions are crucial, as they influence our understanding of stellar populations and their distributions across different scales. The IMF, in fact, is not only a fundamental aspect of stellar astrophysics but also serves as a key parameter in models that describe the chemical evolution of galaxies and the formation of large-scale structures in the universe.

The plan of this White Paper is as follows. In Sect.~\ref{sec:clouds}, we explore the interconnections between the substructures that are found in MCs, the formation of stars in them, and the origin of the IMF. In Sects.~\ref{sec:res_IMF} and \ref{sec:unres_IMF} we review, respectively, direct (star counts) and indirect methods (including the analysis of integrated light from unresolved stellar systems, gravitational lensing techniques, and mass-to-light ratios of galaxies) in deriving the stellar mass function. Section~\ref{sec:cosmic} is entirely devoted to novel approaches that rely on cosmic census measurements to place constraints on the IMF. In Sect.~\ref{sec:models}, we discuss the implementation of the IMF in simulations of galaxy formation and evolution, highlighting its significance in understanding various physical processes that govern the life cycle of galaxies. In Sect.~\ref{sec:plots}, we further discuss some useful diagnostic plots. Finally, in Sect.~\ref{sec:conclusions} we draw our conclusions and provide an outlook for future investigations.

\section{Cloud fragmentation and mass function on small scales}
\label{sec:clouds}

\begin{figure*}
\centering
\makebox[\textwidth][c]{\includegraphics[width=0.85\textwidth]{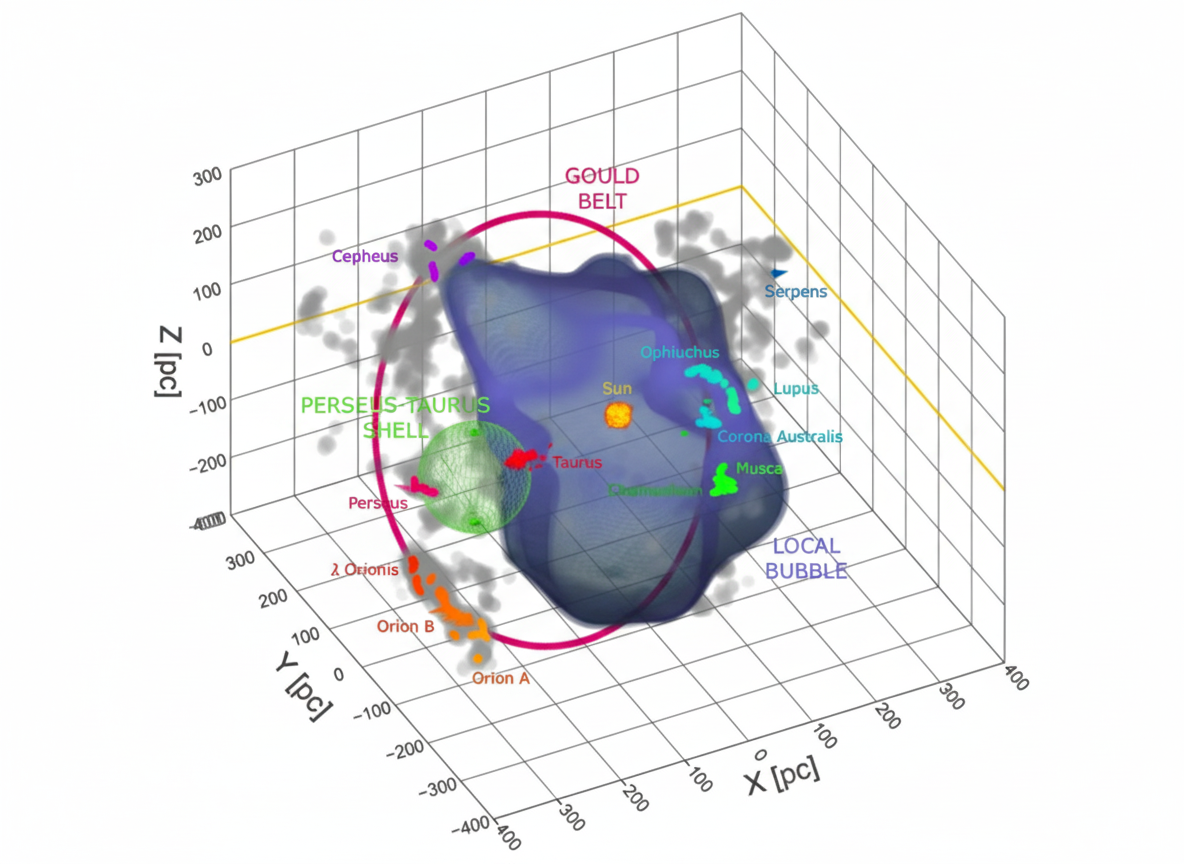}}%
\caption{Three-dimensional visualization of dense gas and young stellar clusters (colour-coded according to name labels in the map) in prominent MCs on the Local Bubble’s surface. The Perseus-Taurus super bubble (green sphere), dust clumps (gray blobs), and a schematic representation of the Gould's Belt 
(magenta circle)
are also shown. Adapted from \cite{Zucker2022}.
  }
\label{fig:LB}
\end{figure*}

As stars are known to form in molecular clouds (MCs), tight links between the structure of MCs and the origin of the IMF may be expected. Indeed, determining the mass distribution of cores (the gas mass reservoirs used for the formation of each star or binary system) in star-forming regions is a fundamental step towards the comprehension of the origin and environmental dependence of the IMF. Such a formidable task requires a non-obvious definition of what a ``core'' is \citep[see, e.g.,][]{Louvet2021} that also depends on whether the star formation is a quasi-static or dynamic process. Moreover, it requires the sampling of different cloud environments.

In this section, we first review the properties of small-scale structures observed in both low- and high-mass star-forming clouds in the Galaxy. Then, we present the cornerstones of the hydrodynamical theory of cloud evolution leading to star formation.

\subsection{Nearby, low-mass star-forming clouds and connection to filaments}
\label{sec:clouds_low-mass}

The expansion of the Local Bubble swept up the ambient interstellar medium (ISM) into an extended shell of cold neutral gas and dust that has fragmented and collapsed into a number of prominent MCs. Nearly all of the star-forming complexes in the solar neighborhood lie on the surface of the Local Bubble  (see Fig.\ref{fig:LB}) and that's where young low-mass stars, by far more abundant than their massive counterparts, can be observed, at distances as small as a few hundred pc.

Within about 400~pc from the Sun, MCs and their substructures can be studied in great detail. Their three-dimensional (3D) magnetic field properties can also be characterized through various methodologies and techniques \citep[e.g., ][]{Sullivan2021,Shane2024}, which helps to establish the significance of magnetic fields in the condensation of dense structures out of the diffuse ISM on physical scales from about 1 to 10~pc.

\subsubsection{Main types of substructures in molecular clouds}
\label{sec:clouds_substructure}

At least three groups of relevant substructures can be distinguished in MCs: clumps, dense cores, and filaments \citep[e.g.,][]{Williams2000,Bergin2007}. {\it Clumps} are coherent regions in position-velocity space with typical masses in the range of $\approx$\,10--1000~$M_\odot$ and sizes between $\approx$\,0.3 and $\approx$\,3\,pc. They may themselves contain significant substructure. When sufficiently massive and actively star-forming, they represent the reservoirs of gas mass out of which star clusters originate in MCs. {\it Dense cores} are smaller individual cloud fragments which correspond to local overdensities, i.e., local minima in the gravitational potential of a MC. They have typical masses in the range of $\approx$\,0.1--10~$M_\odot$ and sizes between $\approx$\,0.01 and $\approx$\,0.1\,pc. A starless core is a dense core with no associated protostellar object, while a {\it prestellar core} is a dense core which is both starless and gravitationally bound. In other words, a prestellar core is a self-gravitating condensation of gas and dust 
which may potentially form an individual star (or small system) by gravitational collapse, but not a star cluster \citep[][]{Ward-Thompson1994,Ward-Thompson2007,Andre2000,DiFrancesco2007}. Molecular clouds have also long been recognized to be filamentary \citep[e.g., ][]{Barnard1907,Schneider1979}, but the realization that {\it filaments} are truly ubiquitous in the cold ISM and play a central role in the star formation process came only more recently with the results of {\it Herschel} Space Observatory \citep{Pilbratt2010} imaging surveys \citep[e.g.,][]{Andre2010,Molinari2010}. In particular, {\it Herschel} studies of nearby ($d < 500$\,pc) MCs have shown that most ($>$75\%) prestellar cores lie in ``supercritical'' filaments for which the mass per unit length exceeds the critical line mass of nearly isothermal cylinders \citep[e.g.,][]{Inutsuka1997}, $M_{\rm line, crit} = 2\, c_{\rm s}^2/G \approx 16$~$M_\odot$~pc$^{-1}$, where $c_{\rm s} \approx 0.2$~km~s$^{-1}$ is the isothermal sound speed for molecular gas at $T \approx 10$~K \citep{Konyves2015, Marsh2016}, the typical temperature of the bulk of metal-rich MCs \citep{Blitz1993}. Remarkably, all nearby molecular filaments share approximately the same half-power width, $\approx 0.1$\,pc \citep{Arzoumanian2011, Arzoumanian2019, KochRosolowsky2015}. While there has been some controversy about the reliability of this result \citep[e.g.,][]{Panopoulou2022}, tests performed on synthetic data \citep{Roy2019, Andre2022} suggest that {\it Herschel} width measurements are not significantly biased, at least for high-contrast nearby filaments.

Importantly, many clumps, at least those of sufficient mass and density, are structured in the form of {\it hub-filament systems} \citep[e.g.,][]{Myers2009,Kumar2020}. Hub-filament systems consist of a central hub surrounded by a converging network of filaments. They have been proposed to be the main sites of star clusters and, when sufficiently massive, high-mass star formation \citep[e.g.,][]{Kumar2022}.

\subsubsection{Mass spectrum of molecular clouds and their substructures}
\label{sec:low-massCMF}

\begin{figure*}
\centering
\includegraphics[width=0.75\textwidth]{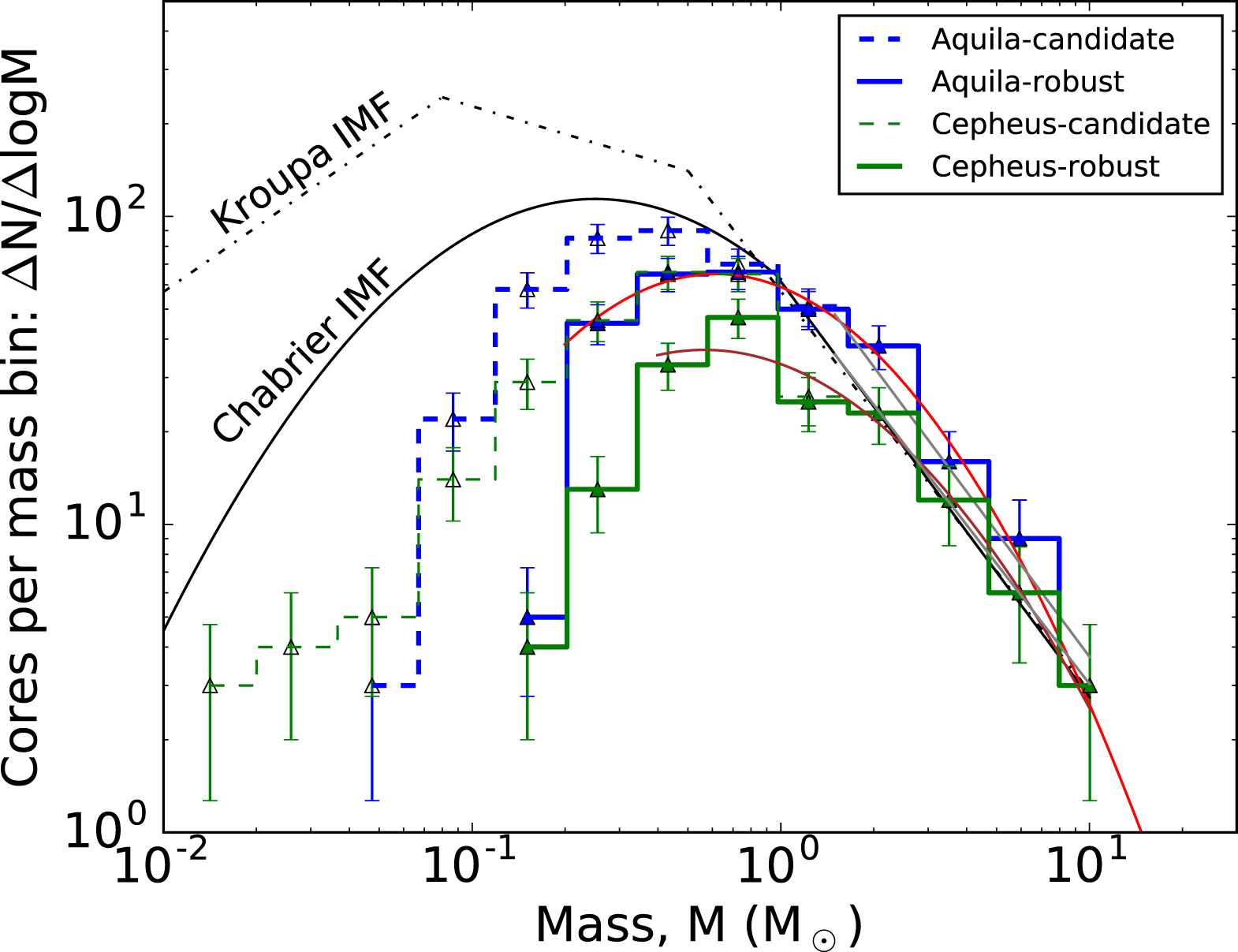}
\caption{Mass functions of candidate (dashed histograms) and robust (solid histograms) prestellar cores in the Cepheus clouds (green) and the Aquila Rift (blue). Lognormal fits to the robust prestellar mass functions of Cepheus and Aquila, excluding the likely incomplete lowest mass bins, are also shown (brown and red lines, respectively) and compared to different stellar IMFs (dot-dashed and solid lines) scaled by a factor of $10^3$. We note that the here shown ``Chabrier IMF'' is the ``system IMF'' claimed in a conference proceedings by \cite{Chabrier2005}, which is based on incomplete star-counts and, therefore, underestimates the numbers of stars below $0.5\,M_\odot$ (cf also the recently published estimate of the cIMF by \citealt{Kirkpatrick2024}). We remark that it is incorrect to use the "system IMF" constructed by merging all multiple systems in the Solar neighborhood because in star forming regions the multiplicity fractions are significantly larger and these also depend sensitively on the degree of dynamical processing of the populations studied such that if the actual binary (or multiplicity) fraction for a particular population is not known then the variation thereof masquerades as an apparent variation of the sIMF (Sec.~\ref{sec:resIMF_binaries}). An update on the sub-stellar sIMF is provided in Sec.~\ref{sec:resIMF_BDs}.
Figure from \cite{DiFrancesco2020}.
}
\label{fig:CMF}
\end{figure*}

The mass function of MCs and CO clumps within MCs is known to be rather shallow,  $\Delta N / \Delta \log M \propto M^{-0.6\pm0.2}$ {\citep[e.g.,][]{Solomon1987,Blitz1993,Kramer1998,Rice2016}}, and significantly shallower than the Salpeter IMF. This implies that most of the molecular gas mass in the Galaxy resides in the most massive MCs and within the MCs themselves in the most massive CO clumps. In contrast, the mass distribution of nearby self-gravitating prestellar cores, or prestellar core mass function (CMF), broadly resembles the stellar IMF in both shape and mass scale \citep[e.g.,][see Fig.~\ref{fig:CMF}]{Motte1998,Johnstone2001,Alves2007,Nutter2007,Konyves2015,Marsh2016, DiFrancesco2020,Pezzuto2021}. The similarity between the prestellar CMF and the system IMF is consistent with an essentially one-to-one correspondence between core mass and stellar system mass ($M_{\star \, \rm sys} = \epsilon_{\rm core}\, M_{\rm core} $), with a typical core-to-star formation efficiency $\epsilon_{\rm core} \approx 0.2$--0.4 \citep{Alves2007,Konyves2015}. This efficiency factor may be attributable to mass loss due to the effect of outflows during the protostellar phase \citep{Matzner2000,Guszejnov2016,2021MNRAS.507.2448M}. The difference in shape between the observed mass distribution of MCs or CO clumps and that of prestellar cores may a priori arise from the use of different tracers, typically CO for clouds or clumps and dust continuum for prestellar cores. However, millimeter/submillimeter dust continuum studies have also reported mass functions shallower than the Salpeter IMF for both small MCs/large clumps \citep[e.g.,][]{Ellsworth2015} and unbound starless cores \citep[e.g.,][]{Marsh2016}. The mass functions of the latter types of cloud structures therefore appear to genuinely differ from the IMF and prestellar CMF.

\begin{figure*}
\centering
\includegraphics[width=\textwidth]{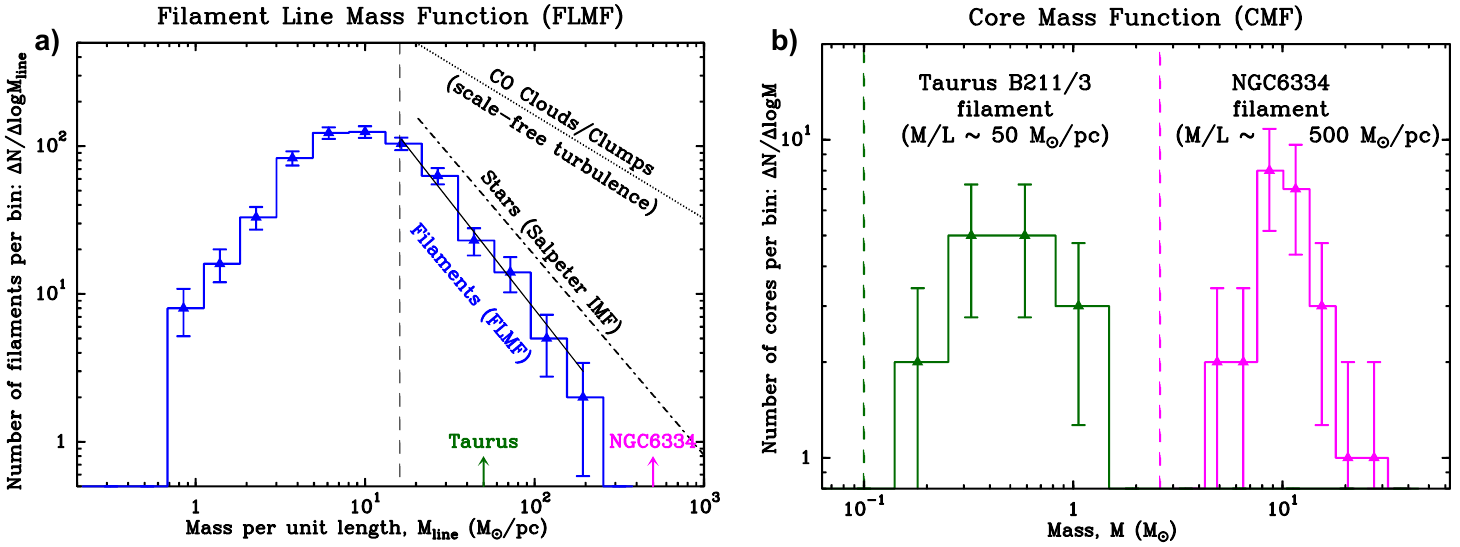}
 \caption{\small 
Potential role of filaments in the origin of the prestellar CMF and stellar IMF. As higher-mass cores form in higher-line mass filaments [panel (b), adapted from \citealp{Shimajiri2019} and \citealp{Marsh2016}], 
the Salpeter slope of the global CMF and IMF may be inherited from the filament line mass function [panel (a), adapted from \citealp{Andre2019}].
 }
 \label{flmf_CMF}     
\end{figure*}

The simple picture of a one-to-one connection between prestellar core mass and star (or stellar system) mass has been challenged on the grounds that dense cores are not well-defined entities in numerical simulations of isothermal gravo-turbulent fragmentation \citep{Hennebelle2018,Louvet2021} and typically fragment almost indefinitely, down to the opacity limit for fragmentation, $\approx 0.03\, M_\odot $. This does not appear to be the case, however, in real observations of nearby, low-mass dense cores. Indeed, interferometric observations of dense cores identified with {\it Herschel} exhibit very little, if any, subfragmentation \citep[][Andr\'e et al., in prep.]{Dunham2016,Sadavoy2017, Pokhrel2018, Maury2019}.

Recently, a good estimate of the filament mass function (FMF) and filament line mass function (FLMF) 
in nearby MCs has been derived using a comprehensive study of filament properties from 
{\it Herschel} Gould Belt survey observations \citep{Arzoumanian2019,Andre2019}. The FLMF is well fit by a power-law distribution in the supercritical mass per unit length regime (above 16~$M_\odot$~pc$^{-1}$), $\Delta N / \Delta \log M_{\rm line} \propto M_{\rm line}^{-1.6\pm0.1}$ (see Fig.~\ref{flmf_CMF}a). 
The FMF is very similar in shape to the FLMF and also follows a power-law distribution at the high-mass end (for $M_{\rm tot} > 15 \, M_\odot$), $\Delta N / \Delta \log M_{\rm tot} \propto M_{\rm tot}^{-1.4 \pm0.1}$, 
which is significantly steeper than the MC mass function. Both the FLMF and the FMF are reminiscent of the form of the IMF at the high-mass end ($m \geq 1 \, M_\odot$), which scales as the Salpeter power law, ${\rm d}N / {\rm d}\log m \propto m^{-1.35}$ in the same format. Thus, molecular filaments may represent the key evolutionary step in the hierarchy of cloud structures at which a steep, Salpeter-like mass distribution is established (see also Sect.~\ref{sec:gas-to-star_theory}). The FMF differs in a fundamental way from the MC mass function in that most of the filament mass lies in low-mass filaments. In particular, this result implies that most of the mass of star-forming filaments lies in thermally transcritical filaments with line masses within a factor 2 of the critical value, $M_{\rm line, crit}$. 

\subsubsection{A filament scenario for the origin of the IMF?}
\label{sec:filaments}

The most massive prestellar cores identified with {\it Herschel} in nearby clouds (with masses between $\approx 2$ and 10~$M_\odot$) tend to be spatially segregated in the highest column density parts/filaments 
of the clouds, suggesting that the prestellar CMF is not homogeneous within a given cloud, but depends on the local column density (or line mass) of the parent filaments (\citealp{Konyves2020}; see also \citealp{Shimajiri2019}). In Orion~B, for instance, there is a marked trend for the prestellar CMF to broaden and shift to higher masses in higher density areas \citep{Konyves2020}. This supports the view that the global prestellar CMF results from the superposition of the CMFs produced by individual filaments \citep{Lee2017, Andre2019}. 

The close link between the FMF (or FLMF) and the prestellar CMF may be understood as follows. As already mentioned, the thermally supercritical filaments observed with {\it Herschel} in nearby clouds have a typical inner width $W_{\rm fil} \approx 0.1\,$pc. They are also virialized with $ M_{\rm line} \approx  \Sigma_{\rm fil} \times W_{\rm fil}  \approx M_{\rm line, vir} \equiv 2\, \sigma^2_{\rm tot} / G$, where $\sigma_{\rm tot}$ is equivalent to the effective sound speed \citep{Fiege2000, Arzoumanian2013}. This implies that the effective Jeans or Bonnor-Ebert mass  $M_{\rm BE, eff}  \approx 1.3\, \sigma_{\rm tot}^4 /(G^2 \Sigma_{\rm fil} $) scales roughly as $\Sigma_{\rm fil}$ or $ M_{\rm line} $ in supercritical filaments. Thus, higher-mass cores may form in higher $M_{\rm line} $ filaments, as indeed suggested by observations  \citep[][see Fig.~\ref{flmf_CMF}b]{Shimajiri2019}. If the CMF produced by a single supercritical filament were a narrow $\delta$ function peaked at $M_{\rm BE, eff} $, then there would be a direct correspondence between the FLMF and the prestellar CMF \citep[cf.][]{Andre2014}. In reality, the prestellar CMF generated by a single filament is expected to be broader than a $\delta$ function \citep[][]{Inutsuka2001}, and observationally it appears to broaden as $ M_{\rm line} $ increases \citep[][]{Konyves2020}, although for statistical reasons, this is difficult to constrain accurately. The global prestellar CMF therefore results from a convolution of the {FLMF} with the CMFs produced by individual filaments \citep{Lee2017}. It can be shown, however, that the high-mass end of the global CMF is primarily driven by the power-law shape of the FLMF in the supercritical regime and depends only weakly on the breadths of the individual CMFs \citep[cf. Appendix B of][]{Andre2019}.  

Filaments and  prestellar cores within them therefore appear to be fundamental building blocks of the star formation process and the evolution toward the IMF. In particular, filament fragmentation may be the dominant physical mechanism generating the broad peak of the prestellar CMF, and by extension the IMF for masses $0.1 \leq m/M_\odot \leq 10$. 

\subsection{High-mass star-forming regions}
\label{sec:clouds_high-mass}

\begin{figure*}
\centering
\includegraphics[width=\textwidth]{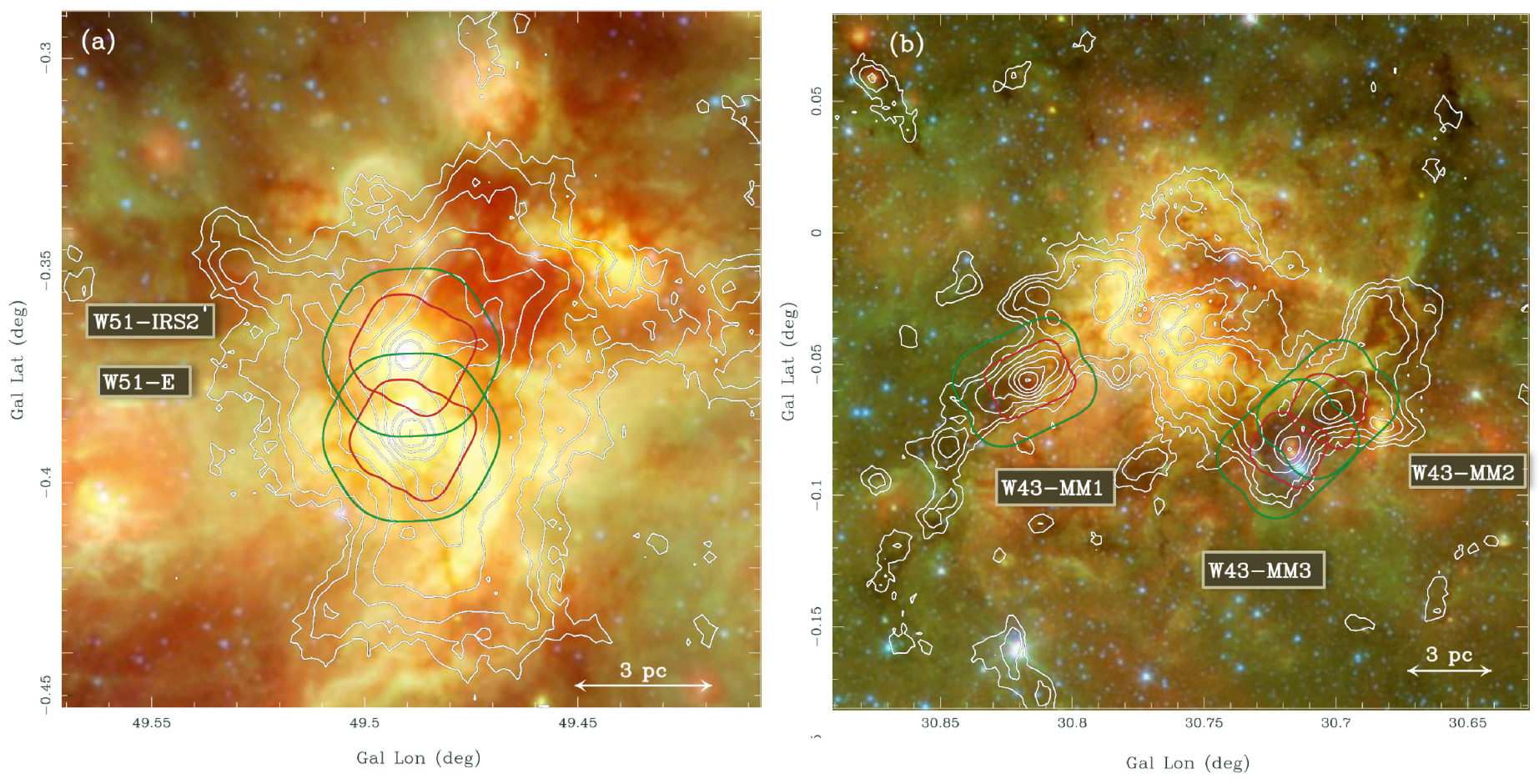}
\caption{Examples of ALMA-IMF high-mass protoclusters and their surroundings. ATLASGAL 870~$\mu$m emission contours (white) are overlaid on {\it Spitzer} three-color images. Green and red contours refer to the primary beam response of the ALMA 12~m array mosaics down to 15\% at 3 and 1.3~mm, respectively. Adapted from \cite{Motte2022}. 
  }
\label{fig:ALMA-IMF}
\end{figure*}

In the last three decades or so, nearby star-forming regions have been extensively studied. However, these regions do not capture massive star formation, nor are representative of the variety of physical and environmental conditions that can be encountered in the Galaxy. Dedicated surveys at millimeter/submillimeter and far-infrared wavelengths, such as the Atacama Pathfinder Experiment (APEX) Telescope Large Area Survey of the Galaxy \citep[ATLASGAL, ][]{Schuller2009}, the {\it Herschel} InfraRed Galactic Plane Survey \citep[HiGAL, ][]{Molinari2010HiGAL}, and the Bolocam Galactic Plane Survey \citep[BGPS, ][]{Aguirre2011}, have provided complete samples of 0.1--1~pc clumps towards the inner Galactic plane at distances up to $\approx$8 kpc, also including (with irregular coverage) regions of the outer Galaxy \citep{Ginsburg2013,Csengeri2014,Konig2017,Urquhart2018,Elia2021}. 

While images of the Gould Belt clouds show a good correspondence between the CMF and the stellar (Salpeter) IMF \citep[][see Sect.~\ref{sec:low-massCMF}]{Alves2007,Andre2010}, analyses of 1.3~mm observations of the high-mass protocluster W43-MM1 at the tip of the Galactic bar ($d \approx 5.5$~kpc) obtained with the Atacama Large Millimetre/submillimetre Array (ALMA) unambiguously point to a CMF whose shape differs from that of the IMF, challenging either the direct relation between the CMF and the IMF or the IMF universality, or both \citep{Motte2018}. The unexpectedly large proportion of high-mass star-forming clouds found in W43-MM1 is particularly noteworthy in light of the fact that this region may be more representative of actively star-forming Galactic-arm environs where most Milky Way stars are born than local cloud complexes.

Recent studies take advantage of the unprecedented capabilities of ALMA to measure the CMF in massive star-forming clouds with a resolution down to a few thousand au \citep[namely, hundredths of a pc; e.g.,][]{Cheng2018,Cheng2024,Lu2020,Kinman2024}. The ALMA-IMF Large Program \citep{Motte2022,Pouteau2022,Nony2023,Louvet2024} has observed 15 high-mass protoclusters (see Fig.~\ref{fig:ALMA-IMF}) covering the (2.5--33)~$\times 10^3 \, M_\odot$ mass range, distances from 2 to 5.5~kpc, and evolutionary stages from young to evolved \citep[based on the amount of dense gas in the cloud that has been impacted by H{\small II} regions; see][]{Motte2022}. Nearly 700 gravitationally bound cores are extracted from the protoclusters; the CMF is derived for a sub-sample of cores above the completeness limit \citep{Louvet2024}. It is concluded that not only the CMF is top-heavy in high-mass protoclusters, but also it varies both spatially and temporally (the CMF is flatter at younger ages). If a self-similar mapping is assumed between the CMF and the IMF, the above results imply that the 15 high-mass protoclusters imaged by the ALMA-IMF program will generate atypical IMFs.

Yet, it is worth recalling that some common assumptions may affect the derivation of the CMF and its relationship to the IMF, namely, the assumed one-to-one correspondence between the mass of the core and the mass reservoir for the star (neglecting possible core sub-fragmentation\footnote{Recent observations suggest that fragmentation takes place below 1000~au in some high-mass protostellar cores \citep{Izquierdo2018,Olguin2022}, while in others it is not observed \citep{Girart2018,Olguin2023}.}), the assumption of a uniform gas-to-star mass transfer (with no dependency on the core’s density/environment), and the adoption of the same lifetime for all cores \citep[$\approx$1--2~Myr for prestellar cores in local clouds, ][]{Konyves2015,Konyves2020}.

\subsection{Theoretical frameworks for the origin of the IMF}
\label{sec:gas-to-star_theory}

The formation and evolution of large-scale ISM structures are predominantly governed by turbulent motions. While these structures are usually transient, smaller-scale over-densities generated by converging flows or compressive shocks can be withheld by gravity and start to collapse. Due to the nature of gravity in structures of different geometry, sheet-like structures tend to develop most rapidly. Following that, fragmentation within sheets leads to the formation of filaments, which then fragment to form cores. This is the general picture of hierarchical structure formation that eventually leads to star formation at the last level of mass concentration. Such behavior of dimensional evolution is reflected by the fractal nature of the star-forming gas: The fractal dimension of interstellar gas as measured from the size distribution of MCs is $D \simeq$~2--3, which suggests that the 3D space is not uniformly filled with mass \citep[e.g., ][]{ElmegreenFalgarone1996,Fleck1996,SanchezAlfaro2008}. This is also reflected by the mass function of MCs and star-forming clumps \citep{Larson1981,HennebelleFalgarone2012}. On the other hand, stars in clusters show fractal dimensions close to unity \cite[e.g., ][]{Gomez1993,Larson1995,Simon1997}, indicating that the stars, although distributed in the 3D space, are placed along some invisible lines. These lines have been revealed by {\it Hershel} observations of dust continuum, and are referred to as interstellar filaments \citep[][see Sect.~\ref{sec:clouds_substructure}]{Arzoumanian2011,Konyves2015}. It is thus very natural to draw the conclusion that stars form within filaments, where the mass is concentrated (see also Sect.~\ref{sec:filaments}).

Whether the stellar mass is determined by a gravitationally-bound local gas reservoir or through highly-dynamic stochastic accretion processes has remained a hotly debated issue for decades. It has now become clear that both effects likely play a role. The discussion here will focus on the mass reservoir paradigm, in order to shed light on the link to the observed dense structures in star-forming clouds. \cite{Inutsuka2001} first introduced the Press-Schechter ``cloud-in-cloud'' formalism to calculate the CMF in the context of star-forming gas characterized by highly non-linear, scale-dependent turbulence. Other models based on the same idea \citep{PadoanNordlund2005,HennebelleChabrier2008,Lee2017} show remarkable differences in the choice of the underlying gas density distribution, including its scale dependence and its evolution in time under the effect of self-gravity \citep{Kritsuk2011,Kainulainen2009,HennebelleFalgarone2012,LeeHennebelle2018a}. There are also different criteria in literature for the selection of the mass entities which eventually form self-gravitating cores across various scales, with either energy arguments or simple density thresholds being considered. These different model ingredients inherently result in variations in the shape of the predicted CMF.

While there is overall a remarkable resemblance between the shape of the CMF and that of the IMF \citep[][but see Sect.~\ref{sec:clouds_high-mass}]{Andre2014,Pineda2023}, the mapping from the CMF to the IMF should be discussed in two regimes: the high-mass power law and the peak around the characteristic mass. Mapping from the core to the star requires consideration of different processes -- accretion, sub-fragmentation, and stellar feedback -- which are all highly dynamic and could skew the IMF slope away from that of the CMF. At the low-mass end, the CMF could have a peak value that is inherited from the large-scale cloud structure \citep{InutsukaMiyama1992, HennebelleChabrier2008}. One should be aware that observational determinations of the CMF peak are affected by several uncertainties and assumptions (see, e.g., end of Sect.\ref{sec:clouds_high-mass}). 
Moreover, in isothermal simulations, the CMF peak keeps moving towards smaller masses as the numerical resolution of the simulations increases \citep[cf.][and references therein]{Louvet2021}. At variance with the hierarchical molecular gas fragmentation that can be well approximated as an isothermal process, the formation of protostars needs a proper thermodynamic description, which sets a lower limit for fragmentation, closely related to the adiabatic first hydrostatic core\footnote{Theoretical work pioneered by \cite{Larson1969} has revealed that the formation of a stellar core can be unfolded in four main stages. During the initial, almost isothermal collapse, gravity is only counteracted by the magnetic field. This leads to the formation of a pseudo-disc supported by magnetic forces \citep{Galli1993a,Galli1993b}. As gas accumulates, the densest region becomes optically thick and a pressure-supported region forms as the temperature rises: this is the first hydrostatic core, famously known as the ``first Larson core''. Further mass accretion then raises the temperature to the H$_2$ dissociation point, which softens the equation of state and lets the collapse proceed further. When the temperature evolution becomes adiabatic again, a new pressure-supported region forms that is named the ``second Larson core''. These stages have been thoroughly analyzed by means of numerical studies using either grid-based \citep[e.g., ][]{Bodenheimer1968,Stahler1980,Masunaga2000,Commercon2011,Bhandare2018} or smoothed particle hydrodynamics methods \citep[e.g.,][]{Bate1998,Stamatellos2007,Wurster2018}. Whether the first Larson core exists at low metallicities (where a steeper equation of state exists due to less efficient cooling in collapsing MCs), however, remains an open question.} formation, or to the protostellar radiation. The IMF peak is floored by this local physics that is related to the gas intrinsic properties and is independent of the large-scale cloud structure \citep{Larson1985,Elmegreen2008,Offner2009,Bate2009,Krumholz2011,Guszejnov2016,LeeHennebelle2018b,ColmanTeyssier2020}. If the CMF has a peak lower than this floor value, it will never be observed in the IMF since all smaller cores will not follow a collapse process similar to those that go through a first Larson core stage. Different proposed empirical shapes of the IMF can be reconciled here, where there is a statistical spread around the cutoff mass, below which a different collapse channel for brown dwarfs leads to a different mass function power law.

We conclude this section by stressing that the problem of star formation is a complex one, where many physical processes, including gravity, magnetohydrodynamics, atomic and molecular physics, radiation, stellar physics, and feedback, combine and affect each other. Its solution ultimately requires the development of cutting-edge simulations that resolve individual star formation, integrate stellar dynamics (ideally on the scale of binary separations, to track the appearance of stellar multiplicity), follow the evolution of the ISM metallicity and magnetic field, and allow for radiative gas cooling, while also implementing all possible stellar feedback channels \citep{2021MNRAS.506.2199G,hennebelle2024}.

The emergence of the stellar IMF is one of the accomplishments expected from any star formation theory. For testing the IMFs springing from models against integrated observations of stellar clusters (or galaxies), it is useful to clarify how the IMF depends on the star-forming gas global properties and to keep in mind that, instead of a non-universal IMF, we might be seeing different manifestations of a general (universal) IMF. This could be due to the global parameters of the star-forming clouds, such as the virial parameter (defined as twice the ratio of turbulent kinetic energy to gravitational energy), the sonic Mach number (proportional to the internal velocity dispersion of the clumps divided by the local speed of sound), and the Alfv\'en Mach number (proportional to the internal velocity dispersion of the clumps divided by the Alfv\'en speed) \citep[see, e.g.,][]{Bertoldi1992,2005ApJ...630..250K,Padoan2011,Federrath2012}, coupled to truncation at the lower-end according to thermodynamics (in turn influenced by metallicity). This is where stellar evolution and chemical enrichment models should come in. Finally, as already mentioned the gIMF is a convolution between the cluster IMFs and the cluster mass function across the galaxy evolution. Depending on the slopes of the two mass functions, it is possible, in principle, that the IMF is ``overwhelmed'' by the cluster mass function, so that the gIMF is dominated by the large-scale structure dynamics rather than by local processes.

\section{The IMF from resolved observations}
\label{sec:res_IMF}

Resolved stellar populations that reach low-mass stars come two-fold: the ensemble of stars nearby to the Sun (Sec.~\ref{sec:solarneighb}) and the populations of stars in star clusters (Sec.~\ref{sec:resIMF_stcl}).

\subsection{The Solar neighbourhood}
\label{sec:solarneighb}

Concerning the first, the aim is to estimate, from the observed distribution of stars, the IMF, $\xi_{\rm c}(m)$, where ${\rm d}N = \xi_{\rm c}(m)\,{\rm d}m$ is their number in a complete volume with initial masses in the mass range $m$ to $m+{\rm d}m$. Note that this definition of the IMF is that of the {\it composite IMF} (cIMF, see Sect.~\ref{sec:IMFtypes}) and it differs from that given at the beginning of Sect.~\ref{sec:intro} because here the ensemble of stars used is not restricted to those stars born together in one star formation event. This needs to be kept in mind when discussing the IMF, and an interesting question emerges therewith, namely, is the $\xi_{\rm c}(m)$ deduced from star counts in the Solar neighbourhood compatible with the {\it stellar IMF} (sIMF), as introduced at the beginning of Sect.~\ref{sec:intro} and defined in Sect.~\ref{sec:IMFtypes}? Technically, the ``IMF'' extracted from resolved stellar populations such as the Solar neighbourhood field and star clusters is never the IMF since the assessed population of stars is not in its initial state. Nevertheless, since stars of $m<1\,M_\odot$ evolve very slowly compared to a Hubble time and once corrections for various biases (e.g., unresolved multiple systems) have been applied, the obtained mass function of stars is customarily referred to as the IMF. It is only with the advent of the IGIMF theory \citep{KroupaWeidner2003, Jerabkova18} that the correspondence between the true sIMF and the cIMF gardened from the local Galactic field population has become quantified.

The full details of how star counts are transformed into an estimate of $\xi_{\rm c}(m)$ can be found in \cite{Scalo1986, Kroupa+1993, Kroupa+2013, Hopkins2018}, with the concept of the sIMF possibly being an optimal distribution function being introduced in \cite{Kroupa+2013}. Here a few critical issues are pointed out. 

The most accessible stellar sample at disposal for constructing a first estimate of $\xi_{\rm c}(m)$ are the stars nearby to the Sun, as was indeed used by \cite{Salpeter1955}. Obtaining $\xi_{\rm c}(m)$ from Galactic field star counts is a complex problem requiring detailed  knowledge of stellar evolution, of stellar dynamical processes, of the multiplicity properties of stars and the evolution thereof, and of Galactic structure needed to connect late-type and early-type star count surveys. The problem is also complicated because only the stellar luminosities are observable, and from these the stellar masses need to be inferred. Analysis of star counts in terms of obtaining $\xi_{\rm c}(m)$ are commonly restricted to main sequence stars because the pre- and post-main sequence evolutionary stages are short and stars in these phases can thus be safely ignored (an exception are low-mass stars with about $m<0.4\,M_\odot$ for which the pre-main sequence contraction times are long compared to a Galactic orbit such that this phase needs to be accounted for when analysing star counts). 

Assuming all stars are detected (i.e. all members of all multiple stellar systems are known and observed), if $\Psi_{\rm c}(M_{\rm V})$ is the luminosity distribution function of main sequence stars in the photometric $V$-band (as a proxy for any photometric band) then ${\rm d}N=\Psi_{\rm c}(M_{\rm V})\,{\rm d}M_{\rm V}$ is the number of stars in
the absolute magnitude interval $M_{\rm V}$ to
$M_{\rm V}+{\rm d}M_{\rm V}$ (in a volume or per unit volume), and 
\begin{equation}
\xi_{\rm c}(m) = - \left( {\rm d}m/{\rm d}M_{\rm V} \right)^{-1}\,\Psi_{\rm c}(M_{\rm V}) \, .
\label{eq:IMF_LF}
\end{equation}
Note that this is also true if the calculation of stellar masses is performed on a star-by-star basis from their photometric properties because we need to construct the distribution function $\xi(m)$. Corrections of the star counts for loss of stars through stellar evolution need to be applied.

The most direct approach to constraining $\Psi_{\rm c}(M_{\rm V})$ is by using an ensemble of main sequence stars known to be volume-complete down to some limiting mass. Three such samples exist: 

\begin{enumerate}
\item[(I)] The Solar neighbourhood sample based on stellar distances available through trigonometric parallax measurements.

\item[(II)] The wider Solar neighbourhood sample based on stellar distances available through photometric parallax measurements for each star.

\item[(III)] Stars in star clusters for which the star counts yield $\Psi_{\rm s}(M_{\rm V})$ because open and globular star clusters are largely thought to be single-age populations formed in one embedded cluster.
\end{enumerate}

\begin{sidewaysfigure*}
\vspace{13.6cm}
\makebox[\textwidth][c]{\includegraphics[width=0.95\textwidth]{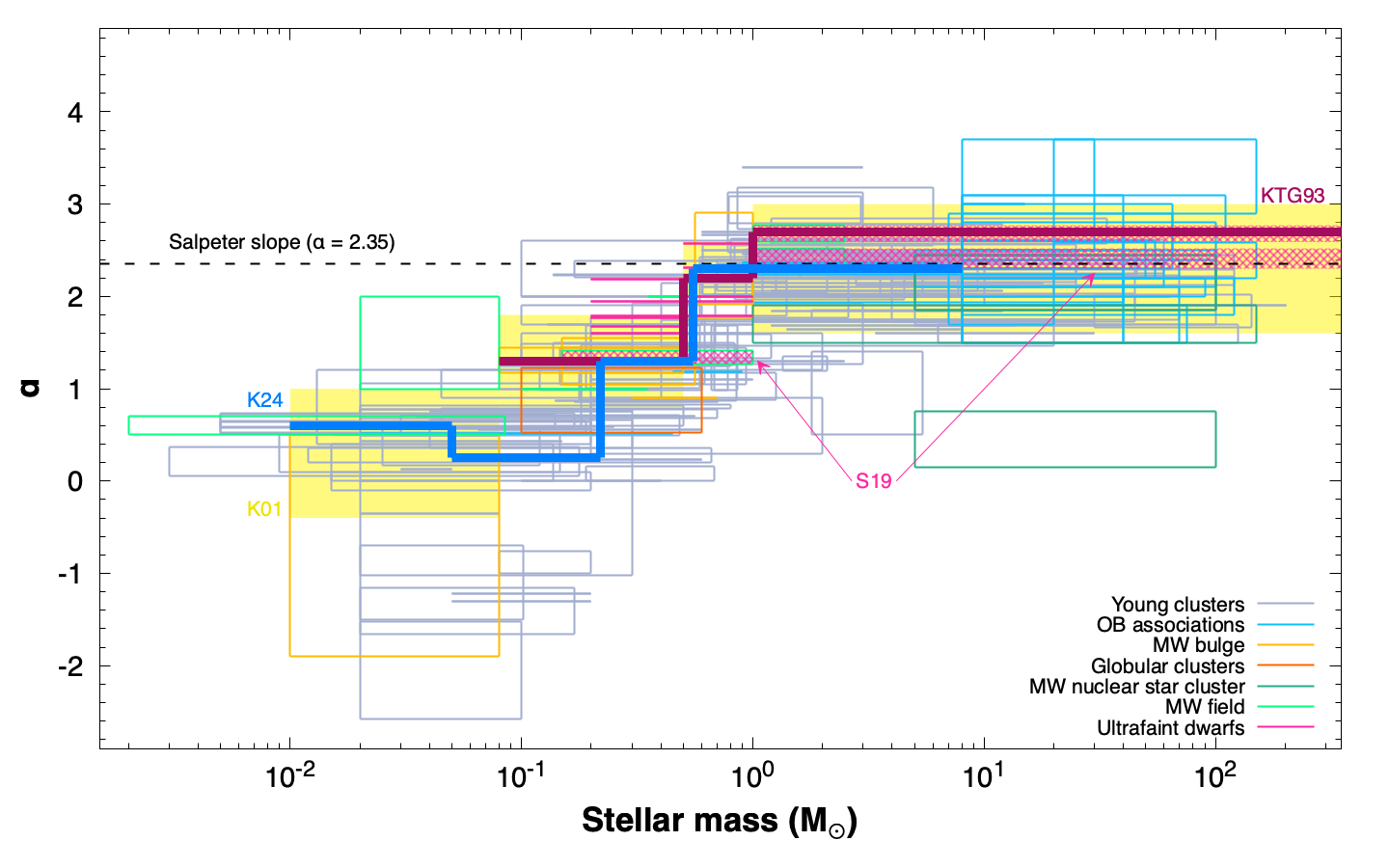}}%
\caption{The IMF slope, $\alpha$, as a function of the stellar mass on the zero-age main sequence. Coloured empty boxes show power-law fits, $\alpha \pm 1\,\sigma$, over specific mass ranges for different targets (legend in the bottom right) as collected from the literature by \citet[][]{hennebelle2024}. Some of the data are estimates of sIMFs as they are for young star clusters and OB associations. The thick lines/patterned rectangles labeled ``KTG93", ``K01", ``S19" and ``K24" show the cIMF fits to resolved star count data for the Solar neighbourhood from, respectively, \cite{Kroupa+1993}, \cite{Kroupa2001}, \cite{Sollima2019}, and \citet{Kirkpatrick2024}. See Sec.~\ref{sec:res_IMF} for further information. Notice the two alternative solutions proposed by \cite{Sollima2019} above 1~$M_\odot$, which depend on the adopted resolution of the star formation history. 
}
\label{fig:slopes}
\end{sidewaysfigure*}

All three approaches have advantages and disadvantages, and all three are constantly benefiting from the  improvement of observational data. The three samples are discussed in turn in Sects.~\ref{sec:resIMF_trigphot} and \ref{sec:resIMF_stcl}. All three stellar samples~(I)--(III) have the same three problems: (a) The stellar mass required to construct the sIMF or cIMF needs to be calculated from the photometric properties of the star. (b) The star counts need to be corrected for stellar evolution since stars change their photometric properties as they evolve to, along and off the main sequence. Both of these problems are related in that they rely on the theory of stellar structure and evolution, which is uncertain (see Sect.~\ref{sec:resIMF_MLR} below). (c) The star counts, and notably those based on samples~(II) and~(III), need to be corrected for unresolved multiple systems (see Sect.~\ref{sec:resIMF_binaries} below), since not detecting a faint companion to a star affects the deduced form of the IMF \citep{Kroupa1991}. To visualize this: if a survey contains 100~stars of which, unknown to the observer, 50 are binary systems, 15 are triple and two are quadruple systems (see, e.g., the overview by \citealt{Offner+2023}), then the observer misses~86 stars! If the observer would instead know that among the 100 systems 67 are multiple and would approximate this with a binary fraction of $f_{\rm bin} = 67/100=0.67$ then the star-counts would receive a correction by~67 missing stars. This shows that assuming higher-order multiples to be counted as binaries is a good first approximation (within 10~per cent) to obtain an estimate of the IMF. 

The data and available cIMFs as derived from these and as described next are compiled and compared in Fig.~\ref{fig:slopes}. Section~\ref{sec:resIMF_final} documents the information available currently on the canonical IMF which is representative of star formation in most of the MW and the variation of the sIMF as deduced from resolved stellar populations.

\subsubsection{The trigonometric (I) and photometric (II) parallax samples of stars}
\label{sec:resIMF_trigphot}

\begin{figure*}
\centering
\makebox[\textwidth][c]
{\includegraphics[width=0.85\textwidth]{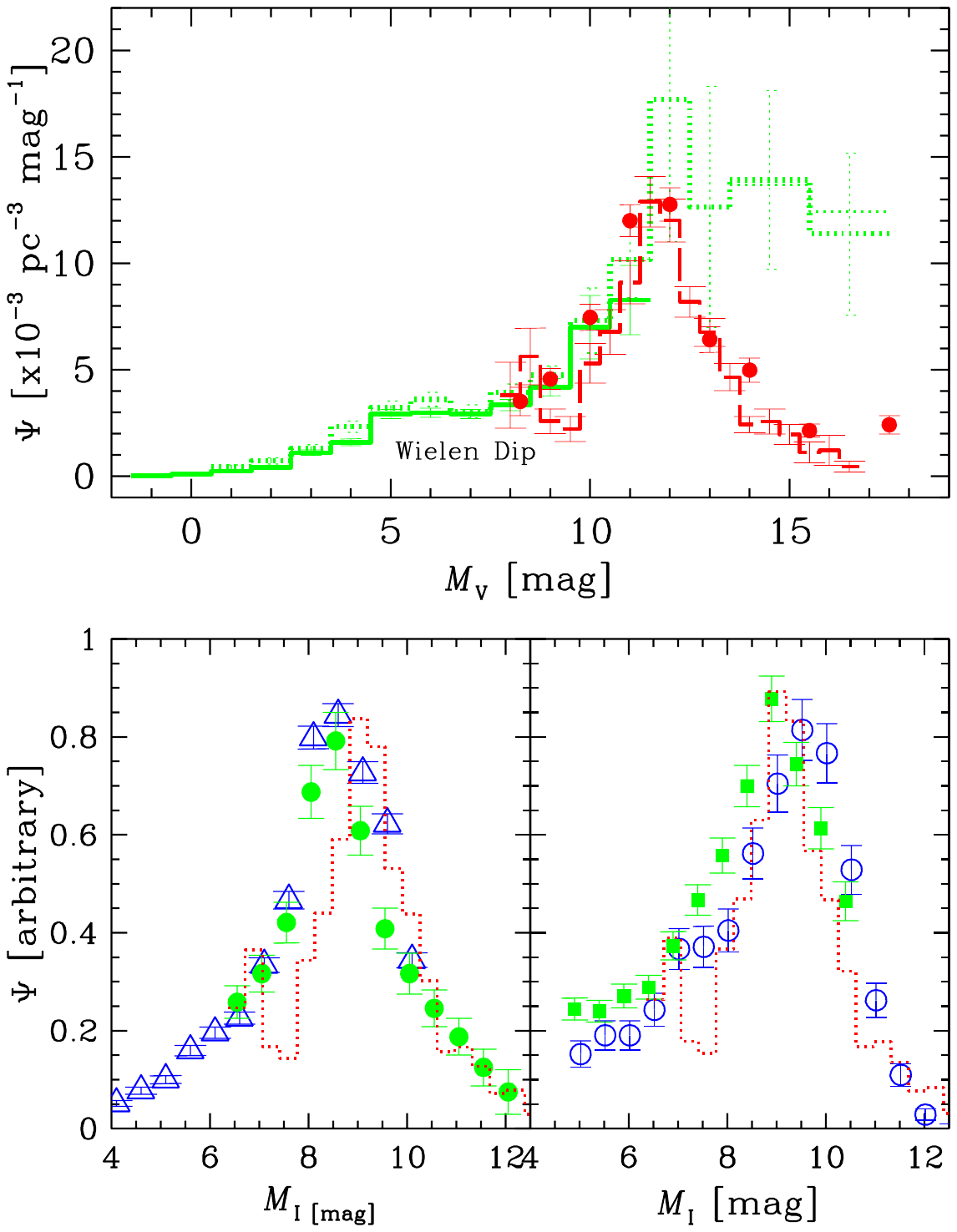}}
\caption{Stellar luminosity functions of late-type stars in the photometric $V$- and $I$-bands demonstrating that stellar luminosity functions have a sharp maximum at $M_{\rm V}\approx 11.5, M_{\rm I}\approx 8.5$. This feature has nothing to do with the stellar IMF but is a consequence of the stars to the left of the peak having a radiative core while stars to the right of it being fully convective. 
{\it Top panel}: The trigonometric-parallax based Solar neighbourhood ensemble~(I) of stars is shown as  $\Psi_{\rm c}(M_{\rm V})$ by the green histograms (solid for the sample of stars complete to about~20\,pc, dotted for the sample complete to about 5\,pc distance from the Sun). Note the ``Wielen Dip'' which comes from structure in the stellar mass--luminosity relation due to the increasing role with decreasing $m$ of the H$^-$ ion providing an opacity source in the outer stellar envelope. 
The samples of stars~(II) in pencil beam sky surveys out to distances of 100~pc based on photometric parallax distance estimates are shown by the red data and histograms. The difference of the star counts to the right of the peaks in $\Psi_{\rm c}$ comes from faint stellar companions of multiple stars not being counted in the deep surveys. 
{\it Left bottom panel}: The luminosity function, $\Psi_{\rm s}(M_{\rm I})$, in the photometric $I$-band in the globular cluster~M15 (blue triangles) and the globular cluster NGC\,6397 (green solid circles). These comprise examples of samples~(III). 
{\it Right bottom panel}: As left bottom panel but for the young Pleiades star cluster (blue open circles) and the globular cluster 47\,Tuc (green solid squares). The red dotted histogram in both bottom panels equals the solid red histogram in the upper panel (scaled appropriately). 
For more details see figure~1 in \citep{Kroupa2002}.
}
\label{fig:resIMF_stellarLFs}
\end{figure*}

Concerning (I), the stars are, by virtue of their closeness, largely well observed, such that their multiplicity properties are reasonably well known, allowing a true assessment of the IMF. However, the sample is very limited in number and is shown by the green histograms in the upper panel of Fig.~\ref{fig:resIMF_stellarLFs} which the cIMF-solution by \cite{Chabrier2003} is based on. In contrast, improving these star counts with the photometric-parallax data~(II) to obtain  more reliable statistical results leads to the ``KTG93'' or ``K01'' c/sIMF solutions (these are identical for $m<1\,M_\odot$) plotted in Fig.~\ref{fig:slopes}, respectively, by \cite{Kroupa+1993} and \cite{Kroupa2001}. This work explicitly takes into account that parallax measurements have uncertainties such that corrections for Lutz-Kelker bias are applied. Also, an empirically-gauged stellar mass--luminosity relation (Sect.~\ref{sec:resIMF_MLR}) is applied to obtain estimates of a volume-complete $\Psi(M_{\rm V})$. The latter work \citep{Kroupa2001} incorporates data from young star clusters and OB associations leading to estimates of the sIMF. 

As the trigonometric-distance-based star counts rely on the nearest stars within~20~pc for G-dwarfs down to  5~pc for faint M-dwarfs, the census of all individual stars is reasonably complete albeit small in number (most binary and triple components are included). Modern assessments of the trigonometric sample based on the Gaia astrometric space mission \citep{Gaia2016} significantly extend the distance range over which the star counts are complete and are available in \cite{Sollima2019} and \cite{Kirkpatrick2024} (``S19'' and ``K24'' in Fig.~\ref{fig:slopes}, respectively). These, and in particular the latter, are very detailed star-count studies but do not apply corrections for the Lutz-Kelker bias. The former relies on using theoretical stellar mass-luminosity relations which are highly problematic for $m<0.5\,M_\odot$, causing spurious structure in the calculated cIMF or sIMF (Sect.~\ref{sec:resIMF_MLR}), while the latter is based on empirically-gauged such relations but without explicitly taking into account the convection--H$_2$ inflection point near $0.33\,M_\odot$ (Sect.~\ref{sec:resIMF_MLR}). This is a highly critical physical property of the mass--luminosity relation and, if not dealt with correctly, it leads to spurious features in the calculated IMF and even an entirely incorrect slope. Since these Gaia-based trigonometric data extend to larger physical distances, the correction for unresolved multiple systems becomes a larger issue and will bias any calculated IMF towards flatter solutions for $m<1\,M_\odot$. The information on the cIMF for sub-stellar objects, i.e. brown dwarfs (BDs), depends on their rate of dimming with time that relies on theoretical models, their distribution in phase-space in the Galactic disc that depends on the physics of their formation in contrast to that of stars, and the binary corrections that are significantly different to those of stars. Below $m<0.15\,M_\odot$ the cIMFs of field stars and of BDs remain very uncertain, despite the latest Gaia data being available. 

Concerning~(II), photometric-parallax-based samples of stars allow a significant increase in the number of late-type stars since survey cones along many line-of-sights in the Galaxy can be added, thus significantly reducing the Poisson uncertainty of the low-mass stellar sample. Since photometric parallax surveys are flux-limited surveys but stars with the same $m$ can have different $M_{\rm V}$ due to different age, abundance and spin, the star counts need to be corrected for Malmquist bias. 
The combination of the star counts~(I) and~(II), uniquely achieved by \cite{Kroupa+1993} that assess the population of late-type stars in the Galactic disc, provides a robust estimate of $\xi_{\rm c}(m)$ down to $m\approx 0.1\,M_\odot$. 

Concerning the combination of the late-type star counts in~(I) and~(II) with the much rarer population of early-type stars in order to obtain information on the whole stellar mass range, we note that this is non-trivial and has been dealt with in greatest detail in the seminal work by \cite{Scalo1986}. The ``KTG93 IMF'' (actually the ``KTG93 cIMF'') is based on adapting Scalo's results together with the detailed star-count analysis combining~(I) and~(II). The following complications arise from the evolution of stellar orbits in the Galaxy: When dealing with field star counts, Galactic structure enters the problem by late-type stars typically having ages extending to more than 10~Gyr, while the rare early-type stars have ages extending from a few to a hundred~Myr.  Statistically significant samples of late-type stars are readily obtained from a relatively small volume around the Sun (from a few to 100~pc), while significantly larger volumes out to some~kpc distances need to be probed to assess the early-type stellar population. Most late-type stars have thus orbited the Galaxy many times and their orbits have been heated reducing their space density in the mid plane, while early type stars remain largely confined to the mid plane. Ejections of massive stars from their birth embedded clusters affect this statement though (e.g. \citealt{OhKroupa2016}). The star counts thus need to be converted to Galactic disc-plane surface densities to obtain correct estimates of the relative numbers of late- and early-type stars being born in the Galaxy. This calculation requires knowledge of the vertical scale height and the age of the Galactic disc (for details see \citealt{Scalo1986, Kroupa+1993}) and of the distribution of its star-forming MCs and their masses as low-mass, low-density clouds do not produce massive stars \citep{Hsu+2012}. As the Sun orbits the Galaxy the local SFR is likely to vary leading to apparent changes in the observed relative number of late- to early-type stars \citep{ElmegreenScalo2006}. The ``cIMF'' deduced from the Galactic field spanning $m\approx 0.1\,M_\odot$ to $m>100\,M_\odot$ thus remains uncertain in shape and normalization for $m$ larger than a few $M_\odot$. A guiding principle adopted by \cite{Kroupa+1993} was to assume that low-mass stars and massive stars always form together (unless the cloud clump that spawns an embedded cluster is of too low a mass to accommodate a massive star) such that the true sIMF must be continuous over the stellar mass range. 

In this sense, independent star counts based on populations that resemble individual star formation events are valuable. These are the star clusters, discussed next. 

\subsection{Star clusters (III)}
\label{sec:resIMF_stcl}

While~stellar ensembles~(I) and~(II) above comprise stars with a large range of ages and metallicities, star clusters~(III) have the advantage that they are composed of stars of the same age and metallicity and, thus, allow us to assess the sIMF. The disadvantage of~(III), however, is the need to combine the star-count analysis with detailed stellar-dynamical studies, because star clusters preferentially loose their low-mass stars through various astrophysical and stellar-dynamical processes and the binary population changes with time and differs from one cluster to another, even though it may have been as invariant as is the sIMF for star-formation in most of the Galaxy (more on this in Sect.~\ref{sec:resIMF_binaries}).

Thus, (i)~the early gas expulsion from a very young compact embedded cluster can have a major effect on the shape of the mass function of stars in the cluster after it revirialises if the cluster was born highly mass segregated, because the gas expulsion leads to a significant loss mostly of the low-mass stars. This effect was predicted by \citep{Haghi+2015} and has been detected in a sample of open star clusters \citep{Alfonso+2024}. A further complication is related to the criterion used to set the end of the star formation episode, which sets the conditions for the construction of the sIMF. And the energy-equipartition process driven by the many weak gravitational encounters between the stars in the re-virialized cluster leads (ii) to the massive stars sinking to the cluster's core such that energetic encounters preferentially eject them from the cluster on the mass-segregation time scale \citep{BanerjeeKroupa2012, OhKroupa2016} and~(iii) to low-mass stars being preferentially evaporated over the two-body relaxation time scale \citep{BaumgardtMakino2003}. In massive star clusters radial gradients of the stellar mass function develop through the energy-equipartition process leading to additional handles on inferring the sIMF \citep{WebbVesperini2016}. Constraining the sIMF on the basis of star counts in an observed star cluster must thus take all the above processes into account as otherwise apparent variations of the deduced sIMF emerge that are not correct. This may be the origin of some claims that the sIMF shows stochastic variations \citep[e.g., ][]{Dib+2017}.
If star formation exhibits complete stochasticity, star clusters originating from the same initial conditions may undergo vastly different long-term evolution due to the random distribution of massive stars \citep{WJ2021}.
Conversely, in a highly regularized star formation scenario with the sIMF being an optimal distribution function, minimal variation in the number and spatial distribution of massive stars would be expected among star clusters with identical initial masses. The tidal tails that are composed of stars that drift away from their cluster of origin allow a mapping of the evaporation history from the cluster and bear information on the sIMF as well \citep{WJ2021, Wirth+2024a}.

The seeming advantage of dealing with mono-age and mono-abundance stellar populations, as are available in star clusters, is thus significantly compromised by the dynamical evolution of the stellar population in any star cluster.  These processes have been studied and have unearthed the sIMF to be dependent on the density and metallicity of the embedded clusters \citep[see also \citealt{Dib2023}]{Marks+2012}. The contrary claims by \cite{Baumgardt+2023} and \cite{Dickson+2023} that the sIMF does not show such a variation are discussed in \cite{Kroupa+2024}.

Additional challenges arise through multiple populations of stars when using star clusters to investigate the sIMF. Despite the limited age range, the presence of multiple star formation events is observed in both young star clusters and ancient globular clusters (GCs). For instance, research by \cite{Beccari+2017, Jerabkova+2019} indicates that the Orion Nebula Cluster experiences multiple star formation episodes. The ejection-feedback regulation by massive stars may play a significant role \citep{Kroupa+2018, Wang+2019},
with essentially three embedded clusters forming at the same location one after another and within each other. This is possible if stellar-dynamical ejections remove the ionizing stars from the embedded cluster such that molecular gas can refill the embedded cluster.  Therefore, the IMF of a star cluster may not represent a single star formation event but rather a more extended if not sporadic process of star formation and it will be interesting to investigate if the different populations have compatible sIMFs.

Observations also reveal that GCs exhibit the presence of multiple stellar populations (MSPs), characterized by iron spreads and groups of stars with different element abundances. The iron spreads observed in most GCs (but see \citealt{Carretta2009}; Carretta \& Bragaglia 2025, submitted) allow the reconstruction of the number of core-collapse supernovae (CCSNe) having contributed to the enrichment before star formation ceased and provides additional valuable information on the sIMF in forming GCs \citep{Wirth+2022, Wirth+2024}. Globular clusters display at least two populations with distinct element compositions. Within these populations, features such as Helium variation and anti-correlations in element abundances like C-N, Na-O, and Mg-Al have been observed \citep{Gratton2004,Gratton2012,Bastian2018,Milone2022}. Certain GCs, like NGC\,2808, may even contain more than two distinct populations. In the case of NGC\,6752, which hosts three populations, \cite{Scalco+2024} find their sIMFs to be compatible with each other after taking into account that the individual populations have been differently dynamically modified due to their different spatial locations. The second population is thought to have been born more centrally concentrated \citep[see ][and references therein]{Bastian2018} and the evolution of the multiple populations through the energy-equipartition process between them is thus an interesting topic to study \citep{Decressin+2008,  Livernois+2024}, whereby the early gas expulsion process changes the relative numbers of the populations \citep{Decressin+2010}.

Numerous theoretical studies have been proposed to explain the MSPs, but none have yet been shown to comprehensively account for all observed phenomena \citep[and references therein]{Renzini2008,Wang+2020}. The top-heavy sIMF, as is predicted by its metallicity and density dependence according to \cite{Marks+2012}, is likely to account for the mass-budget problem, namely, the element-enriched population being the dominant component in GCs. It is possible though that the ``multiple populations'' formed at the same time with pockets of enrichment occurring in discrete locations in the forming very young proto-GC due to mergers of massive stars for example \citep{Wang+2020}. There would thus be no temporal evolution of a first to a second generation population. A top-heavy sIMF at low metallicity and high density would then naturally lead to the ``second (i.e., more enriched) population'' being more abundant than the non-enriched fraction of the GC stellar population. Indeed, the relatively long formation times of GCs \cite[about 10~Myr; ][]{Wirth+2022, Wirth+2024} would readily accommodate massive-star mergers in the binary-rich initial population of a forming GC. The formation of MSPs remains an active research problem.

A top-heavy sIMF at low metallicity and high gas density would imply many stellar-mass black holes (BHs) to have formed from the massive stars. Cluster survival is an issue under these conditions \citep{Haghi+2020}. If many remain in the cluster they continue to influence the long-term dynamical evolution. In cases where a large number of BHs remains to be present, GCs evolve to have a low-central density of stars, and are more prone to disruption \citep{Mackey+2008, Wang2020}. Dark star clusters may appear when the stellar-dynamical evolution has evaporated most of the stars leaving a BH-dominated star cluster \citep{BanerjeeKroupa2011,Wu+2024,Rostami+2024}.

\subsection{The mass--luminosity relation of late-type stars}
\label{sec:resIMF_MLR}

\begin{figure*}
\centering
\includegraphics[width=.4\textwidth]{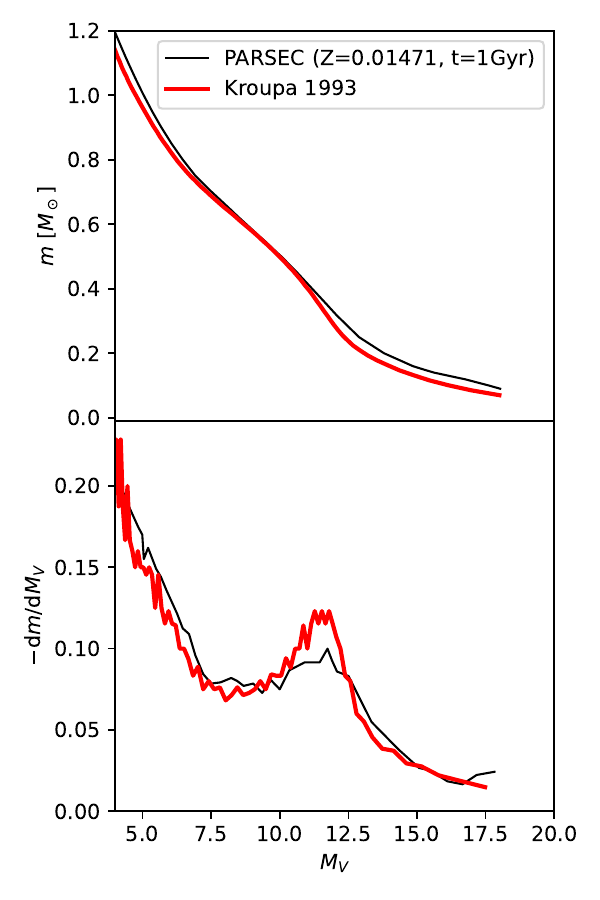}
\includegraphics[width=.4\textwidth]{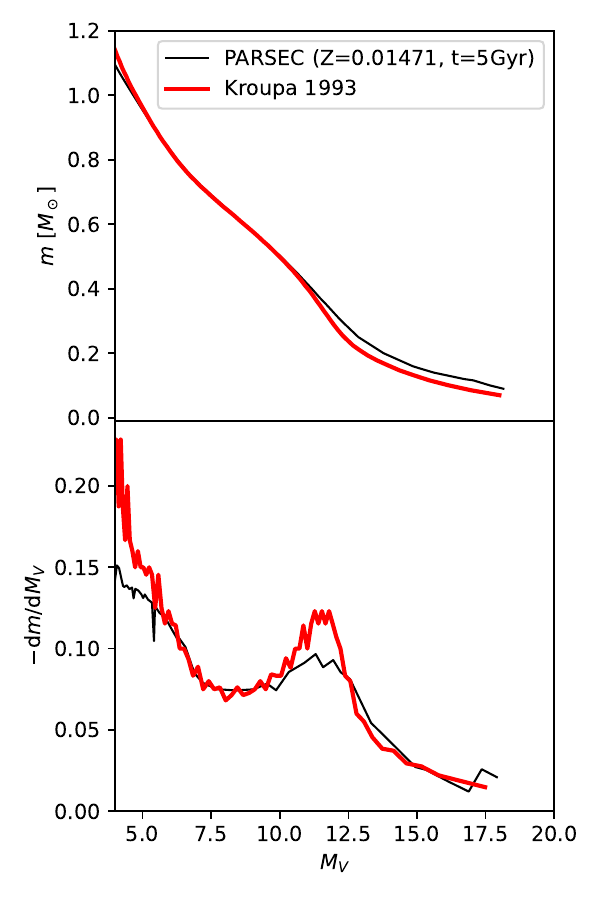}
\caption{Visualization of the problems encountered when transforming stellar luminosities to stellar masses.  The stellar mass--luminosity relation, $m(M_{\rm V})$ (upper panels), and its derivative, $-{\rm d}m/{\rm d}M_{\rm V}$ (lower panels), in the $V$-photometric pass band for a theoretical (black curves) population of stars of age~1~Gyr (left panels) and~5~Gyr (right panels) are shown. In the upper panels, main-sequence brightening of the theoretical (black curves) stars with $m<0.75\,M_\odot$ is evident in the 5~Gyr panel. The black lines are stellar isochrones generated by CMD~3.7 (\url{http://stev.oapd.inaf.it/cgi-bin/cmd\_3.7}), based on PARSEC release v1.2S, for Solar metallicity main-sequence stars ($Z=0.01471$) with an age of 1~Gyr or 5~Gyr, for the UBVRIJHK photometric system \citep{MA2006},
using the YBC version of bolometric corrections \citep{Chen+2019} and assuming a dust composition of 60\% Silicate and 40\% AlOx for M stars, 85\% AMC and 15\% SiC for C stars \citep{Groenewegen2006}, and no interstellar extinction. The red lines are the $m(M_{\rm V})$ relation gauged from Galactic-field data by \cite{Kroupa+1993} subject to strong constraints given by the stellar mass and luminosity data from the orbital solutions of binary stars and the amplitude and width of the bias-corrected stellar luminosity function, $\Psi_{\rm c}(M_{\rm V})$ shown as the red data in the upper panel of Fig.~\ref{fig:resIMF_stellarLFs}.  This relation is valid for a population of Solar-neighbourhood stars of an average age of about 5~Gyr and of solar metallicity. 
  }
\label{fig:MLR}
\end{figure*}

While in the early 1990s the role of the stellar mass--luminosity relation, $m(l)$, has been recognised and studied in much detail (see Fig.~\ref{fig:MLR}), recent researchers, e.g., working on the gIMF of stars in ultra-faint dwarf satellite galaxies, appear to be less aware of this.  A point of utmost importance to be raised here therefore is that the factor $\left( {\rm d}m/{\rm d}M_{\rm V} \right)^{-1}$ in Eq.~\ref{eq:IMF_LF} is highly critical as it constitutes the derivative of the stellar mass--luminosity relation, $m(l)$, which in this case is the mass--absolute-$V$-band magnitude, $m(M_{\rm V})$, relation. Stars with a mass larger than the transition mass of $m \approx 0.33\,M_\odot$ have a radiative core, while less massive stars are fully convective. The mass below which full convection sets in depends on the chemical composition of the star and evolves with time  as the radiative core appears and disappears near the transition mass of $0.33\,M_\odot$ with oscillations of the stellar radius, luminosity and colour that change over time \citep{MansfieldKroupa2021, MansfieldKroupa2023}. The function $m(M_{\rm V}$) is thus not differentiable at this critical mass and $\left( {\rm d}m/{\rm d}M_{\rm V} \right)^{-1}$ has a sharp extremum near this mass. The amplitude of this maximum, its width and location in $M_{\rm V}$ (or any photometric pass band) critically define structure in $\Psi_{\rm s}$ and thus $\Psi_{\rm c}$  which shows a sharp maximum. This pronounced and sharp maximum is evident in all known stellar populations as demonstrated in Fig.~\ref{fig:resIMF_stellarLFs}. 

Now, if a star count analysis relies on theoretical $m(l)$ relations then the deduced sIMF, cIMF or gIMF, $\xi_{\rm s,c,g}(m)$, will be incorrect. This comes about because theoretical $m(l)$ relations cannot capture the true inflection point near $m\approx 0.3\,M_\odot$ correctly as has been explicitly demonstrated by \cite{KT1997}: the peak location in luminosity, the amplitude and maximum of the radiative/convective feature are incorrect because the necessarily simplified calculations of the interior structures of the stellar models in one dimension (1D) cannot encompass the full physics of convection in 3D, stellar rotation, mean-molecular weight and opacity distribution throughout the star and its magnetic activity. This problem is likely the reason why some research groups using a particular set of theoretical stellar models to quantify the $m(l)$ relation obtain sIMFs or gIMFs that systematically deviate from those of others, as is evident in figure~4 of \cite{Yan2024}. 

This problem is demonstrated via Fig.~\ref{fig:MLR} which shows how the theoretical $m(l)$ relation (black curves) leads to a significantly incorrect derivative ($-{\rm d}m/{\rm d}M_{\rm V}$) compared to the correct one (in red) which has been gauged in \cite{Kroupa+1993} by stellar models, star count data and stellar masses from Kepler solutions to binary stars. 

If an observer chooses the theoretical (black) model then the observed amplitude of the stellar luminosity function at $M_{\rm V}\approx 11.5$ needs to be compensated by values of the power-law indices of the IMF, $\alpha_1$ and $\alpha_2$, that are approximately twice larger than when using the red $m(M_{\rm V})$ relation, given that the theoretical (black) $m(M_{\rm V})$ relation has an approximately two times smaller amplitude in $-{\rm d}m/{\rm d}M_{\rm V}$ than the red relation. The IMF derived from star counts will thus be more bottom-heavy in the approximate stellar mass range $0.2<m/M_\odot<0.6$ for the theoretical $m(M_{\rm V})$ relation than the empirically-gauged one, just as is evident for the deviant data in figure~4 of \cite{Yan2024}. 

Since the correctly gauged $m(l)$ relation is only available for the Solar neighbourhood mixture of stars, it will be necessary to develop appropriately gauged $m(l)$ relations at other metallicity and age values to allow a more correct assessment of the IMF in non-Solar-neighbourhood populations of late-type stars. 

\subsection{The binary-star population}
\label{sec:resIMF_binaries}

Star counts need to be corrected for unresolved multiple systems in order to calculate an IMF of all single stars. But each stellar population has its own correction to be applied because the corrections can differ significantly as re-emphasised by \cite{Kroupa2025}, depending on the dynamical age and stage of dynamical processing of the particular population under scrutiny. 
That is, the binary fraction in a low-density embedded cluster, in an open cluster, a globular star cluster or ultra-faint dwarf galaxy, for example, are all different, while the Galactic field population of stars stems from a large number of superimposed star formation events, each providing different binary populations. Thus, assuming the canonical initial binary population as derived by \cite{Kroupa1995a, Kroupa1995b} and updated by \cite{Belloni+2017}, figure~10 in \cite{Marks+2011} informs us that a 1-Myr stellar population formed at a density of $10\,M_\odot/$pc$^3$ \citep[similar to the least-massive embedded clusters known, as observed in the nearest star-forming cloud in Taurus-Auriga, ][]{KroupaBouvier2003, Joncour+2018}, will have a binary fraction of more than 95~per cent, which drops to about 93~per cent by 5~Myr and not much thereafter as such low-mass embedded clusters dissolve on a similar time scale. On the other hand, an open cluster born with a density of $10^{3.7}\,M_\odot$/pc$^3$ will have, at an age of 1~Myr, a binary fraction of 60~per cent and of 50~per cent by 5~Myr. A globular cluster born with a density of $10^{7}\,M_\odot$/pc$^3$ will evolve by about 5~Myr to still contain a binary fraction of about~20~per cent. 

Furthermore, if a dynamically mostly not processed population of 100~young systems has a binary fraction of 80~per cent, then the observer would miss 80 stars, while a dynamically older population might have a binary fraction of 50~per cent such that a low-resolution survey of 100~``stars'' would miss 50 objects. Each unresolved triple and quadruple system hides two and three more stars, respectively. 

Here it is important to understand the difference between ``dynamically old'' (e.g., twenty bulk crossing times old)  and ``astrophysically young'' (e.g., one Myr old): A high-density star-burst embedded cluster can have a low binary fraction because it is already dynamically old with a highly dynamically processed binary population despite its stars being astrophysically much younger than an older T-Tauri association that stems from many low-mass embedded clusters in which the birth binary population has suffered minor dynamical processing therewith being dynamically young. The corrections to be applied to the star counts also depend on the distribution of the mass ratios of the stellar components of the multiple systems and this depends on the history of dynamical processing of a given population. Binaries can exchange their companions in encounters, and systems with a low-mass (e.g. M~dwarf) companion are more likely to be disrupted than systems with similar-mass G~dwarfs, for example. A Galactic field ensemble of stars will have a different dynamical history and thus different properties of binary systems than found in one young star cluster.  There is in fact no single binary-star correction to be applied for all star counts. 

A researcher not fully aware of these differences may calculate different $\xi_{\rm s}(m)$ for different populations and deduce, erroneously, that the IMF differs or varies stochastically. 

In order to bring quantifiable order into this situation, it is first necessary to infer the distribution functions that define an initial population of stars in an embedded cluster, to then study how these evolve due to stellar evolution and stellar-dynamical processing. The concept that all stars form in embedded star clusters (i.e., MC clumps, see Sect.~\ref{sec:clouds}) thus arises in this context, which was born by the need to unify the high binary fraction (of about 100~per cent) observed using the newly developed infrared surveys in the early 1990s in very young pre-main sequence populations in low-mass MCs in comparison to the binary fraction of about 50~per cent in the Galactic field \citep{Kroupa1995a, Kroupa1995b}. These surveys began to uncover that the youngest stars are indeed found in embedded clusters \citep{LadaLada2003}. 

A stellar population is defined by four distribution functions: the IMF, $\xi(m)$, the birth distribution function of periods, $f_{\rm P}(P)$, or equivalently of semi-major axes, of eccentricities, $f_{\rm e}(e)$, and of mass ratios, $f_{\rm q}(q)$, with $q=m_2/m1 \le 1$, where $m_1$ and $m_2$ are the masses of the primary and secondary star in a binary, respectively. 
The latter three distribution functions of orbital elements contain information on the fraction of binaries in the population (e.g., $f_{\rm tot}=1$ if all stars are formed as binaries). While higher-order multiple systems (triples and quadruples) are more common among massive stars — with the multiplicity fraction increasing with stellar mass \citep{Moe+2017, Duchene+2013, Offner+2023} — their role in shaping stellar populations remains under discussion \citep{GoodwinKroupa2005}. In the following, we focus on binary systems, which dominate the multiplicity statistics for low- to intermediate-mass stars and are key to understanding the dynamical and observational properties of young stellar populations.
In order to allow systematic corrections of star counts for the multiplicity problem a major research effort needs to be spent on studying (a)~the properties of the above three birth distribution functions that define an initial stellar population in terms of its binary or multiplicity properties, (b)~how these evolve through astrophysical processes within the binary systems (pre-main sequence and main-sequence eigenevolution\footnote{“Eigenevolution” refers to internal processes within a stellar system that affect the evolution of binaries or multiple star systems (e.g., tides, mass transfer episodes). Pre-main-sequence eigenevolution is the net sum of all system-internal processes that include accretion onto the secondary as it passes through the accretion disc of the primary proto-star, tidal deformation at periastron of the larger pre-main sequence stars, the thereby implied transfer of angular orbital momentum and thus evolution of the orbital elements. Eigenevolution is an essential part of the theoretical description of the distribution of orbital elements of binary systems in various stellar populations \citep{Kroupa1995b}.}
),
(c)~how these distribution functions evolve in different stellar-dynamical environments (stimulated evolution), and (d) how a composite population is to be constructed from individual embedded clusters from which it derives, since the dynamical processing in each embedded cluster differs depending on its density and gas expulsion process. 
 
In addition to unresolved binaries, potential interactions within binary systems (the above-mentioned “eigenevolution”) can significantly alter their evolutionary paths. A substantial fraction of binaries and higher-order systems undergo stellar interactions—such as mergers and mass transfer episodes—that reshape their evolution \citep{Sana+2012, Sana+2013, Naoz2014ApJ...793..137N, Dvorakova+2024}. While the overall interaction rate is estimated to be around 5–10\%, this fraction is strongly mass-dependent and rises to approximately 70\% for massive OB-type stars \citep{Sana+2012, Sana+2013}.
These interactions result in stars with properties distinct from their progenitors due to changes in mass, chemistry and stellar structure (e.g. \citealt{deMink+2014, DeMarcoIzzard2017,2024ARA&A..62...21M,Eldridge2008,Langer2012,Renzo2023}. For example, stellar interactions increase the mass and rotational velocity of the accretors and stellar merger products \citep{Hills+1976, Stryker1993, Bailyn1995, Perets2015}, the latter associated with blue straggler stars \citep{Jadhav2021MNRAS.507.1699J}. 
Donor stars are stripped of a large fraction of their hydrogen envelope (appearing as a helium or Wolf-Rayet stars), which not only reduces their mass but also exposes hotter inner layers and changes their luminosity to mass ratio \cite{2019A&A...629A.134G,2023Sci...382.1287D,2023ApJ...959..125G, McClelland2016}.
These changes impact the shape of the present-day mass function, e.g., due to the systematic transformation of low-mass progenitors into higher-mass products and a non-standard stellar mass--luminosity relation. Increased rotation also causes slight reductions in effective temperature and prolongs the star's main-sequence lifetime \citep{Nguyen2022A&A...665A.126N}. This can lead to errors in mass estimations, uncertainties in age determination, and false identification of the star as an unresolved binary. 
The fact that post-interaction products differs in luminosity and age with respect to single stars also affect the interpretation of integrated spectra of stellar populations  \cite{McClelland2016,2019A&A...629A.134G}.
While recent IMF studies account for unresolved binaries, the effects of post-interaction systems on current observations still require further refinement. It is clear from the available stellar-dynamical modelling that the star-star interactions in binary systems depend on the system mass and are significantly more pronounced in massive systems thereby affecting the massive end of the s,c,gIMF more so than the low-mass end \citep{Schneider+2014, Schneider+2015, Schneider+2016, Schneider+2024}.

\subsection{The form of the stellar IMF and its variation with physical conditions}
\label{sec:resIMF_final}

The multiple efforts to infer the s,cIMF from the various star counts have converged to a functional form that can be described as being the "canonical stellar IMF" (i.e. the canonical sIMF). Any functional form that fits into the yellow region depicted in the alpha-plot (Fig.~\ref{fig:slopes}) can be seen as being adequate as a model for the sIMF. According to the IGIMF Theory (Sec.~\ref{sec:resIMF_IGIMF} below) it can be seen that, by the coincidence that we live in a MW-type galaxy (SFR of a few~$M_\odot$/yr, metallicity about Solar), the gIMF (for the Milky Way as a whole) turns out to be very similar to the sIMF such that the star counts in the neighbourhood of the Sun yield a cIMF that ends up being very similar to the sIMF deduced from star counts in star-forming regions (fig.~2 in \citealt{Jerabkova18}; fig.~2 in \citealt{Haslbauer2024}).
The steeper slope of the cIMF above about $1\,M_\odot$, i.e. the slightly top-light form as deduced from massive-star counts in the vicinity of the Sun (e.g. \citealt{Scalo1986}), is due to the Sun being in an interarm region such that the local cIMF is deficient in massive stars \citep{Kroupa+2024}. 

A useful formulation of the canonical stellar IMF is the mathematically convenient two-part power-law form (eq.~8 in \citealt{Kroupa+2024}), $\xi_{\rm s}(m) \propto m^{-\alpha_{\rm i}}$ with $\alpha_1\approx 1.3$ for $0.1<m/M_\odot \le 0.5$ and $\alpha_2\approx 2.3$ for $0.5<m/M_\odot \le m_{\rm max}(M_{\rm ecl})$. Here $m_{\rm max}=m_{\rm max}(M_{\rm ecl})\le m_{\rm max*}$ is the most-massive-star--embedded-cluster-mass relation as inferred from observational data with $m_{\rm max*}$ being the physical upper stellar mass limit \citep{WeidnerKroupa2006, Weidner+2013, Stephens+2017,   Yan2023, Zhou+2024, Chavez+2025}. The relation follows from optimal sampling (Sec.~\ref{sec:resIMF_optimalSampling}), and observationally, the most massive star that can form appears to have a mass of $m_{\rm max*}\approx 150\,M_\odot$. More massive stars most likely forming from stellar-dynamically-induced mergers of binary components (e.g. \citealt{BanerjeeKroupa2012a, OhKroupa2018}) 
and are therefore probably not formally part of the IMF. While evidence for
a variation of $m_{\rm max*}$  is lacking, it is possible that it increases at very small metallicity.

The $m_{\rm max}=m_{\rm max}(M_{\rm ecl})$ relation constitutes a fundamental constraint for star formation theory and is probably a consequence of the self-regulated nature of the star-formation process. This form of the canonical sIMF accounts for every individual star. A "system IMF", as proposed by e.g. Chabrier, has little physical meaning because the system IMF constructed from the nearby stellar census has no direct relation to star formation where the IMF of multiple systems takes a different shape related to the mass function of molecular cloud cores within which the stars form as binaries and higher-order multiple systems (dashed blue line in fig.~25 in \citealt{Kroupa+2013}). 

The canonical IMF has been widely adopted in the community and contributes to shaping our view of the Universe. \cite{Kroupa2002} tentatively noted that the then available data suggested a metallicity dependency, $\alpha_{1,2} \approx 1.3 + 0.5 \times [\text{Fe/H}]$. This is confirmed by 
a vast survey of about 93000 low-mass stars in the Milky by
\citet{Li+2023}, as well as the available studies of the low-mass IMF in old dwarf galaxies \citep{Yan2024}, who replaced [Fe/H] by the metallicity $Z$ in the formalism. According to these results, the gIMF for low mass stars ($m<1\,M_\odot$) is bottom-light (lacks low-mass stars) when the metallicity is low, and becomes bottom-heavy (dominated by low-mass stars) at super-Solar metallicity. This trend is evident also in many spectroscopic studies of elliptical galaxies (e.g. \citealt{Lonoce+2023}, for reviews see \citealt{Hopkins2018, Smith2020}).  It is important hereby to emphasise that the published bottom-heavy gIMFs in massive elliptical galaxies cannot provide the stars that are needed to enrich the galaxies to the Solar and super-Solar metallicities they are observed to have, especially given their short formation time sclaes. 

In order to obtain an understanding of the possible variation of the top-end of the sIMF, starting from 2009, the detailed analysis of the properties of ultra-compact dwarf (UCD) galaxies and of globular clusters performed in Bonn extracted quantifiable information on the variation of the high-mass part of the sIMF with metallicity, $Z$,  and density, $\rho$,  of the molecular cloud clump forming the embedded cluster. 
There is a threefold of mutually independent evidence that leads to the same result: 
\begin{itemize}

\item[(1)] The large dynamical mass to light ratios of UCDs \citep{Dabringhausen+2009} and 

\item[(2)] the overabundance of low-mass X-ray binaries in UCDs \citep{Dabringhausen+2012} can be explained independently of each other through a top-heavy sIMF, i.e., a surplus of massive stars in the sIMF when the UCDs were forming. The data imply a systematic decrease of $\alpha_3$ for $m>1\,M_\odot$ with the density of the embryo UCD. 

\item[(3)] The observation that GCs with a low present-day density have a deficit of low-mass stars which contradicts their energy-equipartition-driven evaporation was used to calculate how top-heavy the sIMF must have been in each in order to explosively expel the residual gas from the mass-segregated embryo GC so as to unbind the low-mass stars. This leads to a dependency of $\alpha_3$ on the birth density of the GC and on its metallicity that is in close agreement with the results obtained from cases~(1) and~(2) above. 
\end{itemize}

\noindent This consistency of the results (1--3) despite very different objects, methods and physics is quite compelling. 

Independently of the above, modern  analysis of direct star count data in sub-Solar metallicity star burst regions in the Local Group also now yield top-heavy sIMFs. This is notably the case for the 30~Dor region in the Large Magellanic Cloud \citep{Schneider+2018}. This region contains the $\approx 4.5 \times 10^5\,M_\odot$
massive and about 2~Myr old R136 star burst cluster for which previous star count analysis has been showing a canonical sIMF ($\alpha_3\approx\,$Salpeter) for massive stars. N-body models of a realistic binary-rich R136-type cluster predicted it to have ejected a substantial fraction of its massive stars such that adding them back in yields a sIMF that is top heavy with $\alpha_3\approx 2$
\citep{BanerjeeKroupa2012}. This has now been confirmed by proper motion measurements that uncover a large population of ejected massive stars \citep{Sana+2022, Stoop+2024}. The observational data possibly indicate even more ejections than the N-body models predict, which might be due to the massive stars residing preferentially in tight higher-order mutliple systems than the binaries assumed by \cite{BanerjeeKroupa2012}.
\cite{Stoop+2024} confirm the results by \cite{BanerjeeKroupa2012} on the true sIMF but without citing their work. Star counts in the low-metallicity 
$\approx 10^3\,M_\odot$ massive and about 20~Myr old cluster NGC~796 in the Maggelanic Bridge also yield the sIMF to be top-heavy \citep{Kalari+2018}. This cluster would have had a more top-heavy sIMF if its ejected massive stars were to be accounted for, but no such modelling has been done yet. 
Last but not least, the outer regions of the Milky Way disk are forming stars in molecular clouds of low metallicity and the star-count analysis by \cite{Yasui+2023} suggests the two clusters in the region Sh~2-209 to have top-heavy sIMFs\footnote{As for the low-mass domain, more recent work by the same team suggests a sIMF rich in low-mass objects \citep{Yasui2024}.}. Explicit modelling, taking into account the observational uncertainties, confirms these to be consistent with the above formulation of the sIMF$=$sIMF$(\rho, Z)$ dependency 
\citep{Zinkann+2024}.

We thus now have a reasonable formulation for how the sIMF varies, with $\alpha_1=\alpha_1(Z)$, $\alpha_2=\alpha_2(Z)$ and $\alpha_3=\alpha_3(\rho, Z)$.  This variation of the sIMF means that, relative to a canonical sIMF, it becomes more bottom light and more top-heavy with decreasing $Z$, increasingly bottom-heavy and top-light at increasing $Z$, and more top-heavy with increasing $\rho$. Rather remarkable is that this sense of the variation of the sIMF is as predicted by fundamental theoretical arguments related to the cooling of gas and the accretion of gas onto the forming stars as discussed in \cite{Kroupa+2024}. Apart from the variation of the shape of the sIMF with $\rho$ and $Z$, the c,gIMF in addition varies with the SFR of a galaxy due to the existence of the $m
_{\rm max}=m_{\rm max}(M_{\rm ecl})$ relation because galaxies with small SFRs can only form low-mass embedded clusters (the $M_{\rm ecl,max}=M_{\rm ecl,max}(SFR)$ relation, see Sec.~\ref{sec:resIMF_IGIMF}) such that the resulting composite IMF becomes top-light ($\alpha_3>2.3$, \citealt{KroupaWeidner2003}) which implies H$\alpha$-dark star formation \citep{Pflamm+2007}.

A noteworthy new result obtained through the above sIMF$=$sIMF($\rho, Z$) functionality is that extreme star-burst star clusters appear as bright as quasars 
\citep{Jerabkova+2017}
and evolve to ultra-compact dwarf galaxies whose mass consists mostly of neutron stars and stellar mass black holes explaining their high dynamical $M/L$ ratios 
\citep{Mahani+2021}.
The reader is directed to \cite{Kroupa+2024} for an overall assessment and comparison with theoretical notions.

\subsection{Brown dwarfs}
\label{sec:resIMF_BDs}

Brown dwarfs are often postulated to be ``stars" of such low mass (below about $0.08\,M_\odot$) that the fusion of hydrogen into helium cannot sustain their structure and luminosity (e.g. \citealt{Burrows+1989}). Consequently brown dwarfs cool and contract indefinitely. This postulate implies that the sIMF would extend continuously into the sub-stellar regime as e.g. formulated based on observational data by \cite{Kroupa2001, Chabrier2003}. This postulate however turns out to become thoroughly inconsistent with star-count data because of the very high fraction of near unity of star--star binaries in a young stellar population: It is not possible to construct a population of stars and brown dwarfs that has all stars in star--star binaries if brown dwarfs are treated exactly as stars in the mathematical formulation of such a population \citep{Kroupa+2003, Kroupa+2024}. In order to construct a stellar population with realistic binary properties as well as containing brown dwarfs, the latter need to be algorithmically separated out and described by their own IMF, the bdIMF. This is supported by brown dwarfs being extremely rarely observed to be companions of stars (the brown dwarf desert) and brown dwarfs to have a distributon function of semi-major axes that is narrow around 5~AU (while star--star binaries have semi-major axes that span sub-AU to thousands of AU, e.g. \citealt{Kroupa2025}). Also, in terms of a theoretical argument, in order for a molecular cloud core to form, through primary fragmentation, only a brown dwarf it would need to be of very low mass so as not to allow the hydrostatic core to accrete to a mass beyond the hydrogen burning mass limit. In order to be unstable and collapse under self-gravitation, i.e. to ``primarily fragment", this core would need to be very dense. Such dense low-mass cores are rare as their formation in a molecular cloud is unlikely, especially so since star forming occurs in thin cold molecular-gas filaments that are distinctly non-turbulent (for a discussion see \citealt{Kroupa+2024}). The only known natural process to form brown dwarfs is in the outer regions of accretion disks around stars where over-densities of limited mass can form naturally through perturbations \citep{Thies+2011}, as ejected embryos \citep{ReipurthClarke2001} or through the photo-evaporation of the accretion envelope in star-forming regions with a high density of massive stars \citep{KroupaBouvier2003a}. 

The above results thus imply that brown dwarfs constitute a separate population with its own bdIMF (characterized as a power law function with $\alpha_0\approx 0.3$, \citealt{Thies+2007, Thies+2008}) which has some overlap in mass with the stellar sIMF. That is, some massive brown dwarfs can form just like stars from primary fragmentation of a molecular cloud filament, while some very low mass stars can form via peripheral fragmentation in a massive stellar accretion disk. A detailed analysis of theoretical work based on gravo-turbulent clouds and star-count data finds the majority of brown dwarfs observed in the nearby Galaxy to originate as a separate population formed through ``peripheral fragmentation", i.e. the formation of gravitationally unstable over-densities in a proto-stellar accretion disk \citep{Thies+2015}.  Overall, current star-formation activity in the Milky Way gives about one brown dwarf per 3--5 stars \citep[][see also \citealt{Andersen2008}]{Kroupa+2024}.

\subsection{Optimal sampling}
\label{sec:resIMF_optimalSampling}

A key application of the sIMF (and of the cIMF and gIMF) is to use this distribution function to populate a star cluster (or region in a galaxy or the whole galaxy, respectively) with stars. The algorithm of choice has been to do so randomly, i.e., to randomly choose stars from the distribution function and to randomly place them into the system. The sIMF (or c,gIMF) is interpreted to be a probability density distribution function.  This ``stochastic description" of star formation is much applied today, but it fails to account for a number of key observations (see \citealt{Kroupa+2024} for details): the observed small dispersion of $\alpha_3$ values and the existence of the $m_{\rm max}=m_{\rm max}(M_{\rm ecl})$ relation are incompatible with the sIMF being a probability density distribution function. Also, it fails to account for the observation that star-forming disk galaxies have extended UV disks but radially significantly more limited H$\alpha$ emission. Also, the observation that star-forming dwarf galaxies have a smaller ratio of flux(H$\alpha$)/flux(UV) than Milky-Way type disk galaxies cannot be explained by stochastic star-formation since statistically the same ratio of massive ionizing stars (which are detectable through H$\alpha$ emission) to UV-emitting stars (typically B-type stars) would be expected.  Furthermore, stochastic star formation is nonphysical because stars of any mass cannot, in reality, just form anywhere. Physical systems will demand a molecular cloud clump and its cores to constrain the type of stars that can spawn there. 

These empirical and theoretical hints lead to the concept of optimal sampling \citep{Kroupa+2013, Schulz2015, Yan+2017} which rests on the opposite view, namely that star formation is not stochastic but highly self-regulated and subject to the physical boundary conditions. Optimal sampling distributes the mass of a molecular cloud clump into a sequence of stellar masses such that the sIMF constructed from this ensemble of stars in one embedded cluster has zero Poisson dispersion in any stellar-mass bin. 

This notion of optimal sampling immediately solves all the above problematical observations including leading to the observed galaxy-mass--metallicity relation \citep{Haslbauer2024}, and in particular it leads to a theoretical $m_{\rm max}=m_{\rm max}(M_{\rm ecl})$ relation verified by observational data
\citep{WeidnerKroupa2006, Weidner+2013, Stephens+2017,   Yan2023, Zhou+2024, Chavez+2025}. Optimal sampling therewith predicts H$\alpha$-dark star formation, i.e. the existence of molecular clouds and galaxies that are forming only low-mass stars \citep{Pflamm+2007}. 

It does not need to be emphasised that optimal sampling leads to a very deep and an advanced understanding of star-cluster formation and of galaxy evolution. 

\subsection{The IGIMF Theory}
\label{sec:resIMF_IGIMF}

In order to explain why the Galactic field population has a binary fraction near 50~per cent while in low-density star forming regions it is close to 100~per cent it was realized that the stars need to form in dense embedded clusters \citep{Kroupa1995a, Kroupa1995b}. This led to the notion that the stellar field populations in galaxies can be calculated by summing the contents of all the embedded clusters forming in a galaxy after taking into account the astrophysical and stellar-dynamical evolution of these as they dissolve into the field of a galaxy. The implications of this concept on the velocity distribution function of field stars was investigated by \cite{Kroupa2002a} and later developed into the integrated galactic field IMF (IGIMF) by \cite{KroupaWeidner2003}. These authors showed that the gIMF calculated according to the IGIMF Theory by adding the sIMFs of all embedded clusters differs from the sIMF: even if the sIMF is invariant, the existence of the $m_{\rm max}=m_{\rm max}(M_{\rm ecl})$ relation with $M_{\rm ecl}$ in a galaxy ranging up to a maximum value that depends on the galaxy's SFR, $M_{\rm ecl, max} = M_{\rm ecl, max}(SFR)$, implies that the gIMF becomes top-light for galaxies with small SFRs. This explained why the cIMF constructed from the Solar-neighbourhood star counts, e.g. by \cite{Scalo1986, Sollima2019}, is top-light ($\alpha_3 \approx 2.7$ rather than the Salpeter value of~2.3 for $m>1\,M_\odot$ valid for the sIMF). 

The $M_{\rm ecl, max} = M_{\rm ecl, max}(SFR)$ relation \citep{Weidner+2004, Randriamanakoto+2013} is rather important. It is a consequence of the galaxy-wide process of star formation being distributed (apparently optimally) into populations of embedded star clusters distributed according to a power-law embedded-cluster mass function (ECMF) with the total mass in stars in all the new embedded clusters formed within the time $\delta t \approx 10\,$Myr being $M_{\rm tot*}= SFR \times \delta t$. This occurs on a characteristic time scale of about $\delta t = 10\,$Myr which corresponds to the life time of molecular clouds. The ECMF has the form $\xi_{\rm ECMF} = {\rm d}N_{\rm EC}/{\rm d}M_{\rm ecl} \propto M_{\rm ecl}^{-\beta(SFR)}$ for $M_{\rm ecl,min}\approx 5\,M_\odot \le M_{\rm ecl} \le M_{\rm ecl,max}$. For a further discussion see \cite{Kroupa+2024}.

The IGIMF Theory constitutes a powerfull tool despite being based on the very simple concept that the young population of stars forming in a galaxy (or in a region thereof) is the sum of all the populations forming in all the embedded clusters in the galaxy (or in a region thereof). The IGIMF is thus calculated as an integral over the whole galaxy (or a region thereof). This concept, when paired with optimal sampling (Sec.~\ref{sec:resIMF_optimalSampling}), immediately resolves problems with understanding the galaxy-mass--metallicity relation, the radial H$\alpha$ cutoff vs extended UV disks of star forming galaxies, the lack of H$\alpha$ emission vs UV emission in dwarf galaxies. These effects come out because regions of a galaxy with a low SFR density, which is also true for dwarf galaxies in general, produce only low-mass embedded clusters which do not spawn massive stars, such that the corresponding gIMF or cIMF lack massive stars. Incorporation of the systematically varying sIMF with $Z$ and $\rho$ of the embedded clusters leads to the modern formulation of the IGIMF Theory \citep{Jerabkova18, Yan2019,Haslbauer2024, Zonoozi+2025}. A number of  computer codes have been published allowing its efficient use (see \citealt{Kroupa+2024} for a list of these and an overall overwiew of how the IGIMF Theory can clear-up problems in extragalactic astrophysics).

The modern formulation of the IGIMF Theory can be applied to understand the photometric and chemical enrichment history of star-forming disk galaxies \citep{Haslbauer2024, Zonoozi+2025}. It also 
allows the chemical enrichment of the rapidly formed elliptical galaxies to be understood self-consistently \citep{Yan2019, Yan2023}. Applying the IGIMF Theory to the rapid formation of elliptical galaxies and bulges in the early Universe leads straightforwardly to the rapid formation of super-massive black holes and their correlation with host-galaxy properties \citep{Kroupa+2020}. Associated with this is the insight that massive elliptical galaxies are mostly composed of neutron stars and stellar mass black holes which stem from the top-heavy gIMF active during the rapid assembly of the galaxies. This top-heavy gIMF, calculated self-consistently from the embedded clusters and their densities and metallicities, is also needed to synthesize the heavy elements observed in the galaxies \citep{Yan+2021}.

\subsection{Proposals for variable IMFs: consistency with observations?}
\label{sec:resIMF_consistency}

The recent observations with the JWST of apparently massive star-forming galaxies at redshift $z>10$ pose problems for the 
standard LCDM model of cosmology \citep{Haslbauer+2022} which predicts such galaxies to emerge at $z<3$ \citep{McGaugh+2024}. It is tempting to attempt to solve this problem with a non-canonical IMF. Various authors are therefore "playing around" with different IMF forms and variations in the hope of being able to show the standard dark-matter driven structure formation remains viable. Also, computer simulations of star forming molecular clouds at different metallicities are leading to various proposals of how the IMF might be varying.

In view of this and to end this section, a warning needs to be expressed: The IMF is not a distribution function that can be changed at will in order to solve a particular problem. Any formulation of the sIMF and of its variation, whether it comes from analytical arguments or from computer simulations, must comply with a large amount of observational constraints that are today available and that cannot be ignored:
\begin{itemize}

\item The sIMF must be consistent with the observed stellar populations in nearby young star clusters.

\item The cIMF calculated from the sIMF must be consistent with the Solar neighbourhood star counts. 

\item The observed properties of globular star clusters and of ultra-compact dwarf galaxies need to be consistent with the sIMF. 

\item The observed variation of the galaxy-wide gIMF of star-forming dwarf and massive disk galaxies \citep{Lee+2009, Gunawardhana+2011} needs to be consistent with the formulation of the sIMF. 

\item The metallicities of elliptical galaxies and bulges must be explained by this sIMF formulation. 

\end{itemize}

\noindent
Needless to say, the particular formulation of the sIMF in Sec.~\ref{sec:resIMF_final} incorporated into the IGIMF Theory (Sec.~\ref{sec:resIMF_IGIMF}) accounts for these constraints.

\section{IMF from extragalactic observations}
\label{sec:unres_IMF}

Novel approaches for measuring the gIMF in external galaxies have developed rapidly over the past two decades. This includes estimates from unresolved galaxy populations as well as, more recently, using similar techniques applied to integral field spectroscopic data, allowing for investigation into IMF variations within galaxies. These approaches fall into three broad classes: 
(1)~Those that estimate the shape of the low-mass end of the gIMF in passive (elliptical) galaxies \citep[e.g.,][]{2003MNRAS.340.1317V, vandokkum2010,Cappellari12} using IMF-sensitive absorption indices that allow the ratio of the number of giant stars to dwarf stars to be measured pioneered by \cite{KroupaGilmore1994}; 
(2)~Those that estimate the shape of the high-mass end of the gIMF in star-forming galaxies \citep[e.g.,][]{Lee+2009,Gunawardhana+2011}; 
(3)~Mass-to-light ratio estimates, drawing on dynamical mass measurements based on observed stellar or gas kinematics or on gravitational lensing mass estimates, in order to infer a ``mass-mismatch'' metric compared to a reference IMF, for a gIMF \citep[e.g.,][]{Treu2010, Cappellari12}. 

In particular, the combination of multiple of these approaches enables to break degeneracies and promises to provide robust constraints on the gIMF. For example, \cite{Lyubenova+2016} combine stellar population and dynamical modeling of the inner regions of elliptical galaxies to show that both approaches (1 and 3 above) yield fully consistent ``mass-mismatch'' values when adopting a broken power-law IMF with varying slope to account for an excess of low-mass stars compared to a MW-like IMF. At the same time, a single power-law IMF is excluded for most elliptical galaxies, because the excess of low-mass stars would be so high that the predicted stellar mass in their inner regions would be more than the maximum allowed by the dynamical mass measurement.

The review by \citet{Hopkins2018} highlights some key results. In broad terms, elliptical galaxies are found to have an apparent excess of low-mass ($m < 0.5\,M_{\odot}$) stars relative to a MW-like IMF, although no constraints are placed on the high-mass end of the gIMF. More recent work using integral-field and deep long-slit spectroscopy infers that this excess tends to be limited to the central regions of such galaxies \citep[see review by][]{Smith2020}. 

In contrast, star-forming galaxies are found to have a SFR dependence in the abundance of high-mass ($m > 1\,~M_{\odot}$) stars relative to a MW-like IMF, with high SFR systems favouring an excess of high-mass stars \citep{Gunawardhana+2011}, and low-SFR systems a deficit \citep{Lee+2009,Meurer2009}. In these analyses, no constraints are placed on the low-mass end of the gIMF.

Recent work exploring spatially resolved galaxy imaging and spectroscopy has largely focused on elliptical galaxies \citep[e.g.,][]{Nacho2021,Poci2022}, although analyses are now starting to be extended to star-forming galaxies \citep[][Salvador et al., PASA, submitted]{Nacho2024}. The underlying physical processes responsible for the differing gIMF shapes are still poorly constrained, although evidence favours a link to both the density (or surface density) of star formation \citep[e.g.,][]{Gunawardhana+2011}, and also metallicity \citep[e.g.,][]{Nacho2021}.

Recent observations with the \textit{James Webb Space Telescope} (JWST) have provided compelling evidence for a non-universal, top-heavy initial mass function (IMF) in high-redshift, star-forming galaxies. \cite{Hutter+2025} incorporated an evolving IMF into the ASTRAEUS simulation framework, where the IMF becomes increasingly top-heavy as the gas density in a galaxy exceeds a critical threshold. Their model reproduces the ultraviolet (UV) luminosity functions observed at $z = 5$--$15$ and predicts that galaxies with top-heavy IMFs exhibit elevated SFRs, driven by their location in local density peaks that allow for efficient gas accretion. These findings align with earlier predictions by \citet{Jerabkova+2017, Jerabkova18}, who proposed an IMF that varies with SFR and metallicity, naturally leading to top-heavy distributions in environments typical of the early Universe---low in metallicity and high in SFR.

Further observational support comes from recent JWST spectroscopic studies. For example, \citet{Bekki+2023} examined the chemical abundance patterns and compact morphology of the galaxy GN-z11 at $z = 10.6$, concluding that its high nitrogen content and rapid enrichment history can be reproduced with a top-heavy IMF and an extremely short star formation timescale. Similarly, new spectroscopic analyses presented by \citet{Cameron+2024} and \citet{Curti+2024} reveal enhanced $\alpha$-element abundances and signatures of high-mass stellar populations in a range of high-redshift systems, again favoring scenarios where the IMF is skewed toward massive stars (see Sect.~\ref{sec:chemprob} for a detailed discussion about chemical abundance data interpretation and related model uncertainties and degeneracies). Altogether, these measurements provide strong observational backing for the variable IMF framework and underscore the need to move beyond the assumption of a universal IMF in models of early galaxy evolution.

There is, as yet, no clear consensus on the gIMF shapes over the full stellar mass range between early and late type galaxies, or indeed how one may evolve into the other over the course of cosmic history. The approach implemented in the GalIMF model \citep{Yan+2017,Yan2023}, which links both through self-consistent chemical evolution, is an important step in this direction. The role of galaxy mergers and how the gIMFs of pre-merger systems might appear in a post-merger is, as yet, an aspect of this problem that has not yet been explored in any detail. 

There is clearly scope for the development of novel observational gIMF metrics to improve the reach and reliability of direct measurements in external galaxies. The more independent observational constraints that can be applied, the more robust any inferences about gIMF shapes and underlying physical dependencies will be. Recent improvements in population synthesis tools \citep[e.g.,][]{Robotham2024,Bellstedt2024} that allow for flexible IMF implementations or even non-parametric IMF description \citep{MartinNavarro+2024} will be important in supporting such developments.

\section{Cosmic IMF}
\label{sec:cosmic}

The star formation history (SFH) of the Universe and the associated growth of the stellar mass density (SMD) are both dependent on the underlying gIMF, but in subtley different ways. That means that accurate measurements of the SFH and SMD can be jointly used to infer an IMF shape. The IMF in this case, though, is some effective average of the IMF over all galaxy populations at any given epoch. This means that it is a different quantity to either the sIMF or the gIMF, and we refer to it here as a ``cosmic'' or cIMF \citep{Hopkins2018}. This concept was first explored by \citet{BG2003}, and extended by \citet{Hopkins2006}. Both these analyses find evidence for a cIMF which, at higher redshifts ($z>0.5$), needs to have a slight excess in high mass stars ($m > 1\,~M_{\odot}$) compared to a Salpeter slope. This was reinforced by subsequent work \citep{Wilkins2008a,Wilkins2008b}, although later countered in the highly cited review by \citet{MD2014}, who argued for no cIMF evolution and that a Milky Way like IMF shape could explain all the observations. They focused on a carefully selected sample of SFH estimates in the infrared (predominantly at low redshift, $z<2$) and ultraviolet (predominantly at high redshift, $z>2$). The updated measurements presented by \citet{Driver2018} found a similar result. These analyses, though, neglect other work that finds higher SFH estimates at $z>2$ using other techniques \citep[e.g.,][]{Yuksel2008,Kistler2009,Novak2017,Gruppioni2013,RR2016}, including gamma ray burst rates, and far-infrared and radio luminosity estimates of SFR. Also, \citet{chruslinska2020} show that these results can still be consistent with the presence of gIMF variations within the framework of the IGIMF model.

More recent work using {\em James Webb Space Telescope} observations of very high redshift Hydrogen Balmer line and UV luminosities are now beginning to suggest that the original Hubble Deep and Ultradeep Field results for high-$z$ (rest-frame UV) SFH measurements may be underestimated \citep{Harikane+2023, Adamo+2024, Donnan2024, Whitler+2025}. This discrepancy may be further compounded by the emerging evidence for a population of heavily obscured galaxies even at very high redshift \citep[e.g.,][]{vanMierlo2024}. Together, these findings imply a potentially higher high-$z$ SFH, which could in turn exacerbate the existing tension with the observed stellar mass density (SMD). These puzzling observations—many of which rely on rest-frame UV data—highlight the need for a self-consistent re-evaluation of the high-redshift Universe, possibly within a framework that allows for non-universal galactic and cosmic IMF (gIMF and cIMF) evolution.

There is definite scope for extending cosmic census constraints on IMF estimates for galaxy populations as a whole, including to chemical (metallicity) evolution. There is also a need for a careful analysis to integrate a population of gIMF shapes to reproduce the cIMF shape, at a given epoch, in order to better understand how the different quantities are related, and how to unify observational constraints on both jointly.

\section{IMF in models of structure formation and evolution}
\label{sec:models}

The IMF represents a key ingredient in theoretical models of galaxy formation and evolution, as it defines the number of stars that form per stellar mass bin and, hence, the number of supernovae (SNe) and the fraction of baryonic mass locked in long-living stars, for each given star formation episode. Any assumption about the IMF shape and its potential variability in time and space has thus a relevant impact on the predictions of theoretical models, mainly through the amount of energy released by SNe, the chemical enrichment pattern, and the assembly of stellar mass locked in long-living stars, which impacts the predicted $M/L$ ratios.

There is a lively debate around the universality or non-universality of the IMF. However, when it comes to the IMF of the first generations of stars, known as Population~III (Pop~III) stars, there is a general consensus of a remarkable diversity compared to what we observe in the local universe.

\subsection{The primordial IMF}
\label{sec:primo}

Before the first stars shone, the universe was composed mainly of hydrogen and helium. The lack of heavier elements, so-called ``metals'', both in gaseous and dust form, strongly affected the properties of the first-generation of stars, which are expected to be more massive than those observed in the present-day universe \citep[e.g.,][for a recent review]{Klessen2023}. Indeed, from the one hand the lack of efficient coolants reduced the fragmentation process of the primordial gas, causing the formation of protostellar gas clouds 3 orders of magnitude more massive than those formed in present-day conditions \citep[e.g.,][]{Bromm2001,Schneider2002,Omukai2005}. On the other hand, the higher temperature of the pristine gas implied a higher gas accretion rate into the protostars \citep{Omukai2001}. The combination of these two effects likely led to the formation of very massive stars, possibly reaching thousands of solar masses \citep[][]{Tan2008,hirano2015}. Low-mass stars could still form from zero-metal gas \citep{Clark2011,Latif2022}, though at a reduced pace compared to typical present-day star formation conditions.

The emergence of the first stars shaped early cosmic history in ways that crucially depend on their IMF. However, the lack of direct observational constraints makes the primordial IMF elusive. During the last decades, several groups have attempted to determine the IMF of these metal-free, Pop~III stars using numerical simulations \citep[e.g.,][for a review]{greif2015}. Although the different methods and assumptions exploited have provided variegate results, there is broad consensus that the first stars were typically massive and often ended their lives in violent explosions. 
\cite{Jaura2022} stress the importance of high-resolution studies, but also highlight their high computational cost and consequent lack of statistics\citep[see also][for previous studies]{greif2012}. Cutting-edge 3D radiation-magnetohydrodynamics simulations of Pop~III star formation, including non-equilibrium primordial chemistry, turbulence, magnetic fields, and radiation feedback, point to a dominant role of magnetic fields \citep{Sharda2024,Sharda2025}, often overlooked in previous work. 

Magnetic fields would suppress the gravitational collapse in the earliest stages, favoring gas fragmentation and reducing the maximum mass attainable by Pop~III stars. Moreover, jointly to radiation feedback, they would halt the growth of the protostar and set the upper mass cutoff of the Pop~III IMF. In particular, when magnetic fields are considered, it is difficult to create stars more massive than about 100~$M_\odot$, thus reducing the chances for pair-instability supernova explosions in dark matter mini-halos at high redshifts. 
However, these 3D simulations can only follow the evolution of the protostar for a short time compared to the timescale of star formation. This implies that protostellar fragments may eventually merge into the central object \citep{hirano2017}, leading to the formation of Pop~III stars with final masses larger than hundreds of solar
\citep[e.g.,][]{hirano2014,susa2014}

An alternative way to {\it indirectly} study the IMF of Pop~III stars is through the so-called Stellar Archaeology or Near-Field cosmology. This field exploits spectroscopic observations of individual stars in our Milky Way and nearby dwarf galaxies to measure chemical abundances of ancient metal-poor stars \citep[e.g.,][for some of the first and latest works]{christlieb2002, bonifacio2003, aguado2023b, ji2024}. The abundance trends are then interpreted through cosmological models of the Local Group assembly following the chemical evolution from the formation of Pop~III stars down to present-days \citep[e.g.,][for some of the first and latest works]{tumlinson2006, salvadori2007,hartwig2015, koutsouridou2024}.

The searches for the most-metal poor and ancient stars began more than 30 years ago \citep[e.g.,][]{beers1992,bonifacio1998}. Despite of the numerous attempts and dedicated surveys, no metal-free stars have been found yet, nor in the Galactic halo, nor in nearby dwarf galaxies. Some of the most interesting metal-poor stars discovered through the years are the so-called carbon-enhanced metal-poor stars \citep[CEMP-no, e.g.,][]{beers2005,aoki2007,bonifacio2015}. These ancient relics show an over-abundance of C (and other elements) with respect to iron, [C/Fe]$> +0.7$, and since their first discovery have been associated to long-lived stars formed in gaseous environment enriched by intermediate-mass Pop~III stars (a few tens of solar masses on the zero age main sequence) exploding as low-energy ``faint supernovae" \citep{limongi2003,iwamoto2005}. This scenario has been confirmed over the years by different groups, and the fraction and C-excess of these unusual stars exploited by different cosmological models to study the properties of Pop~III stars and limit their IMF \citep[e.g.,][]{salvadori2015,ishigaki2018,rossi2023}.
In particular, the most extreme stellar relics with [C/Fe]$>+1.5$ are consistent for being truly second-generation stars solely imprinted by primordial Pop~III SNe exploding with low-to-normal energy, 0.3--5$\times 10^{51}$ erg, \citep{vanni2023, koutsouridou2023}.

In the last few years, two rare metal-poor stars which are instead consistent for being imprinted by primordial SNe of very high energy, so-called ``hypernovae", have been identified in the Galactic halo \citep{placco2021} and in the nearby dwarf galaxy Sculptor \citep{skuladottir2021}. The progenitors of these hypernovae should have masses on the zero-age main sequence in the range 10--60~$M_\odot$. Thus, these observations demonstrate that intermediate-mass Pop~III stars can evolve as low, normal, or very energetic hypernovae, as predicted by several stellar evolution models \citep{heger2010,nomoto2013}. This implies that state-of-the-art chemical evolution models attempting to limit the properties of Pop~III stars should also account for the unknown energy distribution function (EDF) of primordial SNe. Still, when this is done, cosmological models found that the same observable, such as the fraction of C-enhanced stars as a function of the iron-abundance, can be equally well reproduced by different combinations of the Pop~III IMFs and EDFs \citep{koutsouridou2023}. Stated in different words, there are degeneracies among these two unknowns, which limit the studies of primordial stars with masses ranging between 10--100~$M_\odot$ \citep{koutsouridou2023,vanni2024,Rossi2024b}.

\begin{figure*}
\centering
\includegraphics[width=0.75\textwidth]{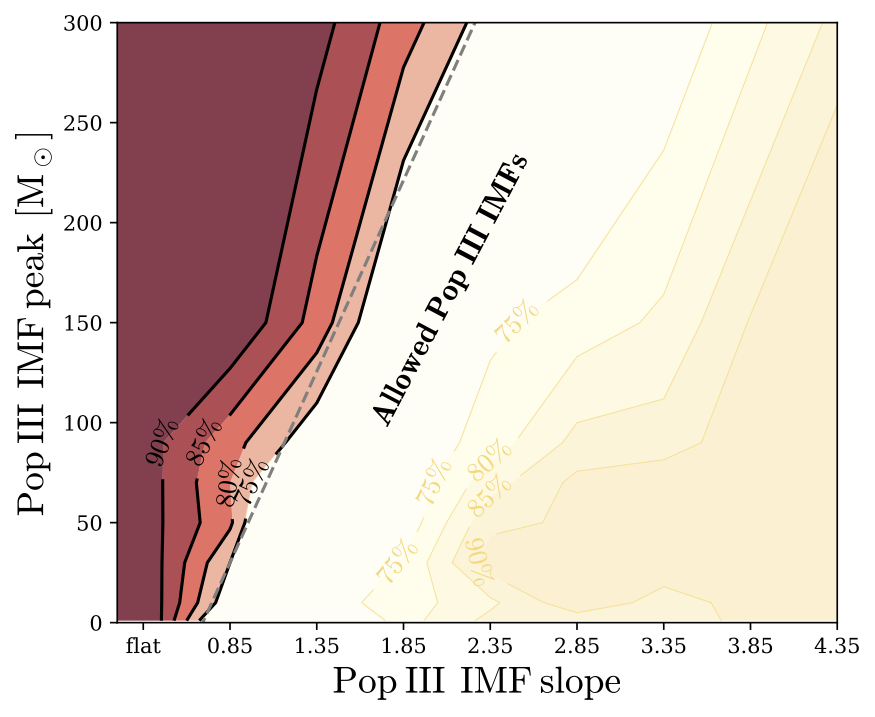}
\caption{Confidence levels at which we can exclude a Pop III IMF with characteristic mass $m_{ch}$ and slope $\alpha$, based on the (i) current non-detection of PISNe descendants 
in the SAGA catalog at [Fe/H]$< -2.5$ (red contours); (ii) possible detection of a 
{\it single} massive PISN descendant at [Fe/H]$\, < -2$ (light-yellow contours). The central area represents the Pop III IMFs that (will) remain possible. Re-adapted from \cite{koutsouridou2024}.}
\label{fig:PopIII_IMF}
\end{figure*}

On the other hand, very massive stars, with masses on the zero-age main sequence in the range 140--260~$M_\odot$ and initial metallicities below some threshold \citep[][but see \citealt{gabrielli2024}]{heger2003}, are predicted to explode as energetic Pair Instability Supernovae (PISNe) with energies that increases with the progenitor mass (from $10^{52}$ erg to $10^{53}$~erg). These destructive events leave no remnants and imprint the surrounding gas with a distinctive nucleosynthetic pattern \citep{heger2002,takahashi2018}. Thus, finding the imprints of PISNe in the chemical inventory of the atmospheres of ancient stars would offer a strong constrain to the high-mass end of the primordial IMF. Recently, \cite{koutsouridou2024} use N-body simulation of MW-like analogues coupled with the {\tt NEFERTITI} semi-analytical model (SAM), to show that {\it even a single} PISN descendant ($100\%$ of its metal from these sources) will be able to strongly limit the large parameter space for the Pop~III IMF, excluding the bottom-heaviest IMFs (see light-yellow area in Fig.~\ref{fig:PopIII_IMF}).

Unfortunately, searches for stars displaying unambiguous signs of PISN pollution have been inconclusive up to now \citep[see][for a new methodology to identify promising candidates]{Aguado2023}. Earlier claims that the very metal-poor star LAMOST~J1010+2358 shows the unique signature of chemical enrichment from a 260-$M_\odot$ PISN \citep[extremely low sodium and cobalt abundances, joint to a peculiar odd–even effect;][]{Xing2023} have been severely questioned by follow-up observations that extended the analysis to key elements such as C and Al \citep{Skuladottir2024,Thibodeaux2024}. Due to the tendency of stars to be born in associations, it is possible that even the most metal-poor stars have been enriched by two or more progenitors, which would dilute the unique signatures of PISN nucleosynthesis \citep{2019ApJ...870L...3H}. Another possibility is that PISN-enriched environments, which have [Fe/H]$\approx -2$ \citep{Karlsson2008,debennassuti2017}, can quickly cool their gas forming normal (Pop~II) stars which, in a few Myrs, start contributing to the chemical enrichment, thus washing out the key signature of PISNe \citep{salvadori2019,vanni2023}.

In their search for hidden Pop~III descendants in the best studied ultra-faint dwarf galaxy (UFD), Bo\"otes~I, \citet{Rossi2024b} spot three candidates: one mono-enriched and two multi-enriched. These stars show the chemical signatures of Pop~III SNe with progenitor mass in the range 20--60~$M_\odot$, spanning a wide range of explosion energies, 0.3--5$\times 10^{51}$ erg. The authors conclude that old stars born from the ashes of a single Pop~III SN are extremely rare - as also shown by more sophisticated N-body simulations for the MW assembly \citep[e.g.,][]{koutsouridou2023} and that, in these low-mass UFDs, stars enriched by a single PISN are even rarer (if they exist). This is mainly due to the expected low binding energy of UFDs, which are predicted to be associated with low-mass minihalos \citep{sf2009, salvadori2015} and thus can easily lose newly produced metals from PISN during these energetic explosions \citep[see also e.g.,][]{bromm2003}. 
In line with the hypothesis that the PISN products might be more easily ejected outside low-mass galaxies, \cite{vanni2024} recently spotted a distant (redshift $z\approx 3$) gaseous absorber in the sample presented by \cite{saccardi2023}, which has a chemical abundance pattern consistent for being imprinted by a PISN. If confirmed by future measurements of additional chemical elements, this object might provide the first observational probe of the existence of PISNe.

In conclusion, although recent JWST observations, and gravitational wave events \citep[e.g.,][]{costa2021}, may suggest the existence of extremely massive primordial stars (up to 10000 solar, \citealt{nandal2024}), we cannot exclude that very massive first stars ending their life as PISNe did not exist at all. 
As a matter of fact, to match the abundance scatter of Galactic halo stars with theoretical models, there is no need to invoke the existence of PISNe: what really matters are the different explosion energies of the first SNe \citep[][see also \citealt{romano2010}]{vanni2023, koutsouridou2023, Rossi2024a}. However, the dearth of C-enhanced stars in the bulge population can be explained by invoking an enrichment from PISNe. Thus, this can be an indirect probe that PISNe existed and enriched more massive and star-forming galaxies \citep{pagnini2023}. Furthermore, the possible pollution of the star surface after its formation due to the presence of an unseen companion should also be carefully evaluated and taken into account \citep[see][]{bonifacio2003}. 

Ultimately, stellar archaeology has provided unique observational-driven constraints on the primordial IMF, showing that the Pop~III IMF is certainly different from the one we observe in present-day star-forming regions. \cite{Rossi2021} were able to constrain at a 99 per cent confidence level the minimum or the characteristic masses of Pop~III stars, $m_{\rm min} >$ 0.8~$M_\odot$ or $m_{\rm ch} >$ 1~$M_\odot$, through the comparison of the predictions of their chemical evolution models with data for four well-studied UFDs, Bo\"otes~I, Hercules, Leo~IV, and Eridanus~II. \cite{koutsouridou2024} were able to exclude a flat Pop~III IMF at a 99 per cent confidence level, and the bottom-heaviest at a 70 per cent, by interpreting the non-detection of PISN descendants among halo stars at [Fe/H]$<-2.5$ with their cosmological {\tt NEFERTITI} model (see red shahed area in Fig.~\ref{fig:PopIII_IMF}. Finally, using different chemical evolution models and observables, several groups agree in suggesting that the peak of the primordial IMF should be $m_{ch}\geq 10 M_{\odot}$ \citep[e.g.,][]{ishigaki2018, pagnini2023, koutsouridou2023, hartwig2024}.

To complement these near-field cosmology studies, we can use direct detection of Pop~III stars, which are now at our reach thanks to the JWST telescope. Indeed, a top-heavy Pop~III IMF can be potentially uncovered in high-z galaxies hosting these pristine stellar populations \citep[e.g.,][]{schaerer2002,zackrisson2011}.
In a different perspective, \citet{lazar2022} propose to use transients produced by Pop~III stars, such as extremely luminous PISNe and bright gamma-ray bursts arising from the collapse of rapidly rotating stars into black holes, to effectively constrain the shape of the primordial IMF, in particular, if it is top-heavy or not. These transients can be surveyed by current and upcoming space facilities such as, for instance, {\it Euclid} \citep{moriya2022}, NASA's {\it Nancy Grace Roman} Space Telescope \citep{moriya2023}, and the Transient High-Energy Sky and Early Universe Surveyor \citep[THESEUS, ][]{amati2021}, a mission concept currently under study by the European Space Agency (ESA).

\subsection{Chemical abundances as gIMF probes}
\label{sec:chemprob}

In the previous section, we provide a glimpse on the usefulness of chemical elements as indirect IMF tracers. In the next paragraphs, we delve deeper into this topic, extending to different cosmic epochs.

Galactic chemical evolution (GCE) models -- intended either as stand-alone bundles or as modules embedded in more complex hydrodynamical simulations, possibly performed in a cosmological context -- follow the evolution of the chemical composition of the gas out of which stars form in galaxies and are routinely used to interpret chemical abundance measurements. The abundance ratios of two elements injected into the ISM by stars that die, respectively, shortly and long after an episode of star formation inform on the galaxy formation timescales \citep{tinsley1979}. If the progenitor stars fall in distinct, well-separated mass ranges, these ratios may provide also important clues on the shape of the prevailing gIMF. Being the end products of, respectively, massive stars exploding as CCSNe on short time scales and low-mass stars in binary systems leading to supernova Ia (SNIa) explosions on longer time scales, Mg and Fe can, in principle, be exploited as sensitive indicators of possible gIMF variations. Back in the nineties, it was suggested that Fe and Mg absorption features in nuclear elliptical galaxy spectra, pointing to [Mg/Fe] ratios exceeding those of the most metal-rich stars in the solar neighbourhood, could be related to different star formation timescales, gIMF variations, and/or differential matter retention in conjunction with galactic-scale outflows \citep{worthey1992,matteucci1994,2011MNRAS.417.2962P,2021MNRAS.503.4474Y,2021MNRAS.502.5935S}. These physical processes could act alone or in combination and, once implemented in GCE models, lead to highly degenerate solutions. Other independent observations, such as the variation of the mass-to-light ratio, $M/L_B$, as a function of the blue luminosity of elliptical galaxies \citep{faber1976}, can be used, in principle, to discriminate among the different processes at play.

Specific CNO isotopic ratios constitute another, even more powerful gIMF indicator. \citet{henkel1993} related the values of the oxygen isotopic ratios measured in the active nuclear regions of a bunch of nearby galaxies, \ce{^16O}/\ce{^18O}~$\approx$~150--200 and \ce{^18O}/\ce{^17O}~$\ge$~8, to an overabundance of \ce{^18O}, which could reflect the creation of a number of massive stars higher than that characterizing the central parts of the Milky Way in starbursts. On the same line of reasoning, \citet{casoli1992} postulated that selective \ce{^12C} production in massive stars could explain the high \ce{^12C} /\ce{^13C} ratios observed in some merging galaxies. A recent theoretical investigation by \citet{romano2017}, though, taking into account the dependence of the stellar yields on the initial mass and metallicity of the stars, proves that the C isotopic ratio is not very responsive to gIMF variations.

\subsubsection{The \ce{^13C}/\ce{^18O} ratio as a gIMF tracer}
\label{sec:13c18oIMF}

After a meticulous analysis accounting for both theoretical and observational uncertainties, \citet{zhang2018} have suggested that the \ce{^13C}/\ce{^18O} ratio in the molecular gas, which can be probed via the rotational transitions of the \ce{^13C}O and C\ce{^18O} isotopologues, is a much better chemical indicator of the shape of the gIMF, suitable to probe both low and high redshift environments. We know from stellar evolution and nucleosynthesis theory that most \ce{^13C} comes from low- and intermediate-mass stars, while \ce{^18O} comes only from massive stars \citep[see][for a review of CNO synthesis in stars]{romano2022}. Massive stars can produce significant amounts of \ce{^13C} when spinning quickly at low metallicities \citep[e.g.,][]{chiappini2008,limongi2018,romano2019}, however, it is currently unclear whether (and how many) massive stars at very low metallicity are fast rotators (although the current trend seems to suggest massive stars rotate faster at lower metallicities -- \citealt{2024ApJ...974...85T}). While magnetic braking by a fossil field can efficiently decrease the initial surface equatorial rotational velocity within the early stages of the main sequence evolution of solar-metallicity massive stars \citep{Keszthelyi2022}, a much less pronounced effect is expected at low metallicity, due to weakening stellar winds that reduce the magnetic braking efficiency \citep{Keszthelyi2024}. It should be noted that the effects of ISM isotopic chemistry (e.g,. fractionation), which can alter isotopic ratios set by nucleosynthesis in massive stars, can also influence conclusions on the IMF derived from isotopic abundance ratios.

Observationally, the abundance ratio of \ce{^13C} to \ce{^18O} can be easily obtained from the intensity ratio of two optically thin lines, \ce{^13C}O and C\ce{^18O}. These lines can be obtained simultaneously with current facilities, thanks to the close spacing of their rest frequencies. As a plus, they have almost identical critical densities and upper energy levels, essentially free from excitation differences. Finally, it is safe to assume that any differential lensing effect between the two lines is negligible, even if the lines are observed in strongly lensed galaxies -- see \cite{zhang2018}, for a detailed discussion of all the above points and, in general, on the use of the $I$(\ce{^13C}O)/$I$(C\ce{^18O}) line intensity ratio as a proxy for the isotopic ratio.

\citet{zhang2018} focused on a sample of four gravitational-lensed submillimeter galaxies at $z \approx$~2--3 observed with ALMA. By adopting the models by \citet{romano2017}, benchmarked on high-quality MW data, they conclude that the gIMF in powerful starbursts is skewed towards massive stars. More recent work by \cite{guo2024}, addressing main-sequence galaxies in the same redshift range and making use of a customized version of the open-source code OMEGA, which is part of the NuGrid chemical evolution package \citep{cote2017}, also points to a top-heavy gIMF. Arguably, more high-redshift systems will be analyzed in the future, providing a more complete and better picture of gIMF variations from cosmic noon to cosmic dawn.

\subsubsection{Zinc abundances in low-metallicity environments}
\label{sec:zincIMF}

\begin{figure*}
\centering
\makebox[\textwidth][c]{\includegraphics[width=\textwidth]{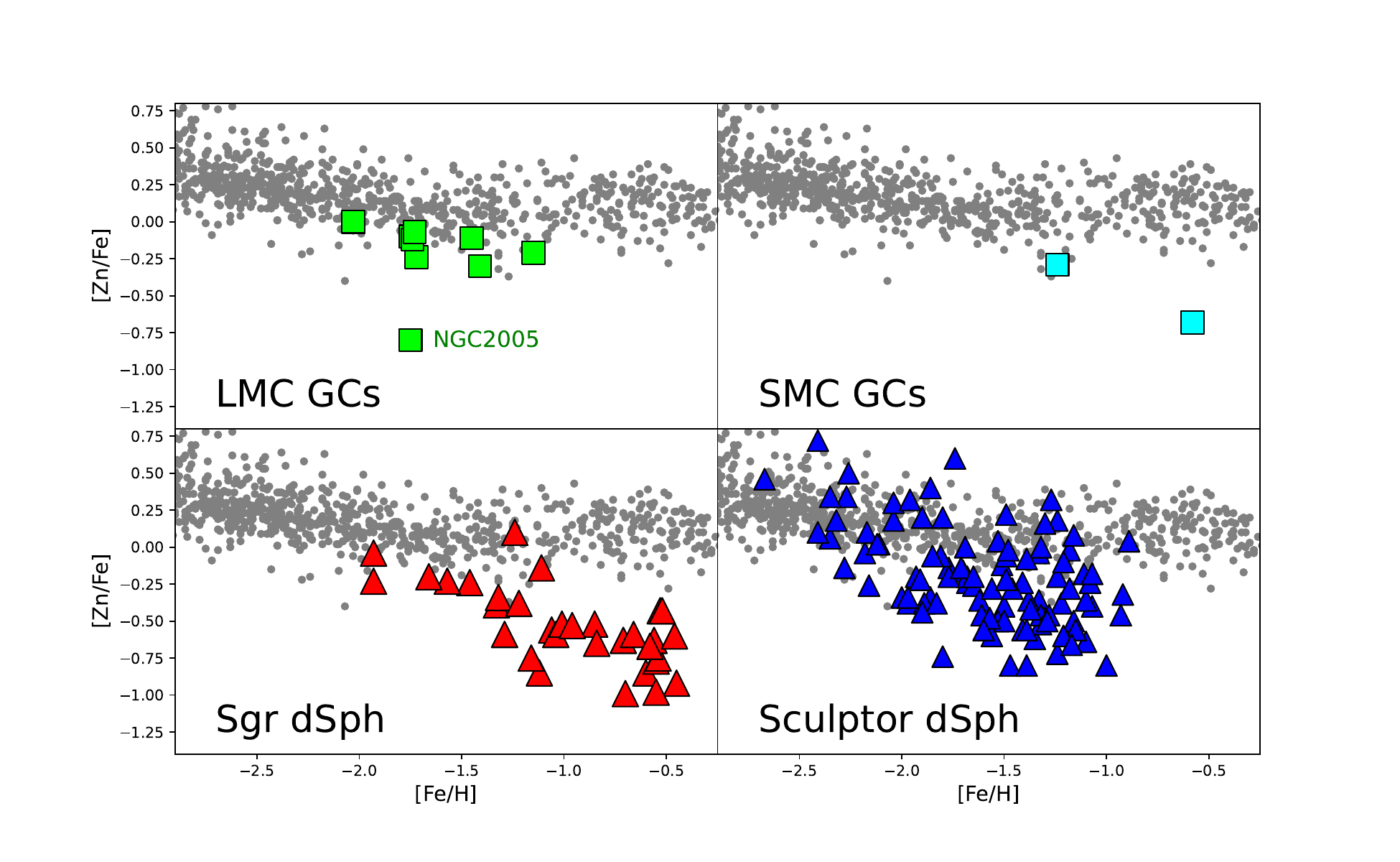}}
\caption{Behavior of the [Zn/Fe] abundance ratio as a function of [Fe/H] for MW field stars \citep[grey circles, SAGA database,][]{Suda2008} in comparison with abundances measured in different environments of the Local Group. {\it Top-left panel:} abundances of old LMC GCs \citep[green squares,][]{mucciarelli2021}. The accreted cluster NGC\,2005 is marked. {\it Top-right panel:} abundances of two intermediate-age SMC GCs \citep[cyan squares,][]{mucciarelli2023}. {\it Bottom-left panel:} abundances of field stars of the remnant of the Sagittarius dwarf spheroidal galaxy (red triangles, Liberatori et al. submitted). {\it Bottom-right panel:} abundances of field stars of the Sculptor dwarf spheroidal galaxy 
\citep[blue triangles, ][]{skuladottir2017}.}
\label{fig:zinc}
\end{figure*}

Another element whose abundance can be affected significantly by IMF variations is zinc (Zn). \citet{umeda2002} found that large [Zn/Fe] ratios, such as those measured in Galactic metal-poor halo stars, result from Pop~III SN models with deeper mass cut, smaller neutron excess, and larger explosion energies. In particular, stellar progenitors with initial mass above 20~$M_\odot$ and explosion energies in excess of $2 \times 10^{52}$ ergs are required to reproduce values as high as [Zn/Fe]$\approx$0.5. Such high-energy \citep[or hypernova;][]{paczynski1998} models explain better the trends of the Fe-peak elements, while also avoiding the overabundance of Ni that would result from deep mass-cuts in normal-energy models.

On the other hand, low [Zn/Fe] ratios have to be expected for [Fe/H]~$< -$1.5 dex if the formation of stars more massive than 20--30~$M_\odot$ is reduced/suppressed, how it might be the case of some small, metal-poor galaxies. Figure~\ref{fig:zinc} provides a summary of [Zn/Fe] abundance ratio measurements in different galaxies of the Local Group characterized by a star formation efficiency lower than that of the MW, namely, the Large and Small Magellanic Clouds (LMC and SMC, respectively), the remnant of the Sagittarius dwarf spheroidal galaxy and the Sculptor dwarf spheroidal galaxy. SMC, Sagittarius and Sculptor exhibit a clear trend, with [Zn/Fe] decreasing by increasing [Fe/H], possibly suggesting a low contribution to Zn production by massive stars besides an increasing Fe contribution from SNeIa as the metallicity increases.

The globular cluster NGC\,2005 in the LMC displays a peculiar [Zn/Fe] ratio, displaced by $-0.65$~dex from the average [Zn/Fe] of other LMC globulars of similar metallicity \citep{mucciarelli2021}. The chemical peculiarity of NGC\,2005 is easily understood if the cluster originated in a substructure \citep[perhaps, a now disrupted LMC satellite; see][]{mucciarelli2021} that experienced a low SFR ($< 5 \times 10^{-4}\,M_\odot$ yr$^{-1}$) and, hence, a gIMF skewed against massive stars \citep{Jerabkova18}. This explanation, resting on GCE models adopting nucleosynthesis prescriptions benchmarked against the MW data \citep{romano2010}, is well founded, but not unique: \citet{salvadori2019} highlight that an under-abundance of [Zn/Fe]~$< 0$ is the key to identify possible descendants of massive (140--260~$M_\odot$) primordial stars that explode as PISNe. Regarding the specific case of NGC\,2005, it is worth noticing that the cluster appears to be deficient also in other elements, such as Si, Ca, Sc, V, and Cu. This is well accounted for \citep[see ][their figure~2]{mucciarelli2021} by a GCE model adopting an upper mass cutoff for the gIMF of its parent stellar system lower than typically assumed,  $m_{\rm up} \simeq$ 40~$M_\odot$ rather than $\sim$100~$M_\odot$.

Zinc can be used also to probe the chemical evolution of damped Lyman $\alpha$ systems (DLAs), using the S/Zn ratio (not affected by differential dust depletion) as a proxy for the $\alpha$/Fe ratio \citep[][but see also \citealt{decia2016}]{centurion2000}. This opens up the possibility of constraining the high-mass slope of the gIMF in DLAs -- a topic that certainly deserves further investigation.

\subsubsection{Other chemical diagnostics}
\label{sec:otherIMF}

Another element that can signal a lack of massive stars is manganese (Mn). At low metallicities, Mn is produced in much smaller amounts in hypernovae ($m \ge$ 20~$M_\odot$) than in normal CCSNe ($m >$ 10~$M_\odot$), making hypernovae necessary to explain the lowest Mn abundances observed in Galactic halo stars \citep[see ][]{romano2010}. Stellar abundances of Mn higher than those measured in a control sample of low-metallicity MW stars, especially in the case of simultaneous detection of low Zn abundances (see previous section), may indicate that the system under scrutiny formed less massive stars than the MW field (see Liberatori et al. submitted, for the case of Sagittarius).

Carbon-to-oxygen and nitrogen-to-oxygen ratios are also possible gIMF gauges. The slightly super-solar C/O ratios derived from absorption features in SDSS spectra of massive elliptical galaxies are consistent with the predictions of GCE models that assume gIMFs skewed towards high-mass stars \citep[][and references therein]{romano2020}. Similarly, the anomalously high N/O and C/O ratios measured in luminous, metal-poor strong N- and C-emitters at high redshifts \citep{bunker2023,cameron2023,isobe2023,deugenio2024,senchyna2024,schaerer2024} can be explained if a large number of massive Pop~III stars ($m \simeq$ 50--85~$M_\odot$) explode as faint SNe in these systems \citep{Rossi2024a}. However, in both cases the explanation is not unique. In the first case, shorter and more intense star formation in elliptical galaxies coupled to galactic winds offers a viable solution \citep{matteucci1994}. In the second case, formation of second-generation stars in proto-GCs from pristine gas and AGB stellar ejecta \citep{dantona2023}, hot hydrogen-burning nucleosynthesis within supermassive stars formed through runaway collisions in environments resembling proto-GCs \citep{charbonnel2023}, intermittent star formation in early systems \citep{kobayashi2024}, pollution from massive fast rotators \citep{nandal2024}, and differential galactic outflows at cosmic down \citep{rizzuti2024} have been invoked in turn to match the observed abundances. The interplay between abundance ratios and the IMF is complex: while ISM abundance ratios can influence the IMF, the IMF-dependent stellar enrichment can, in turn, alter those ratios \citep{2023MNRAS.518.3985S}. Disentangling these interconnected processes remains a key challenge that needs due attention.

\subsubsection{The gIMF of low-mass stars and the mean stellar metallicity within galaxies}
\label{sec:lowIMF}

As discussed hereinabove, GCE provides a unique way of constraining the IMF, alternative to other methods, such as direct star counting and stellar population synthesis. 
Since more massive stars have shorter lifetimes and provide more efficient chemical feedback than low-mass stars, GCE studies have focused mostly on estimates of the IMF for stars more massive than the Sun. As introduced above, this is accomplished by studying the abundance evolution of at least two different elements or isotopes, E$_1$ and E$_2$, that are produced preferentially by stars within distinct mass ranges, making use chiefly of the [E$_1$/E$_2$]--[E$_2$/H] diagnostic plot. With a proper choice of elements/isotopes, the [E$_1$/E$_2$]--[E$_2$/H] track is highly sensitive to changes in the gIMF affecting the relative numbers of stars responsible for the bulk of E$_1$ and E$_2$ production. As a result, the [E$_1$/E$_2$] ratio probes effectively the prevailing IMF on galactic scales.

On the other hand, stars below 0.7--0.8~$M_\odot$ have main-sequence lifetimes longer than a Hubble time and do not affect actively the chemical evolution of a galaxy. Their role is simply that of locking up a fraction of the baryons: the stuck material cools and does not enter the baryonic cycle any further. This notwithstanding, the metallicity census of long-living stars in galaxies 
is affected by the IMF slope in the low-mass regime, due to the different masses and lifetimes of the observed stars. In fact, as a galaxy gets enriched and its chemical composition evolves, stars with lower masses and longer lifetimes tend to have a metallicity biased towards lower values. The probability distribution of the mean metallicity of the observed stars can be calculated with GCE models, which constrains the IMF slope in the mass range probed by the observed stars.

This method has been exploited \citep[see][]{Yan2020} to estimate the low-mass IMF of the UFD Bo\"otes I, using a GCE model implementing an environment-dependent IMF \citep{Yan2019}. Reassuringly, the result, suggesting a mildly bottom-light gIMF for Bo\"otes I, agrees well with stellar population synthesis studies \citep{Yan2024}. Applying this method to galaxies of different ages would provide constraints on the IMF of low-mass stars in different mass ranges, the main uncertainty of the method being, as with any other GCE study, that associated to the adopted stellar nucleosynthesis prescriptions \citep[see][]{romano2010}.

\subsubsection{Uncertainties impacting gIMF determinations from chemical yardsticks}

While the sensitivity of the chemical indicators discussed above (Sects.~\ref{sec:13c18oIMF}--\ref{sec:lowIMF}) to possible gIMF variations is not in question, a number of uncertainties plaguing GCE studies prevents us from  exploiting their full potential. Reviewing them in detail is beyond the scope of this White Paper, nevertheless, in the following we provide a quick rundown of current uncertainties affecting both theory and observations.

\paragraph{Observational uncertainties}

Uncertainties associated with the process of chemical abundance determination must be disentangled into random and systematic components. Concerning the abundance determination in stellar atmospheres, a thorough discussion can be found in the review by \cite{jofre2019}. Taking as an example a common metallicity indicator as Fe, it is important to stress that, while the precision of measurements (namely, how close different measurements are to each other) for giant stars in the MW can reach 0.01~dex, even with low-resolution stellar spectra \citep[e.g., ][]{2017ApJ...843...32T,2020ApJS..249...24S}, the accuracy is typically around 0.25~dex, primarily due to uncertainties in stellar parameters, such as surface temperature, gravity, and microturbulence. The adoption of different model atmospheres -- plane-parallel 1D or 3D models, consideration or neglection of departures from local thermodynamic equilibrium (LTE) conditions\footnote{It is now well-known that {\it ``3D and non-LTE effects are intricately coupled, and consistent modeling thereof is necessary for high-precision abundances''} \citep{lind2024}. Yet, many studies in the literature still rely on 1D and/or LTE modeling.}, and adoption of incorrect log({\it gf}) values may increase significantly the systematic error. 

Regarding the determination of isotopic abundance ratios from molecular lines, line opacity, beam dilution, optical depth, selective photodissociation and chemical fractionation effects as well as neglection of cosmic microwave background corrections are some of the factors that might skew the abundance ratios inferred from the observations from the true ones \citep[see][among many others]{casoli1992,goldsmith1999,roueff2015,colzi2018,zhang2018,nomura2023,sun2024}.

\paragraph{Theoretical uncertainties}

The derivation of the gIMF via comparison of predicted abundance ratios with observed ones presents a myriad of challenges, primarily stemming from the uncertainties associated with stellar yields (i.e., the quantities of various chemical elements ejected by stars during their entire life cycles). IMF-weighted stellar yields, also accounting for the different lifetimes of the stars through detailed GCE modeling, are essential for understanding the chemical evolution of galaxies and the contributions of different stellar populations. However, the complexity of stellar evolution introduces significant uncertainties that can profoundly affect our interpretation in terms of the prevailing gIMF. Stellar convection as well as other mixing processes (thermohaline mixing, atomic diffusion, rotation-driven meridional circulation and shear instability), mass loss during different phases of a star’s life, and magnetic fields play crucial roles in determining the final yields, and each of these processes is subject to considerable uncertainty \citep[see, e.g.,][]{maeder2000,busso2007,charbonnel2007,michaud2015,romano2017,salaris2017,blouin2024,mao2024}. For instance, the mixing mechanisms in stellar interiors are not fully understood, which complicates predictions of how effectively elements are mixed within stars and transported to their surfaces. Similarly, mass loss rates, especially during the asymptotic giant branch phase, can vary significantly depending on factors such as metallicity and pulsation modes, leading to discrepancies in the estimated yields. The location of the mass cut (essentially, the boundary that separates the collapsing core from the outer layers that are ejected), possible mixing and fallback of processed materials, different SN explosion energies, asphericity, etc., further complicate the assessment of the yields for massive stars, making it difficult to ascertain how much material is returned to the ISM and, consequently, deeply affecting GCE model predictions \citep[see, e.g.,][and references therein]{kobayashi2006,romano2010,nomoto2013,limongi2018,kobayashi2020}. Additionally, when stars evolve in close binary or multiple systems, interactions between stars can dramatically alter their evolutionary paths and nucleosynthesis outcomes \citep[e.g.,][]{jose2007,mori2018,farmer2023}. These interactions introduce additional uncertainties that are not present in isolated stellar evolution models, thereby complicating the interpretation of abundance ratios. Compounding these issues are uncertainties in nuclear reaction cross sections, which are critical for modeling the nucleosynthesis processes that occur during stellar evolution. Many of these cross sections remain poorly constrained at astrophysically relevant energies, leading to further ambiguity in the predicted yields of specific elements \citep[see, e.g.,][]{Adelberger2011,Bertulani2016,Descouvemont2020}. This lack of precise data can result in substantial variations in the estimated elemental abundances that emerge from stellar nucleosynthesis. Even if all the above-mentioned uncertainties are reduced to a minimum, grids of stellar yields are computed for finite values of the initial mass, chemical composition and rotational velocity of the stars, making uncertain interpolations/extrapolations necessary \citep[e.g.,][their figure~1]{romano2010}.

\begin{figure*}
\centering
\includegraphics[width=0.75\textwidth]{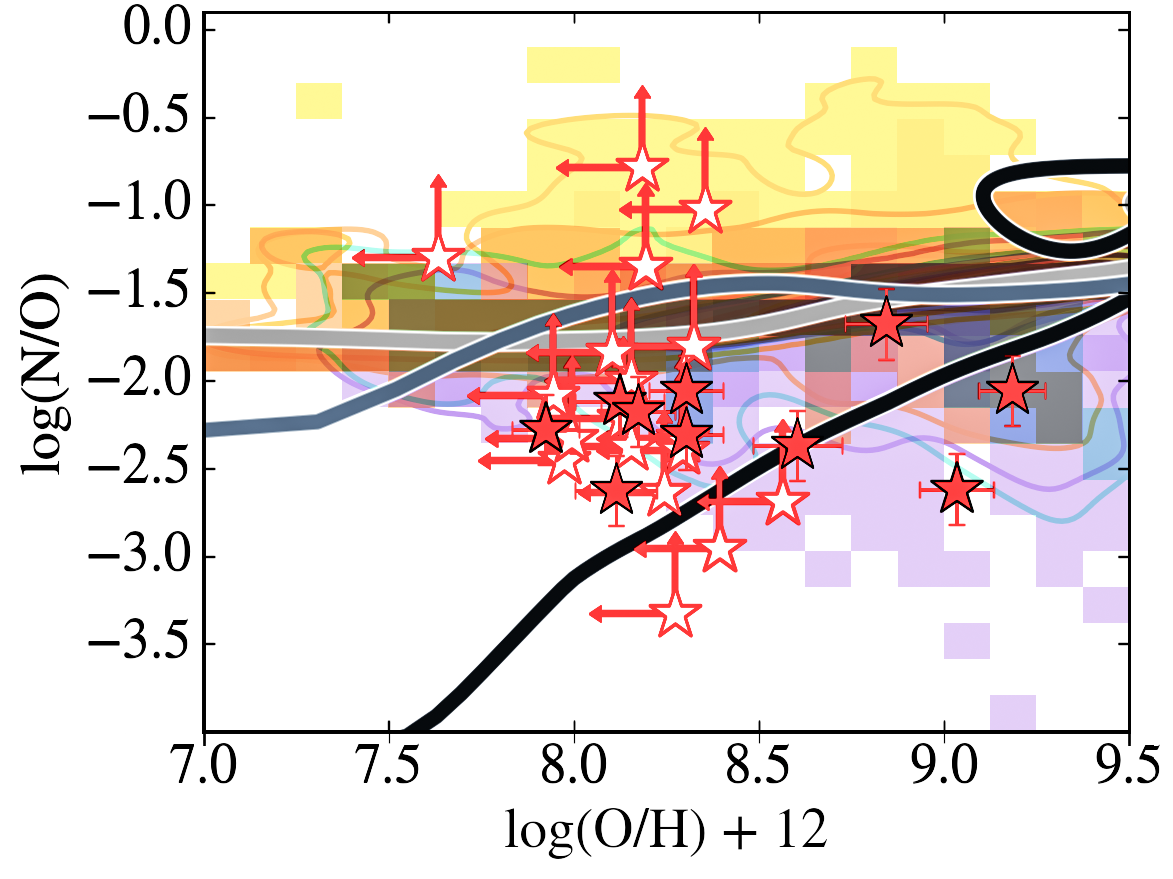}
\caption{ Log(N/O) versus log(O/H)+12 diagram for stellar populations inhabiting the halo of the Milky Way. The continuous lines represent the average trends predicted by homogeneous GCE models that implement yields for either Pop~II massive fast rotators ($v_\mathrm{rot} =$~300 and 150~km~s$^{-1}$, black and dark gray lines, respectively) or non-rotating massive stars ($v_\mathrm{rot} =$~0~km~s$^{-1}$, light gray line). The density map shows the effects of adding the localized contributions of Pop~III SNe that explode with different explosion energies (yellow: faint SNe, orange: normal CCSNe, light blue: high-energy SNe, purple: hypernovae). The star symbols represent stellar abundances homogeneously derived for a sample of halo subgiants. It is clearly seen that inhomogeneous early chemical evolution must be considered in order to reproduce the observed scatter in the data. Models and data from \cite{Rossi2024a}.}
\label{fig:early}
\end{figure*}

Another issue concerns the use of homogeneous GCE models, namely, models that assume complete and instantaneous mixing of stellar ejecta over the simulation volume. While useful for providing a broad overview of chemical enrichment processes, these models face significant limitations when applied to the study of the earliest phases of galaxy evolution. During the first fast, chaotic stages of galaxy formation, the environment is characterized by inhomogeneities due, for instance, to the accretion of satellites and/or fragmentation of early, unstable discs. These factors concur to localized, clumpy chemical enrichment that cannot be accurately captured by homogeneous models assuming a uniform redistribution of the products of stellar nucleosynthesis throughout the galaxy. As a result, inhomogeneous models are necessary to account for the observed scatter in chemical abundances, as they better reflect the complex interplay of star formation, stellar feedback, and gas mixing that occurs in the early, dynamic stages of galaxy formation \citep{Oey2000,Argast2004,Karlsson2005,Cescutti2015}. As an example (among the many that can be found in literature), Fig.~\ref{fig:early} illustrates how the scatter in N abundances of Galactic halo giant stars below the red clump can be nicely explained by accounting for localized pollution by a first generation of massive stars that end up as SNe with different explosion energies \citep[see][and references therein]{Rossi2024a}. In particular, faint SNe/hypernovae characterized by much lower/higher explosion energies than normal CCSNe, are essential to reproduce the highest/lowest observed values of log(N/O). Though the proposed solution is by far not unique, it illustrates fairly well how important is to investigate as widely as possible the parameter space of stellar evolution and nucleosynthesis models, and how deep the impact of these investigations can be.

Finally, the predictions of GCE models are inherently subject to uncertainties due to the approximations made in capturing the complex physical processes that govern the evolution of galaxies. Key processes such as gas inflows and outflows, radial gas flows, and stellar motions are often simplified. Even advanced hydrodynamical simulations, while providing more detailed insights, must rely on subgrid recipes to account for unresolved physics at the smallest scales and, therefore, are not exempt from uncertainties that can significantly influence their final outcomes. As such, a comprehensive understanding of galaxy evolution, including getting insights into the gIMF, necessitates a cautious interpretation of model predictions, acknowledging the limitations of the approximations employed. Luckily enough, some of the processes that regulate the formation and evolution of galaxies affect the abundances of different elements exactly the same way. For instance, the accretion of virgin gas dilute in the same way the abundances of all metals. Therefore, when proper abundance ratios are considered instead of absolute abundances, the effects of some parameter variations cancel out almost completely.

\subsection{Hydrodynamical simulations}

In this and the following section, we consider the IMF of the stellar population always as an input to models, as opposed to higher resolution simulations that can make predictions about the shape of the IMF itself (see Sects.~\ref{sec:gas-to-star_theory} and \ref{sec:primo}). How details of the IMF are realized in simulations of structure formation and evolution has become a focus of both idealized and cosmological galaxy-evolution models during the past 10--15 years. The IMF defines the sources of energy and metals released into the ISM, while the specific treatment of various feedback processes combined with the gas resolution defines how the feedback couples to the ISM and regulates the galactic baryon cycle.

Galaxy formation simulations frequently use IMF-averaged feedback prescriptions, where star particles are assumed to represent single stellar populations that fully sample the IMF. At poor mass-resolution, the IMF-averaged feedback from massive stellar particles typically accounts for a reasonably massive stellar population with tens or hundreds of massive stars. To account for stochasticity, the simplest approach is to discretize the feedback events by, for instance, Poisson sampling the number of SN events originating from each particle \citep{2010MNRAS.408..812S, 2014ApJ...788..121K, 2014MNRAS.445..581H}. With increasing resolution, as highlighted by \citet{2016A&A...588A..21R}, \citet{2020MNRAS.492....8A} and \citet{2021MNRAS.502.5417S}, the IMF-averaged number of SNe and the amount of ionizing radiation can significantly under- or overestimate the equivalent value in a stochastically populated stellar population when mass-resolution subceeds $\approx$1000 $M_\odot$. Beyond this resolution, the stellar masses should instead be explicitly sampled and kept track of to allow for the accounting of changes in evolutionary phase of each star in the simulation \citep{2012ApJ...745..145D}. Explicit IMF sampling leads to local inhomogeneities and more bursty stellar feedback especially when both pre-SN and SN feedback are included and properly resolved \citep{2018ApJ...869...94E,2021MNRAS.502.5417S,2025arXiv250220433B}.

As outlined in \citet{2010A&A...512A..79H} and \citet{2016A&A...588A..21R}, there exists a variety of ways to populate the IMF of a stellar particle in a simulation. One typically draws stellar masses from an input IMF with a stopping condition when the mass reservoir is used up. The \textit{stop before} and \textit{stop after} schemes either discard or keep the last stellar mass that exceeds the total available mass to be sampled. The former will always result in an IMF that favors low-mass stars, especially if the reservoir is small, while the latter will overshoot the mass that was actually available to be sampled. \textit{Stop nearest} will combine these two methods by rejecting or accepting the last sampled mass based on which ever brings the total sampled mass closer to the target mass, in hopes of averaging the effect of rejections and acceptances on a galactic scale. Finally, the stellar mass distribution can be further simplified by Poisson sampling the number of stars along a binned stellar mass distribution \citep{2017MNRAS.466..407S, 2023MNRAS.521.2196A}, thus reducing the number of trials and errors in case a more coarse grained mass distribution suffices.

With increasing resolution, the stopping condition is reached with a lower number of randomly drawn stars. This will end in an artificial deficiency of massive stars in the \textit{stop before} and \textit{stop nearest} methods, while the \textit{stop after} method will violate conservation of mass. If mass is wished to be conserved at least on a global scale while not biasing the IMF in the high-mass end,  the particle mass can be supplemented with an additional reservoir either by accounting for the overshoot mass in the next cycle of star formation (typically resulting in ``action at a distance'', wherein mass anywhere in the galaxy can be consumed, as discussed in \citealt{2016MNRAS.458.3528H, 2021MNRAS.502.5417S, 2024arXiv240508869D}), or from a localized mass reservoir \citep{2019MNRAS.483.3363H, 2021PASJ...73.1036H} drawn, for example, from the nearby star-formation eligible gas. The latter enforces strict mass-conservation of a limited and localized mass-reservoir, thus the number of massive stars can be undersampled in a physically motivated way to produce an IGIMF-like stellar population \citep{2021PASJ...73.1036H, 2023MNRAS.522.3092L}. In simulations that use the sink-particle approach, mass can instead be accumulated cyclically and the sampling done every time a sufficient amount of mass is available to spawn massive stars \citep{2017MNRAS.466.1903G}.

In the majority of the applications mentioned above, only massive stars (typically $>1$--$8$ $M_\odot$) sampled from the IMF are explicitly saved. This allows the study of the main sources of energy and chemical enrichment across a wide range of spatial and temporal scales (e.g. \citealt{2017MNRAS.466.1903G, 2018MNRAS.480.4025F}). The commonly adopted approach of tagging the stellar masses in a stellar population particle may however cause artificial clustering of SNe and increased burstiness and outflows when the particle mass resolution is above 100 $M_\odot$ \citep{2021MNRAS.506.3882S}, i.e. multiple massive stars occupy one particle. Approaches to remedy this can be divided into partly realized IMFs \citep{2019MNRAS.482.1304E, 2019MNRAS.483.3363H, 2020MNRAS.495.1035S, 2021MNRAS.501.5597G, 2024arXiv240508869D, 2025ApJ...980...41B} where massive stars are represented by individual particles while low-mass stars are clumped in population particles, and star-by-star realized IMFs \citep{2023MNRAS.522.3092L} where each star in the simulation occupies its own particle. Cosmological galaxy evolution models can realize single massive stars when the simulation volume and the target galaxy mass are sufficiently small \citep{2021PASJ...73.1036H, 2022MNRAS.513.1372G, 2022MNRAS.516.5914C, 2025ApJ...978..129A}. 

With simulations that realize massive stars as individual particles, the spatial locations of feedback events can be resolved in detail. Effective early stellar feedback has been shown to help SNe regulate star formation and produce a realistic ISM structure \citep{2020MNRAS.495.1035S} as long as the gas-mass resolution is high enough to capture the hot-phase generation of SNe. The impact of runaway massive stars in enhancing galactic outflows has been discussed in \citet{2020MNRAS.494.3328A}, \citet{2023MNRAS.526.1408S} and \citet{2023MNRAS.521.2196A}. The formation of populations of resolved star clusters with mass functions in agreement with observations can now be studied in galaxy-scale \citep{2020ApJ...891....2L, 2024arXiv240508869D} and cosmological scale \citep{2023MNRAS.522.2495G} simulations. Stars can further be treated as objects whose energy and metal output evolves according to detailed stellar evolution models \citep{2023MNRAS.522.3092L}.

The next technical steps in galaxy-scale simulations include introducing accurate small-scale gravitational dynamics methods and improvements to stellar evolution (e.g. protostellar phase, binary stars), following the recent progress in giant molecular cloud (GMC)-scale simulations \citep[e.g.,][]{2020ApJ...904..192W,2021PASJ...73.1057F, 2021MNRAS.501.4464C, 2021MNRAS.506.2199G}. Collisional dynamics will allow the simulated star clusters to undergo core collapse, mass-segregation and dynamical relaxation under the influence of an evolving tidal field, leading to reasonable estimates for the evolution of the internal structure and mass loss of galactic populations of star clusters \citep{2025MNRAS.tmp..325L}. In dense clusters, individual stars can further interact and merge with each other, beyond the initial upper limit of the stellar mass function \citep{1999AA...348..117P}. This can result in collisional growth of the most massive stars and their black hole remnants in star clusters as shown in GMC-scale simulations \citep{2024Sci...384.1488F}. Inclusion of collisional dynamics in future star-by-star simulations of galaxy formation and evolution will thus present novel avenues to study the evolution and impact of the resolved IMF in star-forming regions embedded in their galactic environment. As shown in higher resolution models by, for instance, \citet{2022MNRAS.512..216G}, stellar feedback mechanisms other than winds, UV-radiation and SNe, such as protostellar jets, may be important for regulating the collapse of gas on sub-GMC scales. More detailed, often less energetic modes of feedback should be introduced in larger scale hydrodynamical star formation simulations, however, resolving their impact requires higher gas resolution than typically achieved in such models. Room for improvement in terms of utilizing the IMF in galaxy-scale simulations thus remains.

\subsection{Semi-analytic models}

Semi-analytic models (SAMs) of galaxy formation and evolution have long been established as a complementary approach to hydro-dynamical simulations, in order to follow the redshift evolution of galaxy populations across cosmic times. These tools model the complex network of physical processes responsible for the energy and mass exchanges among the different baryonic components of galaxies by defining a system of differential equations. Each mechanism is described by means of analytical equations whose functional dependences are empirically, numerically or theoretically derived. These prescriptions require the calibration of a fixed number of free parameters (against a well defined set of observational constraints -- see below). SAMs are typically coupled with merger trees extracted from statistical realizations of the evolution of large-scale structure as traced by dark matter halos and provide a flexible tool for studying galaxy evolution in cosmological volumes. In fact, they require only a small fraction of the computational time involved with hydrodynamical simulations of comparable volumes.

The main limitation of this approach lies in the fact that SAMs are not able to follow the spatial distribution of the baryonic component (i.e., the gas distribution inside halos). Therefore, in order to
estimate the best IMF shape for a given star formation episode SAMs typically resort to the integrated properties of each galaxy component at the corresponding cosmic time. This implies that an approach such as the IGIMF is best suited for this theoretical tools, as they provide an estimate for the emerging shape of the gIMF associated with a star formation episode. The impact of an IGIMF approach on the predicted properties of model galaxies has been explored in a series of papers \citep{Fontanot17a, Fontanot18a,Fontanot24} in the framework of the GAlaxy Evolution and Assembly ({\sc gaea}\footnote{\url{https://sites.google.com/inaf.it/gaea/home}}) model \citep[see][for the latest version]{DeLucia24}. It is important to stress that {\sc gaea} is not the only SAM implementing a variable IMF approach \citep[see, e.g.,][]{Gargiulo15}, but at the moment features the largest library of tested variations and the most advanced comparison with observational constraints. We refer the reader to the individual papers for a more detailed description of each individual model implementation. In detail, {\sc gaea} has been coupled with the IGIMF defined in \citet{WeidnerKroupa05} and with the cosmic ray (CR) regulated IMF from \citet[][PP11 hereafter]{Papadopoulos11}. The latter involves numerical simulations for the characteristic Jeans mass in a molecular cloud embedded in a given CR energy density field. Moreover, {\sc gaea} features also an IMF derivation (dubbed CR-IGIMF in \citealt{Fontanot18b}), that combines the previous two approaches: as its main improvement, the CR-IGIMF predicts an evolution of both the high- and low-mass end of the IMF at the same time, as a function of the integrated physical properties of the model galaxies. This is at variance with both the IGIMF and PP11 results, that only predict a high-mass end variability of the gIMF\footnote{In this regard, it is worth mentioning recent work by \cite{bate2025} who, through a suite of radiation hydrodynamical simulations of star cluster formation in clouds of varying metallicity (from 1/100 to 3 times the solar value) receiving different levels of cosmic microwave background radiation (chosen so as to be appropriate to different redshifts, from $z = $~10 to $z = $~0), finds that the IMF becomes increasingly bottom light with increasing redshift and/or metallicity. Based on these numerical results, \cite{bate2025} provides a parametrization to be used to vary the IMF with redshift and metallicity in simulations of galaxy formation.}.

\begin{figure*}
\centering
\makebox[\textwidth][c]{\includegraphics[width=\textwidth]{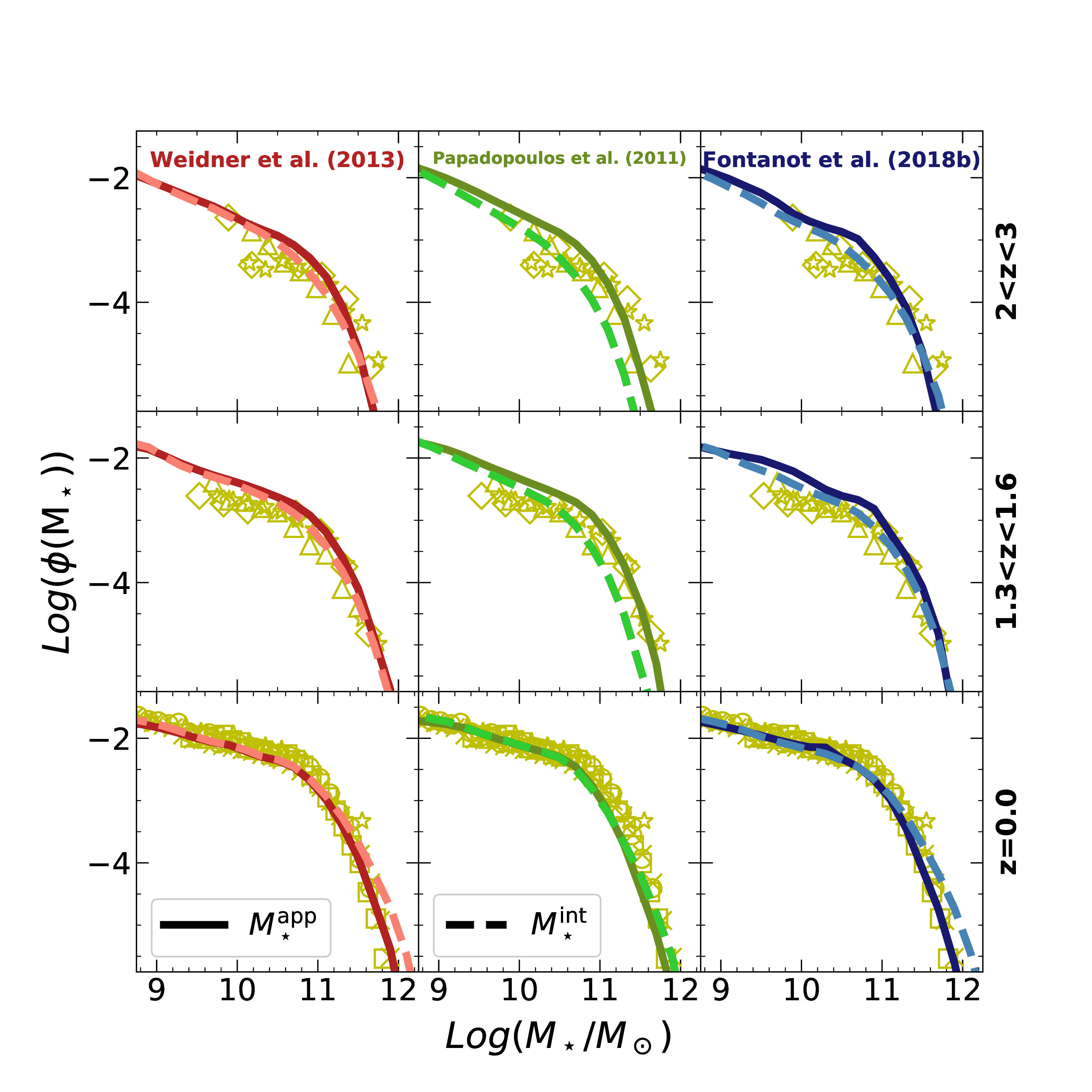}}
\caption{Redshift evolution of the GSMF, as predicted by variable IMF prescriptions coupled with the {\sc gaea} model, namely the \citet{Weidner2013} IGIMF prescription, the \citet{Papadopoulos11} numerical results on the role of CRs as IMF regulators and the CR-IGIMF proposal from \citet{Fontanot18b}. In each panel, solid lines correspond to the GSMFs obtained using stellar mass estimates $M_\star^{\rm app}$ derived from the synthetic photometry self-consistently computed for the variable IMF scenario, and assuming
a universal, MW-like IMF. Dot-dashed lines show the GSMFs obtained using the intrinsic galaxy stellar mass $M_\star^{\rm int}$ predicted by {\sc gaea}. Yellow points correspond to a collection of observational
measurements from \citet{Fontanot2009}.}
\label{fig:fabio}
\end{figure*}

In all three variable IMF scenarios, {\sc gaea} has been modified to account for the differential evolution of the baryonic mass fraction locked into low-mass stars, for the different relative abundance of Type Ia and Type II SNe and their chemical enrichment into the ISM. It is worth noting that in a variable IMF scenario, galaxy stellar mass functions (GSMFs) do no longer represent a viable calibration set for the model: this is due to the fact that all estimates of stellar masses, $M_\star$, in the literature heavily rely on the assumption of a universal MW-like IMF. Therefore, we calibrate our model runs on multi-wavelength luminosity functions (mainly local estimates from the SDSS -- $g$, $r$, and $i$ bands -- plus the redshift evolution from the $K$ and $V$ bands), that we consider a primary observational constrain (in the sense that they do not imply any additional modelling assumption involving the IMF shape). Of course, galaxy luminosities have been derived starting from a set of simple stellar populations computed self-consistently using the assumed variable IMF library. This choice ensures that we are able to associate to each star formation episode with a given IMF, age and metallicity, the correct luminosity. Moreover, this also implies that we can further compare model predictions with observational results by defining what we dubbed ``apparent'' $M_\star$, to distinguish from the ``intrinsic'' $M_\star$ predicted by the model: indeed, we use our synthetic photometry, which is derived self-consistently considering the correct IMF of each star formation event, to estimate $M_\star$ an observer would derive assuming a MW-like IMF.

Results are pretty consistent among the different variable IMF scenarios we tested. The GSMFs at different redshift corresponding to the apparent, photometrically derived, $M_\star$ show systematic deviations from the intrinsic GSMFs, which imply a somehow different interpretation for the rate of structure growth at all redshift  (see Fig.~\ref{fig:fabio}). Moreover, intrinsic and apparent $M_\star$ for individual model galaxies can be used to provide an estimate of the so-called $\alpha$-excess \citep{Cappellari12, Conroy12}: predicted trends qualitatively agree with observations, i.e. they show an increase of the deviations with respect to a MW-like IMF at increasing $M_\star$ and/or mass-to-light ratio. These trends are driven by the predicted IMF variations at the high-mass end and are mainly due to the mismatch between proper and synthetic mass-to-light ratios.

Overall, the metal enrichment of model galaxies reproduces well the evolution of the mass-metallicity relations both in the stellar and cold-gas components, but a steeper than observed relation is predicted for more massive galaxies (i.e., $M_\star>10^{10}$ $M_\odot$): although problematic at low redshift, these
predictions can explain some recent findings of supersolar metallicities for an early type galaxy (ETG) sample at intermediate redshift. Nonetheless, variable IMF models can easily reproduce the trend of increasing [$\alpha$/Fe] with $M_\star$ in local ETGs, which represents a long-standing problem for theoretical models \citep[see][]{matteucci1994}.

The main differences between the three {\sc gaea} realizations in variable IMF scenarios lie in the properties of synthetic spectral energy distributions for model ETGs, estimated as composite stellar populations summing the contribution of individual star formation events, each with its corresponding IMF shape. The CR-IGIMF run in particular is able to reproduce the observed trend \citep{LaBarbera13} in the stellar dwarf-to-giant ratio (i.e. the ratio of the mass contribution in the IMF of stars below 0.6~$M_\odot$ with respect to 1.0~$M_\odot$), which raises as a function of velocity dispersion and/or $M_\star$. The agreement between model and data is at the moment only qualitative, with model predicting trends shallower than observed, but this result shows that the CR-IGIMF is the only variable IMF scenario able to reproduce all different observational indications for a variable IMF, thus strongly supporting the need of IMF variations acting at both ends at the same time. Similar conclusions can be reached by considering the strength of individual spectral features such as $\rm TiO_2$ and $\rm NaD$, that are especially sensitive to IMF variations \citep{LaBarbera17}. Further investigation of alternative variable IMF scenarios predicting variations at both the high- and low-mass ends \citep[see, e.g.,][]{Jerabkova18,bate2025} is promising to improve our understanding of the physical parameters which are mainly responsible for the observed trends in the IMF-sensitive spectral features.

The most recent work incorporating a variable IMF into semi-analytical models is presented by \citet{Hutter+2025}, who introduced an evolving IMF prescription into the ASTRAEUS simulation framework. In this model, the IMF becomes increasingly top-heavy once the local gas density in a galaxy surpasses a critical threshold, reflecting the idea that star formation conditions in the early Universe differ significantly from those at later times. This evolving IMF leads to higher fractions of massive stars in dense, rapidly star-forming environments, which in turn enhance radiative and chemical feedback. The model successfully reproduces the observed ultraviolet (UV) luminosity functions of galaxies across redshifts $z = 5$--$15$, a key observable constraint from JWST. It also predicts that galaxies with top-heavy IMFs tend to reside in local density peaks, where efficient gas accretion drives their rapid growth and elevated SFRs. This implementation offers a self-consistent explanation for several key features observed in early galaxies and provides strong theoretical support for the non-universality of the IMF, in line with the predictions by \citet{Jerabkova18, Yan+2021}.

\section{IMF diagnostic plots}
\label{sec:plots}

With the realization that the sIMF depends on the physical properties of the star-forming molecular cloud clumps and that the observationally constrained gIMFs of star-forming disk galaxies depend on their SFR, it has become necessary to introduce diagnostics that conveniently summarize, capture and quantify the variation. Again, a distinction needs to be made whether an applied diagnostic refers to the sIMF, the cIMF or the gIMF. 

As alluded to in Sec.~\ref{sec:introd_IMFdesignation}, the community studying elliptical galaxies has introduced the ``mismatch parameter", which is a ratio of mass-to-light values arising from integrals over an IMF. In case of dynamical or lensing mass measurements, we caution the community that they can include unseen matter in the form of stellar remnants that are in the galaxy if it had a top-heavy gIMF, or cosmological dark matter that is imposed in an attempt to solve the mass-discrepancy problem arising in hot Big Bang cosmology. 
Diagnostic plots that rely on dynamical mass-to-light ratios to reach inference about the gIMF (e.g. in the central region of an elliptical galaxy) thus need to be used with care. The existence of a dark matter cusp or core becomes critical in this assessment with models of core formation being controversial. For example,  \citealt{GnedinZhao2002} show stellar feedback to never be able to produce enough feedback energy to redistribute the innermost dark matter content of a dwarf galaxy. Related to this matter, the Galaxy appears to have a significant cluster of stellar-mass black holes within its innermost $0.05\,$pc region (the "Star Grinder", \citealt{Haas+2025}) would indicate that the removal of a cusp through some baryonic or dynamical process to generate a core would not be active there as the compact cluster of black holes would also have been removed. Also, the existence of dark matter is being challenged through the Chandrasekhar dynamical friction test \citep{Kroupa2015, OehmKroupa2024}.
On the other hand, when these maximum-mass values are combined with other independent stellar-mass estimates from stellar population models with a varying gIMF, we can in principle hope to obtain robust constraints on possible unseen mass, including in super-massive black holes in the very centers of elliptical galaxies \citep[e.g.][]{Thater+2023}

Diagnostics that are invariant to a possible contribution of unseen matter are based purely on assessing the shape of the sIMF, cIMF or gIMF. Various possibilities exist \citep[see, e.g.,][]{Ferreras+2013, Martin-Navarro2019,Thater+2023}, and in \cite{Kroupa+2024} the parameters $\zeta_{\rm I}$ and $\zeta_{\rm II}$ are suggested.  The gIMF extracted from an observational analysis allows these to be easily calculated once the canonical gIMF (the canonical two-part power-law form) and the gIMF are normalised to the value $\xi(m)=1$ at $m=1\,M_\odot$, $\zeta_{\rm I}$ is the ratio of the number of stars between the hydrogen burning mass limit ($\approx 0.08\,M_\odot$ and $m=1\,M_\odot$ in the gIMF and the canonical IMF (eq.~41 in \citealt{Kroupa+2024}). Similarly, the parameter $\zeta_{\rm II}$ is the ratio of the number of stars between $1\,M_\odot$ and $150\,M_\odot$ (their eq.~42). The former parameter is thus a measure of the bottom-heaviness ($\zeta_{\rm I}>1$) or bottom-lightness ($\zeta_{\rm I}<1$) of the gIMF relative to the canonical form, and the latter parameter is a measure of the top-heaviness ($\zeta_{\rm II}>1$) or top-lightness ($\zeta_{\rm I}<1$) of the gIMF relative to the canonical form \citep[see also][]{Ferreras+2013,Martin-Navarro2019}.  This quantification can of course also be applied to the sIMF. 

Another useful diagnostic is a number that informs us whether a gravitationally bound stellar population is likely to survive as a bound object after stellar evolution mass loss. For the canonical sIMF, about 30~per cent of mass is lost over a~Gyr and longer through the astrophysical evolution of its stars (e.g. fig.~1 in \citealt{BaumgardtMakino2003}). A top-heavy sIMF will lead to more mass loss, and the population will significantly expand to the point of dissolution if more than about 50~per cent of its initial mass in stars is astrophysically evolved away. The quantity $\eta$ (eq.~29 in \citealt{Kroupa+2024}) quantifies the mass in stars more massive than $10\,M_\odot$ in a population relative to the total initial mass in the population. 

Other diagnostics can be introduced, and the authors of such are urged to exactly define these and to make an attempt to also use diagnostics that are insensitive to the presence or absence of cosmologically relevant dark matter.

\section{Conclusions and Perspectives}
\label{sec:conclusions}

The stellar initial mass function (IMF) remains one of the most influential and debated components of modern astrophysics. Its shape and potential variability impact fields as diverse as stellar evolution, galaxy formation, gravitational wave astrophysics, and cosmic chemical enrichment, with relevance also in the context of astrobiology, through setting the number of stars that could host habitable planets. This White Paper has revisited the foundational questions around the IMF and consolidated observational, theoretical, and computational findings from diverse subfields.

A major theme that emerged from this synthesis is the realization that the IMF of stars
forming in a molecular cloud clump needs not be the same as the IMF of freshly hatched stars in a region or an entire galaxy, which entails clearly distinguishing between different formulations of the IMF: the stellar IMF (sIMF) defined for individual star-forming events, the composite IMF (cIMF) emerging from larger regions, the galaxy-wide IMF (gIMF) relevant to entire galaxies, and the cosmic IMF that pushes the IMF concept to the cosmological scale of ensembles of galaxies. While historically often treated as invariant, growing evidence supports the notion that the sIMF varies with physical conditions such as metallicity and gas density. In his introductory speech at the conference {\it The initial mass function 50 years later}, Edwin Salpeter commented: {\it ``although I was hoping that my IMF was roughly right on the average, I expected that it would vary extremely strongly with varying conditions. For instance, I (and others) thought that massive stars would be strongly favored in regions of strong turmoil and possibly in regions of high gas column density in general and the young Galaxy in particular''} \citep{Salpeter2005}. In an attempt to achieve an overall consistent description from molecular cloud clump scales up to the scale of galaxies, the investigation of how the sIMF varies with physical conditions and how this transports to variations of the gIMF has made major strides forward. It is remarkable that the simple ansatz of constructing the cIMF and gIMF by adding all sIMF=sIMF($\rho$, $Z$) forming per 10-Myr epochs naturally leads to changes of the gIMF that do agree with some empirical evidence, while also accounting for the MW data. Indeed, IMF variations on cluster scales propagate to galactic scales, where IMF-sensitive diagnostics can differ across environments and epochs. This White Paper covers possible evidence of the variation of the IMF as gleaned from indirect observables, namely the photometric and elemental abundance properties of stellar populations. The over- or under-abundance of particular elements in a population, for example, could suggest that the gIMF must have had a particular form. The spectrophotometric data have uncovered massive elliptical galaxies to have significantly bottom-heavy gIMFs, with the amount of bottom-heaviness positively correlating with the metallicity  and mass of the galaxy. Extremely massive globular clusters and ultra compact dwarf galaxies, if formed monolithically, could have had such a top-heavy sIMF that their present-day masses are dominated by stellar remnants. Massive elliptical galaxies consist, by implication, mostly of stellar remnants with the visible stars being merely a ``sprinkling on the cake''. This has a deep implication for the allowed content of cold dark
matter particles. The chemical enrichment of galaxies automatically produces the observed galaxy-mass–metallicity relation because low-mass galaxies had small SFRs therewith experiencing less enrichment due to the top-light gIMF. It must be acknowledged, however, that some degeneracy in the solutions still exists.

Theoretical and computational models are increasingly able to simulate complex, multi-physics environments in which stars form, offering predictions that can be tested against empirical data. Yet, bridging the gap between simulations and observations remains a challenge, requiring attention to biases, calibration, and the subtleties of interpreting star counts and integrated-light signatures. The early formation phases of globular clusters---the early gas-embedded phase and the associated violent relaxation as the cluster emerges from this phase---need to be readdressed, also in the light of the extraordinary view of proto-globular cluster formation at high redshift provided by JWST. It would also be appropriate to perform $N$-body modelling of star-burst clusters with measured sIMFs in order to assess the correction that needs to be applied to the sIMF due to the ejected massive stars. Simulations of Pop~III star formation have become highly sophisticated and can now include simultaneously turbulence, magnetic fields and radiation feedback --- the key components that shape the formation of stars. The returned maximum mass with which Pop~III stars may be born has profound implications for our understanding of early chemical enrichment as well as for the rates of transients at high redshift that we expect to unveil with ongoing and future space missions and instrumentation.

Studying the stellar IMF remains one of the most challenging and interdisciplinary problems in astrophysics, reflecting the inherently multi-scale and multi-physics nature of star formation, which spans processes from atomic-scale chemistry in molecular clouds to galaxy-wide dynamics and cosmological context. Addressing such complexity requires cross-disciplinary approaches and coordinated efforts between different areas of expertise. The Sexten workshop, which inspired this work, underscores the importance of fostering these collaborations. By bringing together researchers from observational astronomy, theory, simulations, and laboratory astrophysics, in a setting that encourages discussion, exchange, and collaboration, we can build bridges that lead to synergies across the community.

\subsection{Outlook and Future Directions}

Looking forward, several priorities can help guide the next steps in IMF research:
\begin{itemize}
	\item	Standardizing terminology and methodology across the field will improve communication and clarity, especially when comparing results from different environments or scales. This is not an obvious and easy transition, as this White Paper itself testifies. Yet, stating unambiguously which IMF is studied (sIMF, cIMF, gIMF, or cosmic IMF) and which notation is assumed in future publications would help.
	\item	Developing community-driven, flexible, portable, open-source tools that allow for consistent IMF modeling across a broad range of applications will increase accessibility and reproducibility and accelerate our progress as scientific community.
    \item	Investing in multi-scale modeling efforts, connecting star formation processes in individual molecular clouds to the properties of galaxies and their evolution, will help unify different theoretical perspectives.
	\item	Leveraging the full potential of current and around-the-corner surveys and instrumentation---including Gaia, ALMA, JWST, ELTs, and large spectroscopic campaigns, such as 4MOST and WEAVE---offers an unprecedented opportunity to place robust empirical constraints on the IMF across a wide range of environments and cosmic epochs. Gaia continues to revolutionize our view of the local Universe by enabling detailed studies of resolved stellar populations, open clusters, and tidal structures. At the same time, ALMA pushes its eye beyond the curtains of dust that hinder our comprehension of star formation in far away regions of our Galaxy and in powerful starbursting galaxies at cosmic noon. At higher redshifts, JWST is already uncovering stellar populations and star formation conditions in the early Universe. The next generation of ground-based facilities—such as the Extremely Large Telescope (ELT) and the Thirty Meter Telescope (TMT)—will deliver deep, high-resolution spectroscopy critical for probing unresolved populations and star formation in distant galaxies. Looking further ahead, future missions and observatories like the Nancy Grace Roman Space Telescope (RST), Euclid, the Square Kilometre Array (SKAO), and the Large Synoptic Survey Telescope (LSST, now the Vera C. Rubin Observatory) promise to extend these studies even further. These facilities will provide deep, wide-field imaging and spectroscopy, sensitive radio observations of cold gas and star-forming environments, and massive time-domain datasets. Together, they will allow us to test IMF variations in regimes previously inaccessible, across cosmic time, galactic environments, and spatial scales.
	\item	Recognizing the value of focused, interdisciplinary meetings, like the Sexten workshop, which allow researchers to move beyond disciplinary silos, confront assumptions, and identify synergies between methodologies. As an example, the complexities involved in stellar evolution, combined with the unpredictable nature of stellar interactions and the limitations of nuclear physics data, significantly hinder the ability to confidently model and interpret the chemical evolution of galaxies. Consequently while, in principle, abundance ratios can provide valuable insights into stellar populations, the inherent uncertainties must be carefully considered and accounted for in GCE models to avoid misleading conclusions about the underlying IMF. In the case of integrated stellar populations, there is an ``abundance-ratio/IMF degeneracy'' that affects the IMF inference and, therefore, varying IMF-sensitive indices should be used simultaneously. At the same time, the uncertainties affecting the spectroscopic data that are used to constrain the models themselves should be carefully evaluated. This clearly requires an honest and multidisciplinary approach.
\end{itemize}
Regarding the Gaia mission and the wealth of data about the kinetic properties of Milky Way stars it provided, significantly enhancing the detection rate of young open star clusters and tidal tails, it is worth spending a few words of caution. The Gaia mission presents a remarkable opportunity to advance studies on the IMF in star clusters and deepen our understanding of the relationship between IMFs and star-forming environments. For instance, by examining the long-term evolution of a large sample of low-mass open clusters, it may be possible to determine whether the IMF follows a stochastic or self-regularized (optimal sampling) pattern.
However, as detailed in Sect.~\ref{sec:res_IMF} of this White Paper, any star-count data analysis based on a parallax-limited sample must apply the Lutz-Kelker bias correction as otherwise the deduced stellar volume densities are systematically in error affecting the amplitude of the calculated cIMF. Also, purely theoretical models, namely, isochrones of stars, cannot be relied on to calculate the stellar masses from their luminosities because these will lead to wrong mass estimates and incorrect stellar number densities as explained in Fig.~\ref{fig:MLR}. Instead, carefully empirically-gauged stellar mass-luminosity relations need to be used taking into account the pre-main sequence, main sequence and post-main sequence evolution of the stars. Furthermore, while Gaia data for individual stars are reliable, the possibility nevertheless exists that some of the stars are multiple systems. Missing to count companions biases the star counts and thus the calculated cIMF.

\subsection*{A Final Note on Collaboration}

The complexity of the IMF—its origins, its variation, and its role in shaping galaxies—requires an equally complex and collaborative response. No single dataset, method, or perspective will be sufficient. Instead, continued progress will depend on building shared language, shared infrastructure, and a culture of openness across the community.

This White Paper is not a conclusion, but a step toward that collaborative future. By linking the physics of star formation to the grand structure of the cosmos, the IMF remains a powerful thread connecting many domains of astrophysics—and it is through interdisciplinary dialogue and joint effort that we will continue to unravel its mysteries.

\newpage
\section*{Acronyms}
\begin{table}[h]
\fontsize{10pt}{10pt}\selectfont
\renewcommand{\arraystretch}{1.2} 
\begin{tabular}{@{}ll}
BH(s)    & black hole(s) \\
ccIMF    & cumulative composite IMF \\
CCSNe    & core-collapse supernovae \\
cIMF     & composite IMF \\
CMF      & core mass function \\
CR(s)    & cosmic ray(s) \\
DLA(s)   & damped Lyman $\alpha$ system(s) \\
ESA      & European Space Agency \\
ETG(s)   & early-type galaxy(ies) \\
FLMF     & filament line mass function \\
FMF      & filament mass function \\
GC(s)    & globular cluster(s) \\
GCE      & galactic chemical evolution \\
gIMF     & galaxy(-wide) IMF \\
GMC      & giant molecular cloud \\
GSMF(s)  & galaxy stellar mass function(s) \\
IGIMF    & integrated galactic IMF \\
IMF      & initial mass function \\
ISM      & interstellar medium \\
LMC      & Large Magellanic Cloud \\
LTE      & local thermodynamic equilibrium \\
MC(s)    & molecular cloud(s) \\
MSPs     & multiple stellar populations \\
MW       & Milky Way \\
PISN(e)  & pair-instability supernova(e) \\
Pop~III  & Population~III \\
SAMs     & semi-analytic models \\
SFH      & star formation history \\
SFR(s)   & star formation rate(s)\\
sIMF     & stellar IMF \\
SMC      & Small Magellanic Cloud \\
SN(e)    & supernova(e) \\
SN(e) Ia & type Ia supernova(e) \\
UFD(s)   & ultrafaint dwarf galaxy(ies) \\
\end{tabular}
\end{table}

\noindent {\bf{Affiliations}}
\par$^{6}$ European Southern Observatory, Karl-Schwarzschild-Straße 2, 85748 Garching bei M\"unchen, Germany
\par$^{7}$ National Institute for Astrophysics, Astronomical Observatory of Trieste, Via Tiepolo 11, 34143 Trieste, Italy
\par$^{8}$ IFPU - Institute for Fundamental Physics of the Universe, Via Beirut 2, 34151, Trieste, Italy
\par$^{9}$ School of Mathematical and Physical Sciences, 12 Wally’s Walk, Macquarie University, NSW 2109, Australia
\par$^{10}$ Argelander-Institut für Astronomie, Universität Bonn, Auf dem Hügel 71, D-53121 Bonn, Germany
\par$^{11}$ Max-Planck-Institute für Astrophysik, Karl-Schwarzschild-Straße 1, 85740 Garching bei München, Germany
\par$^{12}$ Department of Earth Sciences, National Taiwan Normal University, Taipei 116, Taiwan
\par$^{13}$ Center of Astronomy and Gravitation, National Taiwan Normal University, Taipei 116, Taiwan
\par$^{14}$ Physics Division, National Center for Theoretical Sciences, Taipei 106, Taiwan
\par$^{15}$ Department of Physics and Astronomy, University of Bologna, Via Gobetti 93/2, 40129 Bologna, Italy
\par$^{16}$ Department of Physics and Astronomy, University of Firenze, Via Sansone 1, 50019 Sesto Fiorentino, Firenze, Italy
\par$^{17}$ National Institute for Astrophysics, Astrophysical Observatory of Arcetri, Largo Enrico Fermi 5, 50125 Firenze, Italy
\par$^{18}$ School of Physics and Astronomy, Sun Yat-sen University, Daxue Road, 519082 Zhuhai, People's Republic of China
\par$^{19}$ CSST Science Center for the Guangdong-Hong Kong-Macau Greater Bay Area, 519082 Zhuhai, People's Republic of China
\par$^{20}$ School of Astronomy and Space Science, Nanjing University, 210093 Nanjing, People's Republic of China
\par$^{21}$ Key Laboratory of Modern Astronomy and Astrophysics, Nanjing University, 210093 Nanjing, People's Republic of China
\par$^{22}$ Astrophysics Research Institute, Liverpool John Moores University, 146 Brownlow Hill, Liverpool L3 5RF, UK
\par$^{23}$ Univ. Grenoble Alpes, CNRS, IPAG, 38000 Grenoble, France
\par$^{24}$ Departamento de Astronomía, Universidad de Chile, Camino El Observatorio 1515, Las Condes, Santiago, Chile
\par$^{25}$ Astronomy Section, Department of Physics, University of Trieste, Via Tiepolo 11, 34143 Trieste, Italy
\par$^{26}$ Leiden Observatory, Leiden University, PO Box 9513, 2300 RA Leiden, the Netherlands
\par$^{27}$ Department of Astrophysics, University of Vienna, T\"urkenschanzstra{\ss}e 17, 1180 Vienna, Austria
\par$^{28}$ Instituto de Astrof\'isica de Canarias, Calle V\'ia L\'actea s/n, 38205 La Laguna, Tenerife, Spain
\\

\noindent {\bf{Contributions}}\\
Tereza Jerabkova and Donatella Romano jointly led the coordination of this White Paper. They were responsible for organizing the workshop, inviting contributors, following up on inputs, and shaping the structure and content of the manuscript. In addition to writing several sections, they harmonized the text across different writing styles and ensured coverage of all key topics, including those beyond their immediate expertise. Their significant editorial and organizational efforts are reflected in their role as first and second authors.

Pavel Kroupa provided a major scientific contribution to the manuscript, offering key insights, literature context, and detailed input across several sections. His long-standing expertise on the topic was instrumental in framing the theoretical perspective throughout the paper.

Specific sections and paragraphs were contributed by Philippe Andr\'e, Martyna Chru{\'s}li{\'n}ska, Fabio Fontanot, Andrew Hopkins, Vikrant Jadhav, Natalia Lah\'en, Yueh-Ning Lee, Alessio Mucciarelli, Stefania Salvadori, Long Wang, and Zhiqiang Yan. These authors are listed alphabetically, as it is difficult to quantify individual contributions given the collaborative and iterative nature of the writing process. In some cases, substantial editorial work was performed to ensure consistency in tone and content.

Additional input, in the form of slides or brief text sections requiring significant integration and editing, was provided by Morten Andersen, Anna Durrant, Fabien Louvet, Mariya Lyubenova, Francesca Matteucci, Piyush Sharda, Glenn van de Ven, and Alexandre Vazdekis.

All authors reviewed and approved the final manuscript.

\begin{acknowledgements}
We thank Jiadong Li for providing the data for Fig.~\ref{fig:MLR}. We thank Jianrong Shi, Maosheng Xiang, and Mingjie Jian for valuable discussions on the observational uncertainties of stellar element abundance measurements. DR acknowledges the Italian National Institute for Astrophysics (INAF) for financial support to the project {\it ``An in-depth theoretical study of CNO element evolution in galaxies''} through Finanziamento della Ricerca Fondamentale, Fu.~Ob.~1.05.12.06.08. 
PK acknowledges support through the DAAD Bonn--Prague Exchange Programme at Bonn University.
TJ acknowledges support from the MUNI Award in Science and Humanities MUNI/I/1762/2023.
VJ thanks the Alexander von Humboldt Foundation for their support. AM and DR acknowledge financial support from the project {\it ``LEGO – Reconstructing the building blocks of the Galaxy by chemical tagging''} granted by the Italian MUR through contract PRIN2022LLP8TK\_001. FF acknowledges support from the Next Generation European Union PRIN 2022 20225E4SY5 {\it ``From ProtoClusters to Clusters in one Gyr''}. PS is supported by the Leiden University Oort Fellowship and the International Astronomical Union -- Gruber Foundation (TGF) Fellowship. SS acknowledges support by the ERC Starting Grant NEFERTITI H2020/804240 (PI: Salvadori) and thanks Ioanna Koutsouridou for providing Fig.~\ref{fig:PopIII_IMF}.
\end{acknowledgements}

\bibliographystyle{aa}
\bibliography{IMF_bibliography}

\end{document}